\documentclass{IEEEoj}

\usepackage{cite}
\usepackage{amsmath,amssymb,amsfonts}
\usepackage{algorithmic}
\usepackage{graphicx,color}
\usepackage{textcomp}

\usepackage{newtxtext,newtxmath}
    \ifCLASSOPTIONcompsoc
      \usepackage[caption=false,font=footnotesize,labelfont=sf,textfont=sf]{subfig}
    \else
      \usepackage[caption=false,font=footnotesize,labelfont=small]{subfig}
    \fi
    \usepackage[utf8]{inputenc}
    \usepackage{ragged2e}
    \usepackage{graphicx}
    \usepackage{tabulary}
    \usepackage{booktabs}
    \usepackage{mathtools}
    \usepackage{balance} 
    \usepackage{tabularx}
    \usepackage{comment}    
    \usepackage{setspace}
    \usepackage{xspace}
    \usepackage[nohyperlinks,nolist]{acronym}
    \usepackage{multirow}
    \usepackage{upgreek} 
    \usepackage{nicefrac} 
    \usepackage{ifthen}
    \usepackage{csvsimple}
    \usepackage{hyperref}
    \usepackage{flexisym} 
    \usepackage[caption=false,font=footnotesize]{subfig}

    \usepackage{titlecaps}
    \usepackage{optidef}
    \usepackage[thinc]{esdiff}
    \usepackage{soul}
    \usepackage{xltabular}
    \usepackage{longtable}
    \usepackage{cuted}
    \usepackage{xtab}
    \usepackage{afterpage}
    \usepackage{lscape}
    \usepackage[linesnumbered,ruled,vlined, noend]{algorithm2e}
    \usepackage{booktabs} 
    \usepackage{letltxmacro}
    \usepackage{array}
    \usepackage[capitalize,nameinlink]{cleveref}
    \usepackage{flushend}
    \usepackage{needspace}

    \crefname{constraint}{constraint}{constraints}
    \creflabelformat{constraint}{#2(#1)#3}
    
 \usepackage{xargs} 
    \usepackage{xcolor}
    \acrodef{5G-NR}{5G new radio}
    \acrodef{5GC}{5G Core}
    \acrodef{BS}{base station}
    \acrodef{C-RAN}{centralized RAN}
    \acrodef{CBR}{constant bit rate}
    \acrodef{CAPEX}{capital expenditure}
    \acrodef{COTS}{commercial-off-the-shelf}
    \acrodef{CU}{centralized unit}
    \acrodef{DU}{distributed unit}
    \acrodef{eMBB}{enhanced mobile broadband}
    \acrodef{FH}{fronthaul}
    \acrodef{FS}{functional splitting}
    \acrodef{GOPS}{giga operations per second}
    \acrodef{gNB}{Next Generation Node B}
    \acrodef{HARQ}{Hybrid Automatic Repeat Request}
    \acrodef{IoT}{Internet of Things}
    \acrodef{LTE}{Long Term Evolution}
    \acrodef{LTE-A}{LTE-Advanced}
    \acrodef{MAC}{medium access control}
    \acrodef{MH}{midhaul}
    \acrodef{MIMO}{multiple-input multiple output}
    \acrodef{MILP}{mixed-integer linear programming}
    \acrodef{MNO}{mobile network operators}
    \acrodef{mMTC}{massive machine-type communications}
    \acrodef{NGFI}{Next Generation Fronthaul Interface}
    \acrodef{NG-RAN}{Next Generation Radio Access Network}
    \acrodef{NR}{new radio}
    \acrodef{OPEX}{operational expediture}
    \acrodef{PP}{processing pool}
    \acrodef{RAN}{radio access network}
    \acrodef{RF}{Radio Frequency}
    \acrodef{RRC}{Radio Resource Control}
    \acrodef{RU}{radio unit}
    \acrodef{TSN}{Time Sensitive Networking}
    \acrodef{UE}{user equipment}
    \acrodef{URLLC}{ultra-reliable low latency communications}
    \acrodef{VC}{virtual configuration}
    \acrodef{VNF}{virtualized network function}
    \acrodef{XH}{crosshaul}
    \acrodef{EMOF}{energy minimization objective function}
    \acrodef{LMOF}{latency minimization objective function}
    \acrodef{BOMF}{bi-objective minimization function}
    \acrodef{HT}{Hierarchical topology}
    \acrodef{MT}{Mesh topology}
    
    \newcommand{\overall}[1][]{\mathcal{O}_{i,j}#1}
    \newcommand{\ppusage}[1][v']{\mathcal{C}_{#1}}
    \newcommand{\linkusage}{\mathcal{R}_{e}}
    \newcommand{\ignored}{\mathcal{N}}
    
    \newcommand{\unitt}{\upsilon}
    \newcommand{\ppaux}{pp_{load}}
    
    \newcommand{\segment}{segment}
    \newcommand{\paths}[1][cand]{R^{#1}}
    \newcommand{\pathh}[1][]{r^{#1}}
    \newcommand{\energy}[1][v]{W_{#1}}
    \newcommand{\rfh}{R^{FH}}
    \newcommand{\rmh}{R^{MH}}
    \newcommand{\pCU}{P^{CU}}
    \newcommand{\pDU}{P^{DU}}
    \newcommand{\pRU}{P^{RU}}
    \newcommand{\xVC}{X^{VC}}
    \newcommand{\placednode}{v^{placed}}
    \newcommand{\allocatedpath}{p^{allocated}}
    \newcommand{\bestpath}{p^{best}}
    \newcommand{\pathscore}{s^{score}}
    \newcommand{\gset}{\mathcal{G} = (\mathcal{V}, \mathcal{E})}
    \newcommand{\gmset}{\mathcal{G}} 
    \newcommand{\jmset}{\mathcal{D}}
    \newcommand{\imset}{\mathcal{I}}
    \newcommand{\kmset}{\mathcal{K}}
    \newcommand{\eset}{e\in \mathcal{E}}
    \newcommand{\jset}{j\in\mathcal{D}}
    \newcommand{\iset}{i\in \mathcal{I}}
    \newcommand{\jjset}{j, j' \in \mathcal{D}}
    \newcommand{\iiset}{i, i' \in \mathcal{I}}
    \newcommand{\kset}{k \in \mathcal{K}}
    \newcommand{\ksetDS}{k \in \mathcal{K}^{DS}}
    \newcommand{\ksetSH}{k \in \mathcal{K}^{SH}}
    \newcommand{\ksetSL}{k \in \mathcal{K}^{SL}}
    \newcommand{\zset}{z \in \mathcal{Z}}
    \newcommand{\vV}[1][v]{#1 \in \mathcal{V}}
    \newcommand{\vPP}[1][v']{#1 \in \mathcal{V}^{PP}}
    \newcommand{\vRU}{v'' \in \mathcal{V}^{RU}}
    \newcommand{\wSW}{w \in \mathcal{V}^{SW}}
    \newcommand{\mV}{\mathcal{V}}
    \newcommand{\mVP}{\mathcal{V}^{P}}
    \newcommand{\mVSW}{\mathcal{V}^{SW}}
    \newcommand{\mVPP}{\mathcal{V}^{PP}}
    \newcommand{\mVRU}{\mathcal{V}^{RU}}
    \newcommand{\mI}{\mathcal{I}}
    
    \newcommand{\mK}{\mathcal{K}}
    \newcommand{\linkC}{C_e}
    \newcommand{\compC}{CPP_{v'}}
    \newcommand{\latMax}[1]{l^{#1}_{i,j}}
    
    \newcommand{\demand}[1][i,j]{\lambda_{#1}}
    \newcommand{\APP}[1][v']{APP_{#1}}
    \newcommand{\ASW}{ASW_{w}}
    \newcommand{\VC}{VC_{i,j,k}}
    \newcommand{\CU}[1][v']{CU_{i,j,k,#1}}
    \newcommand{\DU}[1][v']{DU_{i,j,k,#1}}
    \newcommand{\RU}[1][v'']{RU_{i,j,k,#1}}
    \newcommand{\CD}[1][a,b]{CD_{i,j,k,z}^{#1}}
    \newcommand{\DR}[1][b,c]{DR_{i,j,k,z}^{#1}}
    \newcommand{\PFH}{P^{FH}_{i,j,k,e}}
    \newcommand{\PMH}{P^{MH}_{i,j,k,e}}
    \newcommand{\Pe}[1]{R^{#1}_{i,j,e}}
    \newcommand{\Pep}[1]{R^{#1}_{i',j',e'}}
    \newcommand{\Pz}{P_{i,j,k,z}}
    \newcommand{\Load}[1][]{L_{e}^{#1}}
    \newcommand{\PP}[1][']{PP_{v{#1}}}
    \newcommand{\latS}[1][]{SD^{#1}_{i,j}}
    \newcommand{\PPe}[1]{R_{i,i',j,j',e}^{#1}}
    \newcommand{\latD}[1][]{DD^{#1}_{i,j,e}}
    \newcommand{\latDH}[1][]{DD^{#1}_{i,j}}
    \newcommand{\lat}[1][]{D^{#1}_{i,j}}

    \newcommand{\PPqe}[1]{Y^{#1}_{i,j,e}}
    \newcommand{\PPqex}[1]{X^{#1}_{i,j,e}}

    \newcommand{\EPP}[1][v']{EPP_{#1}}
    
    \newcommand{\WSW}[1][]{W_{w}^{#1}}
    \newcommand{\ESW}[1][]{ESW_{w}^{#1}}
    \newcommand{\nlinc}{n_{w}^{linecard}}
    \newcommand{\nport}{n_{w}^{r_{x}}}
    \newcommand{\FF}{FH-FH}
    \newcommand{\MM}{MH-MH}
    \newcommand{\FM}{FH-MH}
    \newcommand{\MF}{MH-FH}
    
    \newcommand{\FQ}{FH-Q}
    \newcommand{\FSQ}{FH-SQ}
    \newcommand{\MQ}{MH-Q}
    \newcommand{\MSQ}{MH-SQ}
    
    \newcommand{\funTr}[1][]{\ensuremath{\tau^{#1} (\demand[i,j],k)}}
    \newcommand{\funLat}[1]{\ensuremath{\delta^{#1}(i,j,i',j',e)}}
    \newcommand{\funLatij}[1]{\ensuremath{\delta^{#1}(i,j,e)}}

    \newcommand{\funRot}[1]{\ensuremath{\phi(z,#1,e)}}
    \newcommand{\funPrc}[1][]{\ensuremath{\rho^{#1}(i,j,k,\unitt,\demand[i,j])}}
    \newcommand{\slat}[1][]{\delta^{#1}}
    \newcommand{\funSLat}{(\slat[SF](e) + \slat[prop](e))}
    \newcommand{\funSW}{\omega (z,v',u',w)}
    \newcommand{\sumI}{\sum_{i \in \mathcal{I}}}
    \newcommand{\sumJ}{\sum_{j \in \mathcal{D}}}
    \newcommand{\sumK}{\sum_{k \in \mathcal{K}}}
    \newcommand{\sumE}{\sum_{e \in \mathcal{E}}}
    
    \newcommand{\sumPP}[1][v']{\sum_{{#1} \in \mathcal{V}^{PP}}}
    \newcommand{\sumRU}{\sum_{{v''} \in \mathcal{V}^{RU}}}
    
    \newcommand{\sumZ}{\sum_{z \in \mathcal{Z}}}
    
    \newcommand{\pushright}[1]{\ifmeasuring@#1\else\omit\hfill$\displaystyle#1$\fi\ignorespaces}

    \usepackage{makecell}
    \usepackage{collcell}
    \usepackage{stfloats}
    \usepackage{siunitx}
    \usepackage{refcount}

    \newcolumntype{M}[1]{>{\hfil$\displaystyle}p{#1}<{$\hfil}}
    \newcommand{\stepequation}{\refstepcounter{equation}(\theequation)}
    \newcolumntype{E}{>{\stepequation}r}  

    \newcolumntype{Y}{>{\RaggedRight\arraybackslash}X}
    \newcolumntype{S}{>{\RaggedRight\arraybackslash}p{0.16\textwidth}} 

\def\BibTeX{{\rm B\kern-.05em{\sc i\kern-.025em b}\kern-.08em
    T\kern-.1667em\lower.7ex\hbox{E}\kern-.125emX}}
\AtBeginDocument{\definecolor{ojcolor}{cmyk}{0.93,0.59,0.15,0.02}}
\def\OJlogo{\vspace{-4pt}\hskip-4pt\includegraphics[height=18pt]{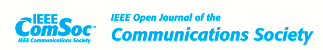}}
\begin{document}
\receiveddate{XX Month, XXXX}
\reviseddate{XX Month, XXXX}
\accepteddate{XX Month, XXXX}
\publisheddate{XX Month, XXXX}
\currentdate{3 July, 2026}
\doiinfo{OJCOMS.2024.011100}

\title{Energy-Latency Optimization for Dynamic Disaggregated Radio Access Networks}

\author{%
    Gabriela~N. Caspa~H.\IEEEauthorrefmark{1}, \IEEEmembership{(Student Member,~IEEE)}, Carlos~A.~Astudillo\IEEEauthorrefmark{1}, \IEEEmembership{(Member,~IEEE)}, AND Nelson~L.~S.~da~Fonseca \IEEEauthorrefmark{1} \IEEEmembership{(Senior Member,~IEEE)}
}
\affil{Institute of Computing, University of Campinas, Brazil}
\corresp{CORRESPONDING AUTHOR: Nelson L. S. da Fonseca (e-mail: nfonseca@ic.unicamp.br).}
\authornote{This work was supported by the São Paulo Research Foundation (FAPESP) under grant number 2023/00673-7, by the Brazilian Federal Agency for Support and Evaluation of Graduate Education (CAPES) under grant number 88882.329100/2014-0 and the National Council for Scientific and Technological Development under grant number 405940/2022-0 }
\markboth{Preparation of Papers for IEEE OPEN JOURNALS}{Caspa \textit{et al.}}

\begin{abstract}
In 5G networks, base station (BS) disaggregation and new services pose challenges for \ac{RAN} configuration, particularly in meeting bandwidth and latency constraints. The BS disaggregation is enabled by \ac{FS}, which distributes the RAN functions in processing nodes and alleviates latency and bandwidth requirements in the \ac{FH}. In addition to network performance, energy consumption is a critical concern for \ac{MNO}, since \acs{RAN} operations constitute a significant portion of their \ac{OPEX}. 
\acs{RAN} configuration optimization is essential for balancing service performance and cost-effective energy consumption.
In this paper, we propose a mixed-integer linear programming (MILP) model incorporating three objective functions: (\textit{i}) minimizing fronthaul (FH) latency, (\textit{ii}) minimizing energy consumption, and (\textit{iii}) a bi-objective optimization that jointly balances both latency and energy consumption. The model determines the optimal \acs{FS} Option, RAN function placement, and routing for \acs{eMBB}, \acs{URLLC}, and \acs{mMTC} slices.
While prior studies have addressed \acs{RAN} configuration from either an energy minimization or a latency reduction perspective, few have considered both aspects simultaneously in realistic scenarios. Our evaluation accounts for different topologies, accounts for variations in aggregated \acs{gNB} demand, explores diverse FS combinations, and incorporates \ac{TSN} modeling for latency analysis, which is critical for RAN performance. Given that \acs{MILP} execution can be computationally intensive, we propose a heuristic algorithm that adheres to RAN constraints while providing near-optimal solutions. The results reveal an inherent trade-off between latency and energy consumption, highlighting the need for dynamic RAN reconfiguration. These insights provide a foundation for optimizing existing and future RAN deployments.
\end{abstract}


\maketitle
\acresetall
\section{INTRODUCTION}


\IEEEPARstart{T}{he} growing demand for mobile traffic poses significant challenges to \acp{MNO}, which must continuously adapt their infrastructures to support heterogeneous services while maintaining operational efficiency. In fifth-generation (5G) mobile networks, three primary service categories are defined: \ac{eMBB}, \ac{URLLC}, and \ac{mMTC} \cite{ITU-T_GSup66_2020-services}. These services exhibit distinct performance requirements and traffic characteristics. Specifically, \ac{eMBB} targets high-throughput applications such as multimedia streaming and augmented reality, whereas \ac{URLLC} supports latency-sensitive and highly reliable services, including vehicular communications and e-health applications. In contrast, \ac{mMTC} is designed to accommodate a massive number of devices transmitting sporadic short packets, as commonly observed in Internet of Things deployments \cite{slices_applications}.

In addition to supporting heterogeneous services, energy-efficient operation of the \ac{RAN} has become a major concern for \acp{MNO}, since the RAN accounts for approximately 75\% of the total energy consumption in mobile networks and nearly 40\% of the associated \ac{OPEX} \cite{rohde2024_energy}. Furthermore, the adoption of wider channel bandwidths, massive multiple-input multiple-output (MIMO) technologies, and dense network deployments has intensified the growth in RAN power consumption \cite{ericsson2022BW}. Consequently, \acp{MNO} require energy-saving mechanisms capable of ensuring efficient utilization of computing and transport resources while preserving service performance requirements.

To improve operational efficiency and mitigate \ac{CAPEX} and \ac{OPEX}, the \ac{RAN} architecture has evolved considerably. In particular, the \ac{BS}, responsible for baseband and radio frequency (RF) processing, has undergone substantial architectural transformations. Early \ac{C-RAN} architectures centralized baseband processing functions within a \ac{PP}, aiming to reduce deployment costs, improve inter-cell coordination, and simplify network management \cite{ORAN_Survey}. In this architecture, the \ac{RF} module is connected to the  \acp{PP} through a mobile \ac{FH} network. However, the separation between \ac{RF} and lower physical-layer functions generates \ac{CBR} traffic, thereby increasing \ac{FH} bandwidth requirements even under low traffic loads. Additionally, stringent latency constraints are imposed on the \ac{FH}, typically limited to 250~$\mu$s due to \ac{HARQ} timing requirements \cite{3GPP_801}.

To alleviate these limitations, 3GPP introduced multiple \ac{FS} options that enable functional disaggregation of the 5G \ac{BS}, referred to as the \ac{gNB}, into three logical entities: the \ac{RU}, \ac{DU}, and \ac{CU}. In this architecture, the \ac{FH} network interconnects the \ac{RU} and \ac{DU}, while the \ac{MH} network connects the \ac{DU} and \ac{CU} \cite{nath2023mastering5g}. Such disaggregation is enabled by virtualization technologies, where the gNB protocol stack is implemented as \acp{VNF} running on commercial-off-the-shelf hardware deployed either at the network edge or in centralized \acp{PP} \cite{wong2023openran}. Meanwhile, the
\ac{RU} remains at the cell site and performs the \ac{RF} processing chain using specialized hardware. The 3GPP standard specifies eight \ac{FS} options, including \ac{C-RAN} as Option~8. The remaining split options provide additional flexibility by: (\textit{i}) adapting \ac{FH} and \ac{MH} traffic loads according to \ac{gNB} demand while relaxing latency constraints; (\textit{ii}) enabling RAN functions to be distributed across multiple \acp{PP}; and (\textit{iii}) allowing a single \ac{PP} to host functions belonging to multiple \acp{gNB} \cite{Matoussi_computing}.

Moreover, latency is a critical requirement in disaggregated \ac{RAN}
deployments. Meeting stringent latency demands generally increases
deployment costs, since processing and transport resources must be placed closer to end users. To address these challenges, packet-switched transport based on Ethernet frame encapsulation has been proposed for disaggregated RAN architectures \cite{IEEE_1914_packet_fronthaul}. This
approach enables the convergence of \ac{FH} and \ac{MH} networks into a unified \ac{XH} transport layer. Within this context, \ac{TSN} provides deterministic low-latency communication capabilities in the \ac{XH} by prioritizing traffic flows according to their service classes \cite{IEEE_TSN_802.1c_2020}. Moreover, packet-based \ac{XH} architectures reduce deployment costs by eliminating the need for dedicated point-to-point fiber links and enabling the reuse of existing Ethernet infrastructures \cite{TSN_ORAN_Garcia}.



MNOs must support heterogeneous services with time-varying demands while reducing \ac{OPEX}, making energy minimization a key objective. In this context, a disaggregated RAN provides the flexibility required to adapt to traffic fluctuations and improve resource utilization efficiency. Such adaptation involves FS selection, the activation and deactivation of PP and network nodes, PP placement, and routing decisions subject to computing, network capacity, and service constraints. These interdependent decisions characterize a RAN configuration problem, which is known to be NP-hard \cite{singh_infocom_nphard}. Moreover, this formulation requires realistic models for computing, energy consumption, and latency performance, which are critical aspects of RAN operation.

Prior studies have addressed the RAN configuration problem from multiple perspectives, including cost reduction, spectrum efficiency, and service admission control \cite{SlicedRAN_Service, Sulaiman_coordinated}. Energy consumption and latency are often analyzed independently or through simplified models \cite{Apt-RAN, klinkowsi2020}. Several works focus on \ac{OPEX} reduction through \ac{FS} selection or \ac{VNF} placement; however, the joint consideration of energy consumption across both \acp{PP} and transport elements remains limited \cite{Moreira, Pires_energy}. Moreover, latency modeling frequently relies on fixed or average values, while queuing delays and routing decisions receive limited attention \cite{DRL_reconf}. Although \ac{MILP}-based formulations can provide optimal solutions, they are computationally demanding and are typically evaluated on small-scale topologies \cite{Coelho}. Consequently, heuristic and AI-based approaches have emerged as practical alternatives for larger-scale scenarios \cite{BiObjective_Pires_2022, Constrained_FS}.

Although more comprehensive formulations, such as \cite{klinkowsi2023}, integrate several of these aspects, including detailed latency models and node deactivation mechanisms, they neglect slicing and energy-related considerations. Overall, the literature still lacks scalable approaches capable of jointly modeling \ac{PP} and transport energy consumption together with realistic latency components in a disaggregated sliced-RAN environment.


In contrast to previous studies, this work addresses the RAN configuration problem by minimizing energy consumption while satisfying heterogeneous service requirements in a 5G sliced-disaggregated \ac{RAN}. Decisions regarding FS selection, function placement, transport routing, and node activation directly influence latency performance, resource utilization, and overall energy consumption under time-varying traffic demands. As a result, a trade-off emerges between energy efficiency and latency performance, which is a critical consideration for MNOs seeking to reduce \ac{OPEX} while meeting service requirements.

The novelty of this work lies in the unified treatment of these interdependent decisions, jointly capturing energy consumption in both \acp{PP} and transport elements while accounting for latency in heterogeneous sliced-RAN scenarios. The main contributions of this work are summarized as follows:

\begin{itemize}

\item An \ac{MILP} formulation that jointly performs \ac{FS} selection, function placement, node activation, and \ac{XH} routing under computing and network constraints. The formulation incorporates deterministic latency estimation together with energy consumption models for \acp{PP} and transport equipment. The execution time remains manageable for medium-sized topologies, although it increases with network size, particularly with the number of edges. Furthermore, the formulation extends \cite{klinkowsi2023} by incorporating network slicing for heterogeneous services and explicitly modeling the energy--latency trade-off.


\item A heuristic approach designed to reduce computational complexity and execution time, thereby enabling the evaluation of larger-scale scenarios.
    

\item A comprehensive performance evaluation of both approaches in realistic scenarios, considering multiple \ac{FS} combinations and diurnal variations in \ac{gNB} traffic demand. The analysis highlights the energy--\ac{FH}-latency trade-off and investigates the variability of \ac{VC} selection across use cases and its impact on resource utilization.

\end{itemize}


The remainder of this article is organized as follows. Section \ref{sec:related_works} reviews the related literature. Section \ref{sec:problem_scenario} describes the system model, discusses the characteristics of \acs{FS} combinations as virtual configurations, and presents the computing and latency models adopted in the \ac{MILP} formulation. Sections \ref{sec:milp} and \ref{sec:heuristic} detail the proposed \ac{MILP} formulation and heuristic approach, respectively. Section \ref{sec:Results} presents the experimental results, and Section \ref{sec:Conclusion} concludes the article. \Cref{tab:acronyms} compiles the main acronyms used throughout the manuscript. 

\begin{table*}
    \centering
    \caption{List of Acronyms}
    \begin{tabular}{ll|ll}
         \hline
    \textbf{Acronym} & \multicolumn{1}{c}{\textbf{Meaning}} &
    \textbf{Acronym} & \multicolumn{1}{c}{\textbf{Meaning}} \\
    \hline

5G    & Fifth-generation &
5GC   & 5G Core \\

BILP  & Binary Integer Linear Programming &
BS    & Base station \\

CAPEX & Capital expenditure &
CBR   & Constant bit rate \\

COTS  & Commercial-off-the-shelf &
C-RAN & Centralized RAN \\

CU    & Centralized unit &
DU    & Distributed unit \\

eMBB  & Enhanced mobile broadband &
EMOF  & Energy minimization objective function \\

FH    & Fronthaul &
FS    & Functional splitting \\

gNB   & Next Generation Node B &
HARQ  & Hybrid Automatic Repeat Request \\

HPF   & High-Priority Flow &
H-PHY & High Physical \\

HT    & Hierarchical topology &
LMOF  & Latency minimization objective function \\

LPF   & Low-Priority Flow &
L-PHY & Low Physical \\

MILP  & Mixed-integer linear programming &
MIMO  & Massive multiple-input multiple-output \\

MNO   & Mobile network operators &
MPF   & Medium-Priority Flow \\

MSE   & Mean squared error &
mMTC  & Massive machine-type communications \\

MT    & Mesh topology &
NGFI  & Next-Generation Fronthaul Interface \\

OPEX  & Operational expenditure &
O-RAN & Open RAN \\

PHY   & Physical &
PP    & Processing pool \\

RAN   & Radio access network &
RF    & Radio frequency \\

RRC   & Radio Resource Control &
RU    & Radio unit \\

SCF   & Small Cell Forum &
TSN   & Time Sensitive Networking \\

UE    & User equipment &
URLLC & Ultra-reliable low latency communications \\

VC    & Virtual configuration &
VNF   & Virtualized network function \\

XH    & Crosshaul &
      & \\
    \hline
    \end{tabular}
    \label{tab:acronyms}
\end{table*}

\section{RELATED WORKS}
\label{sec:related_works}

The literature on RAN configuration has addressed various issues, including OPEX minimization, service admission control, spectrum utilization, energy consumption, and latency. In this section, we review and classify related work according to FS selection, slice resource optimization, and RAN energy and latency management. A summary is provided in \cref{tab:literature_review}, highlighting the key aspects examined in this article.

\subsection{SELECTION OF FUNCTIONAL SPLITTING}

 Alba \textit{et al}. \cite{AnAdaptive_INFOCOM} present one of the earliest works on FS adaptation at runtime.  They highlight the need, objectives, and challenges of dynamic FS selection, emphasizing the limitations of static configurations. The authors implement an adaptive RAN that dynamically adjusts the split and evaluate its impact on packet loss and delay. However, the study considers only two FS Options in a simplified architecture and does not address RAN slicing. Next, focusing on the viability and profitability of the split adaptation, the authors of \cite{DFS_adaptation} propose an MILP to select the optimal FS of a given RAN network. They also introduce a cost-of-flexibility framework to assess the overhead of FS adaptation and propose strategies for deciding when reconfiguration is needed. However, this work does not incorporate network slicing, service-specific requirements, or RAN energy consumption.

The work in \cite{PlaceRAN} and \cite{DRL_Almeida_2022} addresses the joint problem of FS selection, RAN function placement, and path selection using Binary Integer Linear Programming (BILP) and deep reinforcement learning, respectively. Both approaches aim to reduce RAN computing utilization while maximizing centralization. They consider single- and dual-split options for RAN function placement. The deep reinforcement learning approach reduces computational time relative to the optimal BILP solution.

Conversely, Amiri \textit{et al}. \cite{Opt_Load_ORAN} employs FS selection to balance CU computing load and FH transmission load in an O-RAN architecture, aiming to minimize operational costs. They formulate the problem as NP-hard and solve it using a heuristic. Although dynamic FS is considered, energy and latency are not analyzed in detail.

\subsection{RAN SLICING}
FlexRAN \cite{FlexRAN} and Orion \cite{Orion} enable RAN slicing using software defined networking and virtualization. FlexRAN provides real-time hierarchical control, whereas Orion virtualizes radio, computing, and spectrum resources via a hypervisor. Neither framework addresses FS selection nor considers latency or energy consumption.

In \cite{DedicatedPath}, a resilient slice-embedding strategy for URLLC services is proposed, using Dedicated Path Protection and functional splitting. The heuristic minimizes physical resource allocation while ensuring reliability in the presence of failures. Multiple FS options are considered, but the approach focuses exclusively on URLLC and does not address energy or latency optimization.

Ojaghi and colleagues \cite{SlicedRAN_join, SlicedRAN_Service, SO-RAN, MGZ} explore FS and slicing using MILP and heuristics. \cite{SlicedRAN_join, SlicedRAN_Service} aim to maximize network throughput while meeting heterogeneous slice requirements, including customized FS per slice, unit placement, and routing under SLA constraints. In \cite{SO-RAN}, MEC placement is incorporated to provide edge services. In \cite{MGZ}, a heuristic evaluates joint slicing and FS in dense networks. However, the RAN architecture is simplified (DU and RU co-located), and energy consumption is not considered.

In \cite{Coelho}, the authors propose an MILP-based framework that accounts for slice design across different service types and dynamic FS. They focus on the impact of slice isolation levels in the physical network and map new isolation and SLA constraints that influence the FS decision. They seek to minimize the number of active nodes, links, and reconfigurations by evaluating various FS and slice-sharing policies. Although slicing and dynamic FS are addressed, the work does not address latency and energy.

Sen et al. \cite{Sen_slicing} develop an MILP to maximize the RAN centralization degree and reduce the number of processing nodes. They also propose a heuristic as an alternative with lower computational complexity. Their approach selects a specific FS and performs unit placement per slice. The slices provide 5G services such as eMBB, URLLC, and mMTC, but their specific latency and bandwidth requirements are unspecified, and RU configurations remain fixed per 4G standards.
Moreover, latency analysis is simplified by using fixed values for the FH and MH delays.

In \cite{Sulaiman_Multi_Agent}, the authors address the slice-admission control problem to maximize the RAN long-term revenue. Their work involves multi-agent deep reinforcement learning that accepts slice requests based on the availability of computing, bandwidth, and transmission resources, while considering service-specific requirements. Each slice is assigned an FS Option among the standardized FS Options 2, 6, and 7. 
The approach is compared against greedy and node-ranking heuristics, as well as a single-agent method. Latency analysis is simplified, and energy consumption is not addressed.

\subsection{ENERGY AND LATENCY IN RANs}
In \cite{Impact_delay}, FS selection, BBU server allocation, and scheduling are jointly optimized to minimize end-to-end latency in C-RAN.
The model uses an M/D/1 queue to represent BBU processing with deterministic job sizes dependent on the FS, and includes delay contributions from BBU processing, FH transport, and RRH computation. Two objectives are studied: minimizing the total average delay and the maximum average delay across streams. 
While FS flexibility and detailed delay modeling are considered, slicing and energy aspects are excluded.

In \cite{OGREEN_pamuklu}, the authors apply reinforcement learning to minimize RAN OPEX by dynamically selecting both the functional split and energy source. The approach reduces CU and DU energy consumption by leveraging renewable energy and adapting the FS to daily traffic variations. It considers the requirements of eMBB and URLLC slices and assigns an FS to each. However, latency is not analyzed, and the model assumes that the DU and RU are always co-located at the same site and that the gNB placement is already given.

In \cite{Constrained_FS}, the authors propose a constrained deep reinforcement learning model to minimize DU and CU energy consumption by dynamically adapting FS per RU in a virtualized RAN. However, slicing is not explicitly addressed, and latency is not studied in detail. As a continuation of this previous work, the authors in \cite{DRL_reconf} develop a framework that enables dynamic vRAN reconfiguration on-the-fly. They work on an O-RAN architecture and focus on reconfiguring several aspects of the RAN: functional split selection, resource allocation, unit placement, gNB association, and routing. The goal is to minimize the long-term total network cost. Yet, these works neither consider slicing requirements nor account for latency details.

Klinkowski addressed the RAN planning problem using mixed-integer linear programming, jointly considering latency and computational constraints in \cite{klinkowsi2020}, \cite{klinkowsi2023}. In \cite{klinkowsi2020}, the study focuses on the placement of CUs and DUs, while the routing problem is simplified by assuming a single candidate path. Subsequently, in \cite{klinkowsi2023}, an MILP formulation and a simulated annealing–based heuristic are proposed to jointly minimize the number of active processing nodes and the FH latency. The proposed approaches determine CU, DU, and RU placement, as well as path selection. For each RU, the functional configuration allows for either a single or a dual split, employing FS Options 7 or 2. Delay analysis is conducted based on TSN queuing models. However, their proposed optimization model does not account for network slicing requirements associated with 5G heterogeneous use cases, nor the energy consumption of computing and network components.

In \cite{Moreira}, the authors propose an MILP and a heuristic algorithm to minimize RAN power consumption. These algorithms choose an FS Option between Options 2 and 7, perform unit placement, and select paths. The MILP is computationally intensive even for medium-sized topologies, whereas the heuristic achieves near-optimal results with significantly shorter solving times. They consider PP power consumption but ignore network power consumption. Additionally, a brief latency analysis is included from the work \cite{klinkowsi2020}; however, latency is not explored. Slices are considered, but the restrictions of the 5G use cases are not taken into account.   

Pires et al. \cite{Pires_energy} optimize RAN energy consumption, accounting for PPs, network equipment, and VNF migration using MILP and heuristics. Single- and dual-FS architectures are considered, but only single-FS cases are evaluated. eMBB and URLLC traffic profiles are modeled, but slicing and latency optimization are simplified.

In \cite{Akhtar}, the authors propose a two-step approach to minimize both CAPEX and OPEX. The cost model includes backhaul connectivity, fronthaul deployment, energy consumption of processing nodes, network maintenance, and BBU pool rental. They first propose a recursive algorithm to obtain candidate processing pools and optical aggregators, and then use a MILP to determine optimal unit placement and routing. They also analyze the G/G/1 queuing model for RAN delay. However, a fixed FS is assumed (Option 7), and the energy consumption of network equipment and slicing is overlooked.

The work in \cite{Apt-RAN} presents an MILP that jointly minimizes CU power consumption and DU handovers. A heuristic is also proposed to reduce computational time. In the proposed model, DU is assumed to be co-located with RU, and a single FS is employed.
Although it is among the first works to consider RAN energy consumption, it focuses only on CU power consumption and ignores energy models for DU and RU. Besides that, latency and slicing are omitted.

\begin{table*}[!t]
    \centering
    \caption{Literature review}
    \footnotesize
    \renewcommand{\arraystretch}{1.2} 
    \begin{tabular}{
        m{0.07\linewidth}  m{0.14\linewidth}  m{0.17\linewidth}  m{0.15\linewidth}  
        m{0.05\linewidth}  m{0.13\linewidth}  m{0.03\linewidth}  m{0.05\linewidth} 
    }
    \toprule
    \makecell[c]{\textbf{Article}} & 
    \makecell[c]{\textbf{Approach}} & 
    \makecell[c]{\textbf{Objective}} & 
    \makecell[c]{\textbf{Latency}} & 
    \makecell[c]{\textbf{Energy}} & 
    \makecell[c]{\textbf{5G}} & 
    \makecell[c]{\textbf{DFS}} & 
    \makecell[c]{\textbf{Slicing}} \\
    \midrule

    \cite{Moreira} & MILP, Heuristic & Min. PP power consumption & Static, queuing delay & PP, DC & No info & Yes & No \\
    \hline 
    \cite{Pires_energy} & MILP, Heuristic & Min. PP, TN, migration energy & Fixed values & PP, TN, mig. & No info & Yes & No \\
    \hline
    \cite{klinkowsi2020}, \cite{klinkowsi2023} & MILP, Heuristic & Min. active PP \& FH latency & Static, queuing delay & No & 8 MIMO, 32 antenna, 100 MHz & No & No \\
    \hline 
    \cite{SO-RAN}, \cite{SlicedRAN_Service}, \cite{MGZ} & MILP, Benders decomposition algorithm & Max. throughput \& Min. slice cost & Processing, transmission, propagation & No & 100 PRBs, 20 MHz, 2x2 MIMO, 28 MCS & Yes & Yes \\
    \hline
    \cite{Akhtar} & MILP, Recursive clustering algorithm & Min. energy consumption and maintenance cost & Queuing G/G/1 & PP, RU & 10, 20, 40, and 100 MHz & No & No \\
    \hline
    \cite{DFS_adaptation}, \cite{AnAdaptive_INFOCOM} & MILP & Min. OPEX \& Max. spectral efficiency & Propagation, processing & No & No info & Yes & No \\
    \hline
    \cite{DRL_reconf}, \cite{Constrained_FS} & DRL & Min. OPEX (routing and computing) & Fixed values & No & SISO, 10 MHz LTE, 36.6 Mbps & Yes & Yes \\
    \hline
    \cite{Apt-RAN} & MILP, Heuristic & Min. power consumption \& handover & No & CU & No info & No & No \\
    \hline
    \cite{Gao_DRL} & DRL & Min. OPEX: routing, computing & Static & No & 100 MHz spectrum, 4×4 MIMO, 16QAM & Yes & No \\
    \hline
    \cite{DRL_Almeida_2022} & DRL & Min. computing usage, Max. centralization & Fixed values & No & No info & Yes & No \\
    \hline
    \cite{Sulaiman_Multi_Agent} & DRL & Max. long-term revenue & Fixed values & No & 20 MHz, 2x2 MIMO, 64QAM & No & Yes \\
    \hline
    \cite{BiObjective_Pires_2022} & MILP, e-constraint, Pareto & Min. energy consumption - Max. RAN centralization & Fixed values & PP & No info & No & No \\
    \hline
    \cite{OGREEN_pamuklu} & RL & Min. energy consumption & No & PP & No info & Yes & Yes \\
    \hline
    \cite{Coelho} & MILP & Min. CAPEX (slice creation cost) & Fixed values & No & No info & Yes & Yes \\
    \hline
    \cite{Koutsopoulos_queuing} & MILP, Heuristic & Min. 2E2 Latency & Queuing M/D/1, D/D/1 & BBU & No info & No & No \\
    \hline
    \cite{Opt_Load_ORAN} & MILP, Heuristic & MH/FH Load balance & Static & No & 1 user/TTI, 2×2 MIMO, 20 MHz & Yes & No \\
    \hline
    This work & MILP, Heuristic & Min. energy consumption & Queuing \& self-queuing & PP, TN & 100 MHz spectrum, 8×8 MIMO, 64QAM & Yes & Yes \\
    \hline
    \multicolumn{8}{p{251pt}}{\footnotesize \textbf{DFS}: Dynamic Functional Split, \textbf{TN}: Transport Network } \\
    \end{tabular}
    \label{tab:literature_review}
\end{table*}

\subsection{LIMITATIONS OF EXISTING WORK}
Several studies propose MILP formulations that incur high computational complexity and long execution times. To alleviate this limitation, heuristic techniques and learning-based approaches, including deep reinforcement learning, are frequently employed to obtain near-optimal solutions with reduced computational overhead. In some works, the MILP formulation is presented primarily for completeness, while performance evaluation is limited to heuristic solutions. In contrast, our proposal yields optimal solutions with significantly reduced computational time for small- and medium-scale network topologies and introduces a scalable heuristic to efficiently address large-scale scenarios.

Energy minimization in RAN has been widely investigated to reduce \ac{OPEX} costs. However, a large body of existing work focuses primarily on the activation and deactivation of processing nodes, often overlooking the energy consumption of transport network elements such as routers and links. Moreover, some studies omit explicit node energy modeling or provide limited details regarding the adopted computing models and their adaptation to different FSs. Unlike existing literature, the proposed approach jointly considers node activation and deactivation decisions together with processing and transport energy consumption. The computing model employed in this work is designed to be transparent and FS-aware, as it characterizes the computational requirements of each VNF and adapts them according to the selected FS in a disaggregated RAN architecture.

Latency is a key performance metric in RAN design, as access networks must satisfy stringent timing constraints to meet service-level requirements. Nonetheless, several prior studies simplify latency modeling by assuming fixed delay values or considering only propagation delay, thereby neglecting queuing effects that directly impact FS selection. In this work, a detailed latency model based on TSN principles is adopted for a packet-based XH network. The proposed model explicitly captures queuing dynamics and frame prioritization mechanisms. In addition, the resulting FS selection outcomes are analyzed to identify the interplay between latency constraints and power consumption, an aspect that has received limited attention in the existing literature.

Additionally, most existing solutions are evaluated on small-scale network topologies and homogeneous 4G radio configurations, such as fixed channel bandwidth (20~MHz), modulation scheme (64-QAM), and antenna configuration (2$\times$2 MIMO). However, practical 5G deployments are inherently heterogeneous \cite{rochman2024-real_implementation}. Accordingly, this work evaluates both the proposed MILP formulation and the heuristic approach over larger-scale scenarios with diverse 5G radio configurations and realistic-behavior demand.

Existing studies generally address energy minimization, latency modeling, \acs{FS} selection, or slice planning independently. However, practical 5G deployments exhibit strong interdependencies among network slicing, disaggregated RAN architectures, heterogeneous service requirements, and \acs{XH} latency constraints, which jointly determine computing and transport resource utilization and overall energy consumption. Consequently, isolated analyses provide limited insight into the complex trade-offs governing realistic network operation.

In contrast to previous approaches, this work formulates a joint optimization problem for sliced and disaggregated RANs that simultaneously minimizes energy consumption and \acs{FH} latency. The proposed approach jointly optimizes \acs{FS} selection, dynamic node activation and deactivation, processing and transport energy consumption, and \acs{TSN}-aware latency within a unified formulation. It further incorporates realistic deployment scenarios and traffic dynamics to evaluate the impact of network reconfiguration on resource utilization and service-level compliance. By jointly considering these interdependent factors, the proposed formulation enables \acs{MNO}s to identify energy-efficient RAN configurations that satisfy the stringent latency and bandwidth requirements of heterogeneous network services.

\section{SYSTEM MODEL}
\label{sec:problem_scenario}

\begin{figure*}[!t]
    \centering
    \includegraphics[width=1\linewidth]{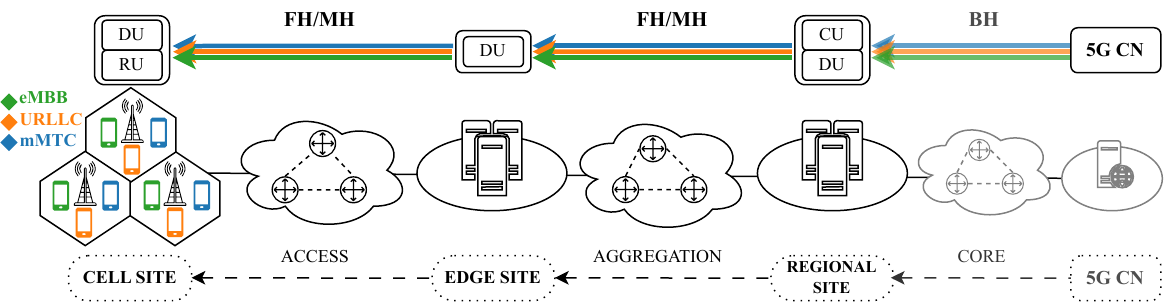}
    \caption{Employed RAN architecture}
    \label{fig:system_model}
\end{figure*}


Modern mobile networks demand a flexible RAN capable of supporting heterogeneous use cases while reducing operational expenses, particularly those associated with energy consumption. This objective can be achieved through a reconfigurable RAN that dynamically adapts to varying traffic demands, thereby improving the utilization of network and computing resources and avoiding over- or under-provisioning. RAN flexibility encompasses FS selection, function placement, and routing. FS selection is a critical configuration, as it directly affects XH bandwidth and latency requirements. Similarly, VNF placement and PP selection can reduce energy consumption by minimizing the number of active nodes. In contrast, strategically placing RAN functions closer to the user equipment (UE) and selecting efficient paths can reduce latency, particularly for FH or URLLC traffic.
To jointly address service heterogeneity and resource efficiency, a sliced RAN is an effective paradigm for concurrently serving eMBB, URLLC, and mMTC services, in which each slice is provisioned with a dedicated configuration to meet its specific performance requirements.


The proposed system model follows the Open RAN (O-RAN) and Next-Generation Fronthaul Interface (NGFI) architectures. It consists of a set of PP sites interconnected through a packet-switched transport network that supports RAN slicing for the three aforementioned service categories.

The PP infrastructure is organized into three layers: regional sites and edge sites equipped with high-capacity PPs, and cell sites, where low-capacity PPs are co-located with the RF equipment, as illustrated in \Cref{fig:system_model}. Regional sites interface with the fifth-generation core (5GC) and are geographically farther from the UEs. In this work, CU placement is restricted to nodes directly connected to the 5GC. Regional and edge sites can host multiple disaggregated gNBs by executing CU or DU functions as VNFs. Cell-site servers are dedicated to RF and baseband processing by hosting the RU. Each cell site supports a single RU serving three slices, namely eMBB, URLLC, and mMTC, where each slice is independently configured and its traffic is modeled as a distinct flow. These flows traverse a transport network composed of TSN routers that carry slice traffic across the PP layers from the 5GC to the cell site in downlink communication.

The analysis begins at the 5GC, which connects to the CU through the backhaul, assumed to be a direct link. Subsequently, the MH, typically located in the aggregation network, connects the CU to the DU through the F2 interface. The FH corresponds to the access network between the DU and RU and employs the eCPRI interface. Depending on the selected VC, either the MH or FH segment may be omitted. Finally, the XH relies on packet-switched routers interconnected through high-capacity fiber links.


\subsection{VIRTUAL CONFIGURATION}
The 3GPP specification in \cite{3GPP_801} defines eight FS options in which the RAN protocol stack is distributed as a chain of functions executed across different PPs. These options relax the stringent transport constraints of early C-RAN deployments based on FS Option 8. In this configuration, the split occurs between the Physical and RF functions, requiring the transmission of digitized in-phase and quadrature (IQ) signals. Consequently, the FH traffic exhibits a CBR, even under low-load conditions, while demanding strict synchronization and signal integrity within a latency budget of 250~$\mu$s.

In practice, deployments have converged to FS Options 2, 6, and 7 \cite{ORAN-WG9_XHaul_transport}.
FS Option 2 splits PDCP and RLC: PDCP performs header compression and security, then forwards PDCP PDUs to RLC for segmentation, concatenation, and reordering to match MAC requirements.
FS Option 6 defines a split between the MAC and the High-Physical (H-PHY) functions. In this option, MAC performs scheduling, resource allocation, and HARQ procedures, and forwards physical resource blocks (PRBs) to H-PHY. Then, H-PHY executes channel coding and modulation mapping for over-the-air transmission. Option 6 follows the Small Cell Forum (SCF) split and is being considered by the O-RAN Alliance for standardization due to its lower XH bandwidth requirement \cite{FS6_ORAN}.

O-RAN further refines 3GPP FS Option 7.2 into FS Option 7.2x. While Option 7.2 scales XH bandwidth with the number of antennas, Option 7.2x scales it with the number of required streams \cite{rawat_FS7.2x}. Table \ref{tab:XH_requirements} summarizes the bandwidth and latency requirements based on \cite{3GPP_801,ORAN-WG9_XHaul_requirements}. Options 2 and 6 tolerate higher latency because their procedures are less time-critical, whereas HARQ retransmissions impose a deadline of up to 4 ms.

\begin{figure}
    \centering
    \includegraphics[width=1\linewidth]{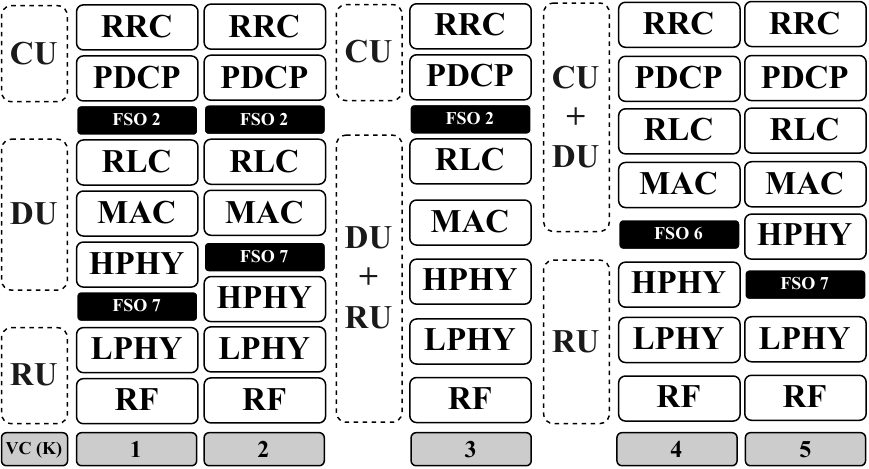}
    \caption{Virtual Configuration Options}
    \label{fig:VC_options}
\end{figure}

The proposed architecture supports both single- and dual-split deployments through the VC options (VCOs) illustrated in \Cref{fig:VC_options}. Each VCO assigns one or two FS options to each disaggregated gNB, thereby determining the number of PP nodes required to host the CU, DU, and RU functions. In all VCOs, the RU remains at the cell site, as it must be physically co-located with the antenna. In VCOs 1 and 2, which correspond to dual-split deployments, the CU and DU are hosted on separate \acp{PP}, resulting in both MH and FH transport segments. In these configurations, the CU--DU split adopts FS Option~2, whereas the DU--RU split employs either FS Option~7.2x or FS Option~6. Single-split VCOs eliminate one transport segment. In VCO~3, the DU and RU are co-located at the cell site and connected to the CU through the MH using FS Option~2. In VCOs~4 and~5, the CU and DU are co-located at regional PPs and connected to the RU through the FH using FS Option~6 and FS Option~7.2x, respectively.

\begin{table}[!t]
    \centering
    \caption{Bandwidth and Latency XH Requirements per FS }
    \begin{tabular}{ccc}
    \toprule
        \textbf{FS} & \textbf{Bandwidth}  & \textbf{Latency} \\
    \midrule
        2 & $\lambda$ & 10 $ms$\\
        6 & $\lambda + CR$ & 250 $\mu s$ \\
        7.2x & $2 \cdot 10^{-9}(1 + c) \cdot  \frac{v_{NL} \cdot PRB \cdot(12N_{mant}+N_{ex})}{Ts^{\mu}}  $ & 250 $\mu s$\\
    \bottomrule
        \multicolumn{3}{m{225pt}}{\footnotesize
    $\lambda$: downlink demand, $CR$: control/schedule signaling rate, $c$: control overhead, $v_{NL}$:supported layers, $PRB$: number of PRBs, $N_{mant}$: mantissa bits, $N_{ex}$: exponent bits, $Ts$: OFDM symbol duration.}   
    \end{tabular}
    \label{tab:XH_requirements}
\end{table}


\subsection{COMPUTING MODEL}
We adapt the models proposed in \cite{Sulaiman_Multi_Agent} and \cite{Wang_computing} to estimate RAN computing demands under FS, following the approach in \cite{Chun_computing}. Each function is virtualized, and its computing requirements are measured in Giga Operations Per Second (GOPS):

\begin{equation}
    C_{RF} = C_{ref} \cdot \frac{BW}{BW_{ref}}\cdot \frac{A}{A_{ref}}
    \label{eq:comp_rf}
\end{equation}

\begin{equation}
    C_{LPHY} = C_{ref} \cdot \frac{BW}{BW_{ref}}\cdot \frac{A}{A_{ref}}\cdot \frac{L}{L_{ref}}
\end{equation}

\begin{equation}
    C_{HPHY} = C_{ref}\cdot \frac{BW}{BW_{ref}} \cdot \left( \frac{A}{A_{ref}}\right)^2\cdot \frac{L}{L_{ref}}
\end{equation}

\begin{equation}
    C_{MAC} = C_{ref} \cdot \frac{BW}{BW_{ref}}\cdot \frac{A}{A_{ref}} \cdot \frac{M}{M_{ref}}\cdot \frac{L}{L_{ref}}
\end{equation}

\begin{equation}
    C_{RLC} = C_{ref} \cdot \frac{A}{A_{ref}}
\end{equation}

\begin{equation}
    C_{PDCP} = C_{ref} \cdot \frac{A}{A_{ref}}
\end{equation}

\begin{equation}
    C_{RRC} = C_{ref} \cdot \frac{A}{A_{ref}}
    \label{eq:comp_rrc}
\end{equation}

The models incorporate the reference computational load per function, denoted by $C_{ref}$, measured in GOPS, as defined in \cite{Debaille_computing}. The parameter $BW$ represents the carrier bandwidth in MHz, $A$ denotes the number of MIMO antennas, $L$ corresponds to the user traffic demand, and $M$ represents the modulation scheme per RU. Similarly, $BW_{ref}$, $A_{ref}$, $L_{ref}$, and $M_{ref}$ denote the corresponding reference values provided in \cite{Desset_computing}.

\subsection{TSN FOR X-HAUL}
\label{sec:TSN_XH}
The IEEE 802.1cm standard \cite{IEEE_TSN_802.1c_2020} provides guidelines for implementing Ethernet bridging in FH and MH networks to satisfy \acs{TSN}.  
It defines configuration profile \textit{A}, which employs the \textit{algorithm of zero}, a strict-priority queuing mechanism that immediately forwards high-priority traffic by preempting pending lower-priority frames. This mechanism prioritizes FH and MH traffic over Ethernet networks.

Within Class~2, the standard considers eCPRI for low-level splits and F1 for high-level splits. Furthermore, Class~2 defines three flow categories aligned with the eCPRI classes of service (CoS) \cite{eCPRI2019}, each associated with a specific latency target. High-Priority Flow (HPF), intended for fast user-plane traffic, requires latency between 25 and 500~$\mu$s. Medium-Priority Flow (MPF), used for slow user-plane traffic, requires latency between 1 and 10~ms. Low-Priority Flow (LPF), associated with control and management traffic, tolerates latency up to 100~ms. Accordingly, this work models FH traffic as HPF and MH traffic as MPF, reflecting their distinct latency requirements.

Configuration Profile A characterizes bridge delays due to queuing and self-queuing (fan-in). It distinguishes two frame types: ``golden'' and ``low-priority'' which are mapped to HPF and to MPF/LPF, respectively.
Queuing delay arises from two effects: \textit{(i)} the residual transmission of a lower-priority frame that was selected before the target frame became eligible, and \textit{(ii)} higher-priority frames queued ahead of the target frame. Therefore, the queuing delay is given by the transmission time $d^{tr}$ of a maximum-size frame.
Conversely, self-queuing delay is caused by other frames with the same priority as the frame in question, arriving simultaneously from different input ports. 



The self-queuing delay can be represented as:

\begin{equation}
    m_{e}^{f} = P_{e}^{f} \land \ P_{e}^{f'}, \quad \forall f, f' \in \mathcal{F}, e \in \mathcal{E}
    \tag{l-1}
    \label{eq:SQ-1}
\end{equation}

First, let's consider $f \in \mathcal{F}$ as XH flows, and $\eset$ as the edges in a graph. $P_{e}^{f}$ and $P_{e}^{f'}$ are binary variables that are set to 1 if flows $f$ or $f'$ pass through edge $e$. Then, binary $m_{e}^{f}$ indicates whether two flows pass over the same edge $e$ in \cref{eq:SQ-1}.

\begin{equation}
    n_{e}^{f} = \neg \left( P_{e'}^{f} \land \ P_{e'}^{f'} \right) \land \ m_{e}^{f}, \quad \forall f, f' \in \mathcal{F}, e \in \mathcal{E}
    \tag{l-2}
    \label{eq:SQ-2}
\end{equation}

Similarly, in \cref{eq:SQ-2}, $e'$ refers to the previous link before reaching $e$, and binary $n_{e}^{f}$ is active only when the previous edges of flows $f$ and $f'$ are different, and they are transmitted over the current link $e$.

\begin{equation}
    SQ_{e}^{f} = d^{tr} \cdot \sum_{i\setminus f}^{\mathcal{F}} n_{e}^{i}, \quad \forall f \in \mathcal{F}
    \tag{l-3}
    \label{eq:SQ-3}
\end{equation}
Finally, the self-queuing delay $SQ_{e}^{f}$ for flow $f$ in edge $e$ is given by all the interfering flows on edge $e$ that come from different ports, and the transmission delay of the packet $d^{tr}$ \cref{eq:SQ-3}.

\subsection{ENERGY CONSUMPTION MODELS}
\label{sec:energy_consumption_models}
This work considers the energy consumption of the PPs and routers. Next, we describe the employed models for both components.
The PPs energy consumption model comprises processing and infrastructure. The latter component comprises the cooling, power supply, and monitoring system. The PP energy consumption is based on the one presented in \cite{DRL_Almeida_2022}:
\begin{equation}
\EPP =
\underbrace{
n_{v'} \cdot \left(\frac{PP_{v'}}{CPP_{v'}}\right) \cdot
\left(WPP_{v'}^{full} - WPP_{v'}^{idle}\right) \cdot T
}_{\text{Processing}}
+
E_{v'}^{infra}
\tag{e-1}
\label{eq:energy_pp}
\end{equation}
where $E_{v'}$ refers to the energy consumption of a PP node $v'$ during a time step, $n_{v'}$ denotes the number of CPUs in the PP. $PP_{v'}$ and $CPP_V$ indicate the computing utilization and total processing capacity of node $v'$ in $GOPS$, respectively. $WPP_{v'}^{full}$ and $WPP_{v'}^{idle}$ represent the power consumption in Watts when the PP is working at 100\% capacity and when it is idle; and $T$ indicates the utilization period.  


The switch power consumption model adopted in this work is based on \cite{kliazovich2015energy} and \cite{energy_SW_Kliazovich}. The model comprises three main components: the chassis, line cards, and active ports. Furthermore, the energy consumption is computed considering the utilization period and is expressed as follows:

\begin{equation}
    \WSW = \WSW[chassis] + \nlinc \cdot \WSW[linecard] + \sum_{x}^{r} \nport \cdot \WSW[r_{x}]
    \tag{e-2}
    \label{eq:energy_switch}
\end{equation}

Where $\WSW[chassis]$ represents the baseline power needed to power the chassis. The second term captures the power used by all active line cards, where $\nlinc$ is the number of line cards and $\WSW[linecard]$ the power per line card. The summation term reflects the dynamic power consumption of ports transmitting at rate $r_x$, with $n_{w}^{r_x}$ denoting the number of such ports and $\WSW[port]$ their individual power demands.
Finally, the energy consumption of the router is given by considering a period of time $T$:
\begin{equation}
    \ESW = \WSW \cdot T
    \tag{e-3}
    \label{eq:energy_switch_2}
\end{equation}

\begin{table*}[!t]
\centering
\caption{Notation Summary}
\small
\begin{tabular}{p{3cm}p{5cm}p{3cm}p{5cm}}
\toprule
\textbf{Symbol} & \textbf{Description} & \textbf{Symbol} & \textbf{Description} \\
\midrule
\multicolumn{4}{c}{\textbf{Sets}} \\
\midrule
$\iset$ & Set of RUs. & $\jset$ & Set of slices: eMBB, URLLC, and mMTC demands/slices, respectively. \\
$\kset$ & Virtual configuration options. & $\gset$ & RAN graph. \\
$\eset$ & Set of edges. & $\vV$ & Set of nodes. \\
$\mathcal{V} = \mathcal{V}^{PP} \cup \mathcal{V}^{SW}$ & Subset of nodes: PPs and SWs. & $\zset$ & Set of paths. \\
\midrule
\multicolumn{4}{c}{\textbf{Variables}} \\
\midrule
$\VC$ & Binary, 1 if VC $k$ is chosen for $i,j$. & $\RU$ & Binary, 1 if the RU of $i,j$ is located on node $v'$ with VC $k$. \\
$\DU$ & Binary, 1 if the DU of $i,j$ is located on node $v'$ with VC $k$. & $\CU$ & Binary, 1 if the CU of $i,j$ is located on node $v'$ with VC $k$. \\
$\Pe{}$ & Binary, 1 if the flow of $i,j$ passes through edge $e$. & $\CD$ & Binary, 1 if the path between nodes $a$ and $b$ of $i,j$ with VC $k$ is selected. \\
$\DR$ & Binary, 1 if the path between nodes $b$ and $c$ of $i,j$ with VC $k$ is selected. & $\Load$ & Transmission load on link $e$ carrying XH traffic of $i,j$. \\
$\latS[FH/MH]$ & Static delay for FH or MH traffic of $i,j$. & $\latS$ & Total static delay in the XH of $i,j$. \\
$\latD[\FQ]$ & FH queuing delay on link $e$ of $i,j$. & $\latD[\MQ]$ & MH queuing delay on link $e$ of $i,j$. \\
$\latD[\FSQ]$ & FH self-queuing delay on link $e$ of $i,j$. & $\latD[\MSQ]$ & MH self-queuing delay on link $e$ of $i,j$. \\
$\latD[FH/MH]$ & Total dynamic delay of FH/MH traffic of $i,j$. & $\lat$ & Total XH delay of $i,j$. \\
$\APP$ & Binary, 1 if node $v'$ is active supporting a PP. & $\ASW$ & Binary, 1 if node $w$ is active supporting a switch. \\
\midrule
\multicolumn{4}{c}{\textbf{Constants}} \\
\midrule
$\compC$ & Computing capacity in GOPS of node $v'$. & $\linkC$ & Transmission capacity in Gbps of edge $e$. \\
\midrule
\multicolumn{4}{c}{\textbf{Functions}} \\
\midrule
\funPrc{} & Processing load (GOPS) of $i,j$ with VC $k$ considering the aggregated RU demand. & \funTr{} & Transmission rate (Mbps) of $i,j$ with VC $k$ in the XH. \\
\funRot{a,b} & Returns if the link $e$ belongs to the path between nodes $a$–$b$ for $i,j$. & $\slat[SF](e)$ & Store-and-forward delay of link $e$ in $\mu$s. \\
$\slat[prop](e)$ & Propagation delay of link $e$ in $\mu$s. & $\funLat{Q}$ & Queuing delay on $i,j$ due to $i',j'$ on link $e$ in $\mu$s. \\
$\funLat{SQ}$ & Self-queuing delay on $i,j$ due to $i',j'$ on link $e$ in $\mu$s. & $\funSW$ & Returns 1 if node $w$ belongs to the path $z$ between nodes $v',u'$. \\
$\EPP$ & Energy consumption of node $v$. & $\ESW$ & Energy consumption of node $w$. \\
\bottomrule
\end{tabular}
\label{tab:Notation_II}
\end{table*}

\section{OPTIMAL FORMULATION}
\label{sec:milp}
This section presents the MILP formulation for the RAN reconfiguration problem. The model jointly determines the VC selection, RAN unit placement, and XH traffic routing for each gNB-slice while satisfying computing, bandwidth, and latency constraints. Three objective functions are considered: \textit{i)} minimization of RAN energy consumption, a major OPEX component for \acp{MNO}; \textit{ii)} minimization of FH latency to support latency-sensitive services; and \textit{iii)} a bi-objective formulation that prioritizes energy efficiency while secondarily minimizing FH latency. 

The RAN network is modeled as a graph $\gset$, where $\mV$ denotes the set of RAN nodes and $E$ denotes the set of edges. The node set $\mV$ is further divided into two subsets: $\mVP$, representing PP nodes, and $\mVSW$, representing network nodes (i.e., XH routers). In turn, the PP subset $\mVP$ is composed of two types of nodes: nodes supporting only upper-layer RAN functions, represented by $\mVPP$, and nodes supporting both upper-layer and RF processing, represented by $\mVRU$. Formally, the sets are defined as follows:
\begin{equation}
\mV = \mVP \cup \mVSW, \qquad
\mVP = \mVPP \cup \mVRU,
\end{equation}
where $\mVPP$ and $\mVSW$ are disjoint sets, i.e., $\mVPP \cap \mVSW = \emptyset$.
The VC Options are represented by the set $\mK$, where $\mK = \{1,\ldots,5\}$ corresponds to the configurations illustrated in Fig.~\ref{fig:VC_options}. Similarly, the set of gNBs is denoted by $\mI$. Each gNB $i \in \mI$ serves three slices $ \jset$, where $\mathcal{D} = \{embb, urrlc, mmtc\}$. A complete description of the notation employed in the model is provided in \Cref{tab:Notation_II}.

The model considers the demand $\demand$ associated with each slice of every gNB, such that the RAN configurations are defined for each gNB-slice pair $(i,j)$. Initially, a VC Option $\VC$ is selected from the set $\mK$. Based on the selected configuration, the corresponding RAN units are allocated to PP nodes in $\mVP$, with the allocation represented by the binary variables $\CU$, $\DU$, and $\RU$. A PP node $\vPP$ may simultaneously host RAN units associated with different gNB-slice pairs $(i,j)$, independently of the selected VC Option. For example, a node $\vPP$ may support the DU associated with slice $embb$ of gNB $a$ under VC 1, while concurrently hosting the CU-DU functions of pair $(b,mmtc)$ under VC 5. In contrast, nodes in $\vRU$ are restricted to serving only a single gNB $i$ and its corresponding three slices.

The computational utilization associated with each RAN unit is determined by the function $\funPrc$, which considers the demand $\demand$, the selected VC Option $k$, and the RAN unit type $\unitt \in {RU,DU,CU}$. Based on these parameters, the function aggregates the computational load generated by the RAN functions assigned to each unit. Consequently, the total processing load allocated to a node must not exceed its available processing capacity $CPP_{v'}$.

Furthermore, a path option $z$ is selected for both the MH and FH segments, whose source and destination nodes are defined by the variables $\CD$ and $\DR$, respectively. These nodes correspond to the placement of the CU, DU, and RU units. The shortest paths associated with each option $z$ are computed using Dijkstra’s algorithm, considering the corresponding RAN units as source and destination nodes. The resulting paths are composed of edges $e \in \mathcal{E}$ and represent fiber links through which one or more gNB-slice pairs may transmit XH traffic, provided that the link transmission capacity $C_e$ is respected. The binary variables $\Pe{MH}$ and $\Pe{FH}$ indicate whether a gNB-slice pair transmits MH or FH traffic through edge $e$, respectively. The function $\funTr$ estimates the XH traffic load according to the selected functional split(s) associated with VC $k$ and the demand of the corresponding gNB-slice pair.

In addition, the model considers two latency components for each flow: static latency $\latS$ and dynamic latency $\latD$. The static latency accounts for propagation and store-and-forward delays and is computed through the functions $\slat[SF]$ and $\slat[prop]$. Conversely, the dynamic latency comprises both queuing delay $\latD[Q]$ and self-queuing delay $\latD[SQ]$. The functions $\funLat{Q}$ and $\funLat{SQ}$ evaluate whether flows $(i',j')$ and $(i,j)$ share the same link and return the corresponding delay based on packet size, as detailed in Subsection~\ref{sec:TSN_XH}.

\begin{table*}[!b]
    \centering
    \caption{MILP}
    \label{tab:MILP_I}

    \begingroup
    \renewcommand{\arraystretch}{1.45} 
    \setlength{\extrarowheight}{0.7pt} 

    \begin{tabularx}{\linewidth}{
      M{0.20\textwidth} 
      M{0.20\textwidth} 
      E                 
      M{0.20\textwidth} 
      M{0.20\textwidth} 
      E             
    }

        \toprule
        \multicolumn{6}{c}{\textbf{Objective function:} energy (\ref{eq:obj_energy_lin}), latency (\ref{eq:OF_latency}), bi-objective (\ref{eq:obj_energy_lin}) and (\ref{eq:OF_latency})} \\
        
        \hline
        \multicolumn{6}{c}{\textbf{VNF Placement}} \\
        \hline

        \multicolumn{5}{@{}c@{}}{$\sumK \VC = 1, \qquad \forall \iset, \jset$}
        & \label[constraint]{eq:vnf_1} \\
        
        \multicolumn{5}{@{}c@{}}{$\sumRU \RU = \VC, \qquad \forall \iset, \jset, \kset$}
        & \label[constraint]{eq:vnf_2} \\
        
        \multicolumn{5}{@{}c@{}}{$\sumPP \DU = \VC, \qquad \forall \iset, \jset, \kset$}
        & \label[constraint]{eq:vnf_3} \\
        
        \multicolumn{5}{@{}c@{}}{$\sumPP \CU = \VC, \qquad \forall \iset, \jset, \kset$}
        & \label[constraint]{eq:vnf_4} \\
        
        \multicolumn{5}{@{}c@{}}{$\quad \qquad \qquad \qquad \RU[v'] + \DU + \CU \leq 1, \quad \forall \iset, \jset, \ksetDS, \vPP \qquad \qquad \qquad \qquad$}
        & \label[constraint]{eq:vnf_5} \\

        \multicolumn{5}{@{}c@{}}{$\RU[v'] + \CU \leq 1, \qquad \forall \iset, \jset, \ksetSH, \vPP$}
        & \label[constraint]{eq:vnf_6} \\

        \multicolumn{5}{@{}c@{}}{$\DU[v''] = \RU, \qquad \forall \iset, \jset, \ksetSH, \vRU$}
        & \label[constraint]{eq:vnf_7} \\ 
        

        \multicolumn{5}{@{}c@{}}{$\RU[v'] + \DU \leq 1, \qquad \forall \iset, \jset, \ksetSL, \vPP$}
        & \label[constraint]{eq:vnf_9} \\
        \multicolumn{5}{@{}c@{}}{$\CU = \DU, \qquad \forall \iset, \jset, \ksetSL, \vPP$}
        & \label[constraint]{eq:vnf_10} \\ 
        

        \hline
        \multicolumn{6}{c}{\textbf{Routing}} \\
        \hline
        
        \multicolumn{5}{@{}@{}@{}@{}c@{}}{$\sumZ \Pz = \VC, \qquad \forall \iset, \jset, \kset$}
        & \label[constraint]{eq:rot_01} \\
        
        \multicolumn{5}{@{}c@{}}{$\CD \leq \Pz, \qquad \forall \iset, \jset, \kset, \vPP[a,b], \zset$}
        & \label[constraint]{eq:rot_02} \\
        
        \multicolumn{5}{@{}c@{}}{$\DR \leq \Pz, \qquad \forall \iset, \jset, \kset, \vPP[b,c], \zset$}
        & \label[constraint]{eq:rot_03} \\
        
        \multicolumn{5}{@{}c@{}}{$\sumPP[a]\sumPP[b]\sumZ\sumK \CD = 1, \qquad \forall \iset, \jset$}
        & \label[constraint]{eq:rot_04} \\
        
        \multicolumn{5}{@{}c@{}}{$\sumPP[b]\sumPP[c]\sumZ\sumK \DR = 1, \qquad \forall \iset, \jset$}
        & \label[constraint]{eq:rot_05} \\
        
        \multicolumn{5}{@{}c@{}}{$\sumPP[b]\sumZ\sumK \CD \geq \sumK \CU[a], \qquad \forall \iset, \jset, \vPP[a]$}
        & \label[constraint]{eq:rot_06} \\
        
        \multicolumn{5}{@{}c@{}}{$\sumPP[a]\sumZ\sumK \CD \geq \sumK \DU[b], \qquad \forall \iset, \jset, \vPP[b]$}
        & \label[constraint]{eq:rot_07} \\
        
        \multicolumn{5}{@{}c@{}}{$\sumPP[c]\sumZ\sumK \DR \geq \sumK \DU[b], \qquad \forall \iset, \jset, \vPP[b]$}
        & \label[constraint]{eq:rot_08} \\
        \multicolumn{5}{@{}c@{}}{$\sumPP[b] \sumZ \sumK \DR \geq \sumK \RU[c], \qquad \forall \iset, \jset, \vPP[c]$}& \label[constraint]{eq:rot_09}\\
        \multicolumn{5}{@{}c@{}}{$
        \sum_{\vPP[a]} \sum_{\vPP[b]}\sumZ \funRot{a,b} \cdot \CD = \PMH, \qquad \forall \iset, \jset, \kset, \eset 
        $}& \label[constraint]{eq:rot_10} \\
        \multicolumn{5}{@{}c@{}}{$
        \sum_{\vPP[a]} \sum_{\vPP[b]} \sumZ \funRot{b,c} \cdot \DR = \PFH, \qquad \forall \iset, \jset, \kset, \eset
        $} &\label[constraint]{eq:rot_11} \\

        \hline
        \multicolumn{6}{c}{\textbf{Link Load}} \\
        \hline

        \multicolumn{5}{@{}c@{}}{$\sumI \sumJ \sumK \PFH \cdot \funTr[FH] = \Load[FH], \qquad \forall \eset$}
        & \label[constraint]{eq:link_1} \\
        
        \multicolumn{5}{@{}c@{}}{$\sumI \sumJ \sumK \PMH \cdot \funTr[MH] = \Load[MH], \qquad \forall \eset$}
        & \label[constraint]{eq:link_2} \\
        
        \multicolumn{5}{@{}c@{}}{$\Load = \Load[FH] + \Load[MH], \qquad \forall \eset$}
        & \label[constraint]{eq:link_3} \\
        
        \multicolumn{5}{@{}c@{}}{$\Load \leq \linkC, \qquad \forall \eset$}
        & \label[constraint]{eq:link_4} \\

        \hline
        \multicolumn{6}{c}{\textbf{Computing Utilization}} \\
        \hline


        \multicolumn{5}{@{}c@{}}{%
        \makecell[c]{%
          $\sumI \sumJ \sumK \big(
            \RU[v'] \funPrc[RU] + \DU[v'] \funPrc[DU] + \CU[v'] \funPrc[CU]\big) $\\[0.2ex]
          $ = \PP, \qquad \forall \vPP$%
        }}
        & \label[constraint]{eq:comp_1} \\
    
        \multicolumn{5}{@{}c@{}}{$\PP \leq \compC \cdot \APP, \qquad \forall \vPP$}
        & \label[constraint]{eq:comp_2} \\

        \bottomrule
        \multicolumn{6}{r}{\textit{Continue on the next page}}
    \end{tabularx}
    \endgroup
\end{table*}

\begin{table*}[!t]
\ContinuedFloat
    \centering
    \caption{MILP (cont.)}
    \label{tab:MILP_I}

    \begingroup
    \renewcommand{\arraystretch}{1.38} 
    \setlength{\extrarowheight}{0.6pt} 

    \begin{tabularx}{\linewidth}{
      M{0.25\textwidth} 
      M{0.20\textwidth} 
      E                 
      M{0.25\textwidth} 
      M{0.20\textwidth} 
      E                
    }
        \toprule

        

        
        \multicolumn{5}{c}{$\sumK \CU \leq \APP, \qquad \forall \iset, \jset, \vPP$}
        & \label[constraint]{eq:comp_3} \\
        
        \multicolumn{5}{c}{$\sumK \DU \leq \APP, \qquad \forall \iset, \jset, \vPP$}
        & \label[constraint]{eq:comp_4} \\
        
        \multicolumn{5}{c}{$\sumK \RU \leq \APP[v''], \qquad \forall \iset, \jset, \vRU$}
        & \label[constraint]{eq:comp_5} \\
 \hline
        \multicolumn{6}{c}{\textbf{Static Latency}} \\
        \hline

        \multicolumn{5}{c}{$\latS[FH] = \sumE \sumK \PFH \funSLat, \qquad \forall \iset, \jset$}
        & \label[constraint]{eq:lat_01} \\
        
        \multicolumn{5}{c}{$\latS[MH] = \sumE \sumK \PMH \funSLat, \qquad \forall \iset, \jset$}         & \label[constraint]{eq:lat_02} \\

        \multicolumn{5}{c}{$\latS = \latS[FH] + \latS[MH], \qquad \forall \iset, \jset$} & \label[constraint]{eq:lat_03} \\

\hline        
\multicolumn{6}{c}{\textbf{Dynamic Latency}} \\
\hline
\multicolumn{5}{c}{$\Pe{FH} = \sumK \PFH, \qquad \forall \iset, \jset, \eset$}  & \label[constraint]{eq:lat_04} \\

\multicolumn{5}{c}{$\Pe{MH} = \sumK \PMH, \qquad \forall \iset, \jset, \eset$}
& \label[constraint]{eq:lat_05} \\

\multicolumn{5}{c}{$\PPe{\FF} = \Pe{FH}\cdot \Pep{FH}, \qquad \forall \iiset, \jjset, \eset$}
& \label[constraint]{eq:lat_06} \\

\multicolumn{5}{c}{$\PPe{\FM} = \Pe{FH}\cdot \Pep{MH}, \qquad \forall \iiset, \jjset, \eset$}
& \label[constraint]{eq:lat_07} \\

\multicolumn{5}{c}{$\PPe{\MM} = \Pe{MH}\cdot \Pep{MH}, \qquad \forall \iiset, \jjset, \eset$}
& \label[constraint]{eq:lat_08} \\

\multicolumn{5}{c}{$\PPe{\MF} = \Pe{MH}\cdot \Pep{FH}, \qquad \forall \iiset, \jjset, \eset$}
& \label[constraint]{eq:lat_09} \\

\multicolumn{5}{c}{$\PPqe{\FQ} \geq \PPe{\FM}, \quad \forall \iiset, \jjset, \eset$} & \label[constraint]{eq:lat_10a} \\
\multicolumn{5}{c}{$\PPqe{\FQ} \leq \sum_{i',j' \in \mathcal{L}} \PPe{\FM}, \quad \iset, \jset, \eset $} & \label[constraint]{eq:lat_10b} \\
\multicolumn{5}{c}{$\latD[\FQ] = \PPqe{\FQ} \cdot \funLatij{\FQ}
\qquad \forall \iset, \jset, \eset$}
& \label[constraint]{eq:lat_10} \\

\multicolumn{5}{c}{$\latD[\FSQ] = \sum_{(i',j') \in \mathcal{H}} \PPe{\FF} \cdot\funLat{\FSQ}, \qquad \forall \iset, \jset, \eset$}
& \label[constraint]{eq:lat_11} \\

\multicolumn{5}{c}{$\PPqex{\MQ} \geq \PPe{\MM}, \quad \forall \iiset, \jjset, \eset$} & \label[constraint]{eq:lat_12a} \\
\multicolumn{5}{c}{$\PPqex{\MQ} \leq \sum_{i',j' \in \mathcal{L}} \PPe{\MM}, \quad \iset, \jset, \eset $} & \label[constraint]{eq:lat_12b} \\

\multicolumn{5}{c}{$\begingroup\setlength{\jot}{-2pt}
\begin{aligned}
\latD[\MQ] = \sum_{(i',j') \in \mathcal{H}} \PPe{\MF}\cdot\funLat{\MQ}
          + \PPqex{\MM}\cdot\funLatij{\MQ}, 
\quad \forall \iset, \jset
\end{aligned}
\endgroup$}
& \label[constraint]{eq:lat_12} \\

\multicolumn{5}{c}{$\latD[\MSQ] = \sum_{(i',j') \in \mathcal{L}} \PPe{\MM}\cdot\funLat{\MSQ}, \qquad \forall \iset, \jset, \eset$}
& \label[constraint]{eq:lat_13} \\

\multicolumn{5}{c}{$\latDH[MH] = \sumE \big(\latD[\MQ] + \latD[\MSQ]\big), \qquad \forall \iset, \jset$}
& \label[constraint]{eq:lat_14} \\

\multicolumn{5}{c}{$\latDH[FH] = \sumE \big(\latD[\FQ] + \latD[\FSQ]\big), \qquad \forall \iset, \jset$}
& \label[constraint]{eq:lat_15} \\

\multicolumn{5}{c}{$\lat[MH] = \latS[MH] + \latDH[MH], \qquad \forall \iset, \jset$}
& \label[constraint]{eq:lat_16} \\

\multicolumn{5}{c}{$\lat[FH] = \latS[FH] + \latDH[FH], \qquad \forall \iset, \jset$}
& \label[constraint]{eq:lat_17} \\

\multicolumn{5}{c}{$\lat = \lat[FH] + \lat[MH], \qquad \forall \iset, \jset$}
& \label[constraint]{eq:lat_18} \\

\multicolumn{5}{c}{$\lat \leq \latMax{max}, \qquad \forall \iset, \jset$}
& \label[constraint]{eq:lat_19} \\

\multicolumn{5}{c}{$\lat[MH] \leq \latMax{MH}, \qquad \forall \iset, \jset$}
& \label[constraint]{eq:lat_20} \\

\multicolumn{5}{c}{$\lat[FH] \leq \latMax{FH}, \qquad \forall \iset, \jset$}
& \label[constraint]{eq:lat_21} \\

\midrule
\multicolumn{6}{c}{\textbf{Energy}} \\
\midrule
\multicolumn{5}{c}{$\sumPP \sumPP[u'] \sumZ \sumK \left( \CD[v',u'] + \DR[v',u'] \right) \cdot \funSW \leq \ASW, \quad \forall \iset, \jset, \wSW $}
 &\label[constraint]{eq:enr_2} \\
        \bottomrule
    \end{tabularx}
    \endgroup
\end{table*}

The latency constraints are analyzed from both the slice and XH perspectives. The slice latency requirement is service-dependent and spans from the 5GC to the RU. 
Meanwhile, the XH latency is determined by the FS requirements in the FH or MH segments, as outlined in Table \ref{tab:XH_requirements}.

The energy minimization objective function (EMOF) minimizes the energy consumption associated with both PPs and RAN network equipment, as defined in (\ref{eq:obj_energy_1}). To achieve this objective, the model deactivates idle nodes and adopts the energy-consumption models proposed in \cite{Apt-RAN} for PPs and in \cite{energy_SW_Kliazovich} for network equipment, as detailed in \Cref{sec:energy_consumption_models}. Furthermore, the infrastructure energy consumption terms $EPP^{infra}_{v'}$ and $ESW^{infra}_{w}$ are also incorporated into the formulation.

\begin{align}
    \min \sum_{v'}^{\mathcal{V}^{PP}} APP_{v'} \cdot \left(EPP_{v'} + EPP^{infra}_{v'}\right) + \notag \\ \sum_{w}^{\mathcal{V}^{SW}} ASW_{w} \cdot \left( ESW_{w} + ESW^{infra}_{w} \right)  \tag{OF-1}  
    \label{eq:obj_energy_1}
\end{align}

Since the objective function contains products of decision variables, a linearization procedure is required. Specifically, the first and second terms in (\ref{eq:obj_energy_1}) are linearized using the Big-M method. In this formulation, $m$ denotes a sufficiently large constant that guarantees the proper activation of the constraints according to the binary variable $n$, which corresponds to either $APP_{v'}$ or $ASW_w$. Moreover, $q$ represents the energy values associated with $EPP_{v'}$ or $ESW_w$. As a result of the linearization process, the auxiliary variable $\epsilon$ is represented by $y_{v'}$ for the product $EPP_{v'} \cdot APP_{v'}$, and by $z_{w}$ for the product $ESW_w \cdot ASW_{w}$, as expressed in Objective Function (\ref{eq:obj_energy_lin}).

\begin{align}
    \epsilon &\leq m * n \label{eq-link-p:1} \tag{l-1}\\
    \epsilon &\leq q  \label{eq-link-p:2} \tag{l-2}\\
    \epsilon &\geq q - (1 - n) \cdot m \label{eq-link-p:3} \tag{l-3}\\
    \epsilon &\geq 0 \label{eq-link-p:4} \tag{l-4}
\end{align}

\begin{align}
\min \sum_{v' \in \mathcal{V}^{PP}} y_{v'} + \sum_{w \in \mathcal{V}^{SW}} z_{w} \tag{OF-E}
\label{eq:obj_energy_lin}
\end{align}

The model is subsequently evaluated using the FH latency minimization objective function (LMOF). The objective consists of minimizing the total FH latency, defined as the aggregate FH delay across all gNB-slice pairs, as expressed in


\begin{align}
   \min: \quad \sumI \sumJ \lat[FH] \tag{OF-L}
    \label{eq:OF_latency}
\end{align}


Finally, a bi-objective minimization function (BOMF) is considered to jointly balance energy consumption and FH latency. This approach adopts a lexicographic optimization method, in which priorities are assigned to the objective functions, and the optimization proceeds sequentially according to these priorities. Initially, the highest-priority objective is optimized. Subsequently, the lower-priority objective is solved while preserving the optimality achieved for the higher-priority objective. In the proposed formulation, RAN energy minimization is prioritized over FH latency minimization because it directly impacts OPEX, whereas the MILP formulation inherently guarantees compliance with the latency constraints.

\begin{align}
\min_{x \in X} \quad f_1(x) = \sum_{v' \in \mathcal{V}^{PP}} y_{v'} + \sum_{w \in \mathcal{V}^{SW}} z_{w} \tag{OF-BE} \label{eq:obj_energy_2}
\end{align}

Let $f_1^*$ be the optimal value of \eqref{eq:obj_energy_2}.

\begin{align}
&\min_{x \in X} \quad f_2(x) = \sumI \sumJ \lat[FH] \tag{OF-BL} \label{eq:obj_latency_2} 
\end{align}

Constraints \eqref{eq:vnf_1}--\eqref{eq:enr_2} (\Cref{tab:MILP_I}) apply to both the mono-objective and bi-objective optimization formulations.

The first group of constraints addresses RAN VNF placement and VC selection.
In \Cref{eq:vnf_1}, a unique VC Option $k$ is set for each pair RU-slice $i,j$ with the decision variable $\VC$. 
\Cref{eq:vnf_2} to (\ref{eq:vnf_4}) configure RAN units (RU, DU, CU) to have a unique VC Option $k$ and node $v$ per pair ($i,j$). In \Cref{eq:vnf_5}, it is ensured that the RAN units are placed in three different nodes for the VC Options $k = \{1,2\}$ where a dual split is selected. Then, the expression (\ref{eq:vnf_6}) restricts that the CU is placed in a different node than the RU, while \Cref{eq:vnf_7} ensures that both DU and RU are placed in the same node. 
Similarly, Constraints (\ref{eq:vnf_9}) and (\ref{eq:vnf_10}) ensure the DU is placed in the same node as the CU, but separated from the RU for $k = \{4,5\}$.


The second group of constraints defines the traffic flow over the transport network links.
In \Cref{eq:rot_01}, a single path option $z$ and a single VC Option $k$ are selected for each slice $j$ associated with RU $i$. Constraints \Cref{eq:rot_02,eq:rot_03} define the FH and MH paths, which connect nodes $a$ to $b$ and nodes $b$ to $c$, respectively, thereby associating the selected VC Option $k$ and path $z$ with the corresponding RU-slice pair $(i,j)$. The constraints in (\ref{eq:rot_04}) and (\ref{eq:rot_05}) ensure that exactly one segment is selected for both the MH and FH paths for each pair $(i,j)$. Subsequently, \Cref{eq:rot_06,eq:rot_07} defines the source and destination nodes associated with the MH segment $(i,j)$ 
where the CU node $a$ is the source and the DU node $b$ is the destination. Similarly, \Cref{eq:rot_08,eq:rot_09} establish the FH path connecting the DU node $b$ to the RU node $c$. Finally, constraints (\ref{eq:rot_10}) and (\ref{eq:rot_11}) activate link $e$ whenever it belongs to a RAN segment, either MH or FH.

The following group of constraints establishes the relationship between routing decisions and link load.
\Cref{eq:link_1,eq:link_2} calculate the total MH and FH transmission load of $i,j$, considering the VC $k$, on link $e$.
In \Cref{eq:link_3}, the total transmission load, considering MH and FH loads conveyed through edge $e$, is estimated. Then, in \Cref{eq:link_4}, that load is constrained to the transmission capacity of the link $C_e$.


The following constraints enforce resource utilization limits at each processing node.
\Cref{eq:comp_1} estimates the total computational load $\PP$ allocated to node $v'$ by considering all RAN functions placed on that node.
 \Cref{eq:comp_2} ensures that the computational capacity of node $v'$ is not exceeded. Subsequently, \Cref{eq:comp_3,eq:comp_4,eq:comp_5} activate node $v'$ through the binary variable $\APP$ whenever at least one slice $j$ is allocated to it.

Moreover, static latency constraints ensure that the propagation and processing delays associated with each VC satisfy their respective limits.
In \Cref{eq:lat_01,eq:lat_02}, the total static latency associated with propagation and store-and-forward delays is computed for both RAN segments, namely FH and MH. Then, \Cref{eq:lat_03} aggregates the total static latency experienced by slice $j$ of RU $i$ along the end-to-end path from the 5GC to the RU.

Dynamic latency constraints additionally account for queuing and transmission delays along the transport path.
For this analysis, the notation of the flow associated with pair $(i,j)$ over link $e$ is simplified through the variables $\Pe{FH}$ and $\Pe{MH}$, which disregard the selected VC Option in expressions (\ref{eq:lat_04}) and (\ref{eq:lat_05}). As discussed in the previous section, computing queuing and self-queuing delays in the XH requires verifying whether multiple flows and their corresponding flow types share the same link $e$. Accordingly, constraints (\ref{eq:lat_06})--(\ref{eq:lat_09}) perform this verification for two distinct flows, namely $(i,j)$ and $(i',j')$. These expressions are conditional and account for all possible coincidences between MH and FH flows. Since the conditional operations involve products of decision variables, the resulting formulation becomes nonlinear. Therefore, the expressions are linearized using the conventional product-linearization approach:
\begin{align}
    z_{i,j} &= x_{i,j} \cdot y_{i,j}, \quad where \ x_{i,j}, y_{i,j} \in {0,1}, \tag{pl-1} \\
    z_{i,j} &\leq x_{i,j}       \tag{pl-2}\\
    z_{i,j} &\leq y_{i,j}       \tag{pl-3}\\
    z_{i,j} &\geq x_{i,j} + y_{i,j} - 1 \tag{pl-4}
\end{align}

The binary auxiliary variable $\PPqe{\FQ}$ indicates whether an MH flow coexists with FH flow $i,j$ on edge $e$. Constraint \Cref{eq:lat_10a} enforces $\PPqe{\FQ}=1$ whenever at least one MH flow is present, whereas \Cref{eq:lat_10b} sets $\PPqe{\FQ}=0$ otherwise. The FH queuing delay of $i,j$ on link $e$ is calculated in \Cref{eq:lat_10}, which takes into account a MH flow coincidence. The self-queuing delay is given in \Cref{eq:lat_11}, considering all the FH flows on the same edge. This latter expression sums all the other coinciding FH flows $(i',j') \in \mathcal{H}$. 

Similarly, the binary variable $X_{i,j,e}^{MH\text{-}Q} \in \{0,1\}$ indicates whether another MH flow coexists with MH flow $i,j$ on link $e$. Constraint \Cref{eq:lat_12a} sets $X_{i,j,e}^{MH\text{-}Q}=1$ whenever at least one distinct MH flow $(i',j') \in \mathcal{L}$ shares the same link, whereas \Cref{eq:lat_12b} enforces $X_{i,j,e}^{MH\text{-}Q}=0$ otherwise.
For MH flows of gNB-slice $i,j$ on $e$, the queuing delay in \Cref{eq:lat_12} consists of all concurrent flows of FH $(i',j') \in \mathcal{H}$ plus a MH flow. Finally, the MH self-queuing delay given in \Cref{eq:lat_13} is estimated by summing the delays of all MH flows sharing link $e$.

The expressions (\ref{eq:lat_14}) and (\ref{eq:lat_15}) represent the total dynamic latency of $i,j$ in the FH and MH segments, respectively. 
Similarly, \Cref{eq:lat_16} and (\ref{eq:lat_17}) show the total latency per segment for FH and MH of $i,j$, accounting for static and dynamic delays. In turn, (\ref{eq:lat_18}) represents the total latency of $i,j$ from the 5GC to the RU, consisting of dynamic and static latencies of FH and MH segments. Then, that latency is first constrained by the slice-latency requirements in \Cref{eq:lat_19}, and by the FH and MH restrictions in \Cref{eq:lat_20} and  (\ref{eq:lat_21}).

Finally, energy constraints account for the power consumption of both processing and network equipment.
\Cref{eq:enr_2} verifies whether a network node $w$ belongs to any FH or MH segment. 


\begin{table}[!t]
    \centering
    \caption{Heuristic Notation Summary}
    \small
    \begin{tabular}{l p{5.5cm}}
    \toprule
    \textbf{Symbol} & \textbf{Description} \\
    \midrule
    $\overall$ & Overall solution structure containing VC, CU, DU, and RU. \\
    $\linkusage$ & Link usage structure containing the total transmission load of $e$. \\
    $\ppusage$ & PP usage structure containing the total computing load of $v$ and its allocated gNB-slices $i,j$.\\
    $ PP[du,cu]$ & Temporal computing utilization. \\
    $\ignored$ & Structure containing no working RAN configurations.\\
    $\mVP_{cand}$ & Set of candidate PPs\\
    $IT$ & Number of iterations\\
    $\pCU$ & PP node supporting CU\\
    $\pDU$ & PP node supporting DU\\
    $\pRU$ & PP node supporting RU\\
    $\xVC$ & Selected VC \\
    $\rfh, \rmh$ & FH and MH paths \\
    $\unitt$ & Set of RAN units CU, DU, RU\\
    $\mVP_{cand}$ & Set of candidate nodes to place CU or DU\\
    $\paths$ & Set of candidate paths for FH or MH\\
    $\placednode$ & Variable containing a placed node of a RAN unit \\
    $\allocatedpath$ & Variable containing list of edges \\
    $\bestpath$ & Variable containing a selected path based on $\pathscore$ \\
    $\pathscore $ & Variable telling how much a path is used \\
    \bottomrule
    \end{tabular}
    \label{tab:Notation_heur}
\end{table}
\section{HEURISTIC APPROACH}
\label{sec:heuristic}

This section presents a heuristic approach as an alternative to the MILP formulation, enabling the RAN configuration problem to be solved with significantly lower computational demand. The proposed heuristic jointly performs VC selection, unit placement, and FH/MH routing while satisfying bandwidth, latency, and resource-utilization constraints. Additional variables are introduced in \cref{tab:Notation_heur} and described throughout this section.
The proposed heuristic consists of three components: a RAN configuration algorithm, a placement and routing module with a configuration handler, and an energy minimization module. First, a feasible solution is constructed incrementally and subsequently refined to reduce energy consumption. During the configuration phase, VCs are selected based on slice service requirements, with computationally intensive slices assigned more distributed configurations to shift processing toward the network edge. Processing units are deployed on the nearest feasible nodes while prioritizing active resources to minimize infrastructure activation. Transport paths are determined using shortest-route selection, favoring already utilized links to promote traffic consolidation. After assigning feasible configurations to all slices, the energy minimization module reallocates gNB-slice RAN configurations to consolidate processing load and enable the deactivation of underutilized PPs.


\begin{algorithm}[!t]
\footnotesize	
\caption{RAN Configuration Algorithm}
\label{alg:Main}
\SetAlgoLined
\SetKwInput{KwIni}{Input}
\KwIni{Network graph $\gmset$, Demand $\demand$, RUs set $\jmset$, Slices set $\imset$, Virtual Configurations set $\kmset$, Number of iterations $IT$}
\KwOut{Placement, VC and routing result in $\overall$,  resources utilization $\ppusage$, $\linkusage$}

\SetKwFunction{Configuration}{Configuration Handler}
\SetKwFunction{Energy}{Energy}

\tcc{First RAN configuration}
$\overall', \ppusage', \linkusage', \ignored \leftarrow$ \Configuration$(\gmset, \jmset, \imset, \kmset)$\label{algline:main-1} \\

\tcc{Re-configure pending gNB-slices}
\While{$\ignored$ is not empty}{\label{algline:main-2}
    $\overall', \ppusage', \linkusage',\ignored \leftarrow$ \Configuration$(\gmset, \jmset, \imset, \kmset, \ignored)$ \label{algline:main-3}
}
\tcc{Energy Optimization}

$\overall[(\pCU, \pDU, \pRU, \xVC, \rfh, \rmh)], \ppusage, \linkusage, \leftarrow$ \Energy($IT, \overall, \ppusage, \linkusage, \ignored$) \label{algline:main-4}
\end{algorithm}

\begin{algorithm}[!t]
\footnotesize	
\caption{Configuration Handler}
\label{alg:Configuration}
\SetAlgoLined
\SetKwInput{KwIni}{Init}
\KwIn{Network graph $\mathcal{G}$, RUs set $\jmset$, Slices set $\imset$, Virtual Configurations set $\kmset$, PP set $\mVP$, PP capacity $\compC$}
\KwOut{Placement, VC and routing result $\overall$,  resources utilization $\ppusage$, $\linkusage$, ignored configurations $\ignored$, Energy consumption }
\KwIni{$\ppusage$, $\linkusage$}
\SetKwFunction{Placement}{Placement}
\SetKwFunction{Routing}{Routing}
\SetKwFunction{Dijkstra}{Dijkstra}

\tcc{RAN Configuration}
\If{$\ignored$ received as argument}{
    $\imset, \jmset \leftarrow$ Retrieve $i,j$ elements from $\ignored$ \label{algline:ran-27-conf}\\
    $\overall^{ign} \leftarrow$ Retrieve selection of VC, placement and routing elements from $\ignored$ to be ignored \label{algline:ran-28-conf}
}
\ForEach{$\iset$}{ 
    \ForEach{$\jset$}{
        $\overall[(\pCU,\pDU,\xVC)] \leftarrow \varnothing$, $\overall[(\pRU)] \leftarrow j$ \label{algline:ran-31}\\
        \tcc{Unit Placement}
            Prioritize virtual configurations $\kmset$ based on slice type and demand. \label{algline:ran-32}\\
            Retrieve candidate nodes $\mVP_{cand}$ for CU and DU for $i,j$ \label{algline:ran-33} \\
            \ForEach{$\kset$}{ \label{algline:ran-34}
                \tcc{Peform placement for both units $\unitt$ CU and DU}
                $ aux^{cu}, aux^{du}, PP[cu], PP[du]\leftarrow$ \Placement($i,j,k, \mVP_{cand}, \unitt, \demand, \compC$) \label{algline:ran-35}\\
                \If{$aux^{cu}$ AND $ aux^{du}$ not empty}{
                    $\overall[(\pCU,\pDU,\xVC)] \leftarrow aux^{cu}, aux^{du}, k$ \label{algline:ran-37}\\
                    $\ppusage \leftarrow PP[cu], PP[du]$  Update placement and computing utilization. \label{algline:ran-38}\\
                }
            }
    }
}
\tcc{Routing}
\ForEach{$\iset$}{ \label{algline:ran-39}
    \ForEach{$\jset$}{ \label{algline:ran-40}
        $\overall[(\rfh,\rmh,SD,DD)] \leftarrow \varnothing$ \label{algline:ran-41}\\
        \tcc{ Perform routing for FH and MH segments}
        $aux^{FH}, aux^{MH}, \Load[FH], \Load[MH] \leftarrow$ \Routing $(\gmset, i,j,k, \segment, \overall[(\pCU,\pDU,\pRU)])$ \label{algline:ran-42}\\
        \If{$aux^{FH}$ AND $aux^{MH}$ not empty}{
            $\overall[(\rfh,\rmh)] \leftarrow aux^{FH}, aux^{MH}$ Update routing  \label{algline:ran-44}\\
             $\linkusage \leftarrow \Load[FH], \Load[MH]$ Update link usage \label{algline:ran-45}\\
        }
        \Else{
            Add configuration $(i,j,k,aux^{FH}, aux^{MH})$ in ignored $\ignored$ \label{algline:ran-47}\\
            Update computing usage $\ppusage$ regarding $i,j,k$ \label{algline:ran-48}\\
            
        }
    }
}
{Calculate static and dynamic $SD, DD$ latency per $i,j$} \label{algline:ran-49}\\
{return $\overall, \ppusage, \linkusage, \ignored$} \label{algline:ran-50}\
\end{algorithm}


The \texttt{RAN Configuration} algorithm, presented in \cref{alg:Main}, takes as input the network graph $\gmset$, traffic demand $\demand$, the sets of RUs $\jmset$, slices $\imset$, and virtual configurations $\kmset$. The algorithm outputs the selected RAN configurations in $\overall$, the processing utilization of PP nodes in $\ppusage$, and the transport-network link utilization in $\linkusage$.
For each gNB--slice pair $(i,j)$, the structure $\overall$ stores the selected unit placement for the CU and DU, represented by $\pCU$ and $\pDU$, respectively, the selected virtual configuration $\xVC$, and the FH and MH routing paths, represented by $R^{FH}$ and $R^{MH}$. Since the RU is always located at the cell site, 
$\pRU$ is fixed and included in $\overall$ for completeness. More formally, the structure can be represented as $\overall[(\pCU,\pDU,\pRU,\xVC,\rfh,\rmh)]$.


In \cref{algline:main-1}, the algorithm invokes the \texttt{Configuration Handler} module, which generates an initial feasible solution by assigning RAN configurations and updating the utilization states of the PP and transport-link resources. Additionally, the function returns the structure $\ignored$, which stores non-configured gNB--slice pairs $(i,j)$ together with their corresponding infeasible configurations, i.e., $\pCU$, $\pDU$, $\xVC$, $\rfh$, and $\rmh$. These configurations are excluded because they violate one or more model constraints and are therefore retained for consideration in subsequent iterations.

Next, in \cref{algline:main-2}, the algorithm iterates over the non-allocated gNB--slice pairs stored in $\ignored$. During each iteration, the \texttt{Configuration Handler} is invoked again in \cref{algline:main-3}, now considering the previously ignored configurations. This iterative process continues until all gNB--slice pairs are successfully configured, thereby producing an initial solution.

After obtaining this initial solution, the algorithm invokes the \texttt{Energy Minimization} function in \cref{algline:main-4}. This function takes as input the network graph, the slice set, the virtual configurations, and the outputs generated in the previous phase: $\overall$, $\ppusage$, $\linkusage$, and $\ignored$. The \texttt{Energy Minimization} function then refines the placement, VC selection, and routing decisions to minimize overall energy consumption while preserving resource constraints. Finally, the optimized solution, including the updated placement and utilization metrics, is returned in $\overall$, $\ppusage$, and $\linkusage$.

The \texttt{Configuration Handler} in \cref{alg:Configuration}  performs virtual configuration selection, CU and DU placement, and FH and MH routing.

The first line verifies whether any ignored configurations, represented by $\ignored$, have been provided as input arguments. If so, \cref{algline:ran-27-conf} retrieves the set of gNB--slice pairs $(i,j)$ to be reallocated from the $\ignored$ structure. Subsequently, \cref{algline:ran-28-conf} retrieves the set of RAN configurations $\overall^{ign}$ that must be excluded from subsequent processing steps.

Next, an iteration is performed for each gNB $\iset$ and slice $\jset$ in order to allocate RAN configurations. In \cref{algline:ran-31}, the $\overall$ structure associated with $\pCU$, $\pDU$, and $\xVC$ is initialized for the pair $(i,j)$. In the case of $\overall[(RU)]$, the corresponding gNB cell site is allocated.

In \cref{algline:ran-32}, the set $\kmset$ is organized according to the requirements of slice $j$. For example, for \acs{eMBB} slices, the set $\kmset$ remains unchanged, since these slices may require up to three PPs due to their high traffic and computational demands. In contrast, for \acs{URLLC} slices, VC options $k$ in which the DU is located closer to the cell site are prioritized. Similarly, \acs{mMTC} slices prioritize single-split configurations because of their lower computational requirements.
Subsequently, \cref{algline:ran-33} retrieves the candidate nodes $\mVPP_{cand}$ for $\pCU$ and $\pDU$ according to both the distance to the cell site and the current node utilization percentage, prioritizing nodes that are simultaneously closer and more highly utilized. Then, \cref{algline:ran-34} iterates over the ordered set of VC options $\kset$ for each slice. In \cref{algline:ran-35}, the \texttt{Placement} function is invoked to obtain a temporary allocation of CU and DU nodes, denoted by $aux^{cu}, aux^{du} \in \mVPP$, as well as the temporary total computational utilization of these nodes, represented by $PP[cu]$ and $PP[du]$. Finally, the temporary allocation is verified for feasibility. If the allocation is feasible, both the overall allocation structure and the computational utilization structures are updated in \cref{algline:ran-37,algline:ran-38}.

The \texttt{Placement} and \texttt{Routing} functions are included in \cref{alg:Functions} and are detailed as follows.
The \texttt{Placement} function, described in Lines 1--\ref{algline:ran-10}, determines the allocation nodes for the CU and DU while ensuring that the computational capacity constraints are satisfied. Initially, the processing load $\ppaux$ required by each RAN unit is computed using the $\funPrc$ function in \cref{algline:ran-2}.
Subsequently, \cref{algline:ran-3} iterates over the set of candidate nodes that can host the DU or CU. For each candidate node, \cref{algline:ran-4} computes the total computational load $\PP[cand]$ by considering both the current utilization and the additional required load $\ppaux$.
Then, \cref{algline:ran-5} verifies whether the resulting computational utilization remains within the node capacity constraints. If the constraint is satisfied, the corresponding RAN unit is allocated to the candidate node in \cref{algline:ran-6}. Otherwise, the node is considered unsuitable, and $\placednode$ is assigned an empty value in \cref{algline:ran-8}. Finally, the selected node $\placednode$ is returned by the function.

The \texttt{Routing} function, defined in \cref{algline:ran-11}--\cref{algline:ran-29}, is responsible for selecting and returning the paths associated with both the FH and MH segments. Initially, in \cref{algline:ran-11}, the source and destination nodes are retrieved from the $\overall[(CU,DU,RU)]$ structure according to the selected segment $\segment$ (i.e., FH or MH). Subsequently, candidate paths are computed using Dijkstra’s algorithm in \cref{algline:ran-12}.
For each candidate path, the static latency $\latS$ is calculated in \cref{algline:ran-14}. If the computed latency is lower than the admissible threshold, \cref{algline:ran-17} computes the required transmission load $\Load[temp]$ for each edge belonging to the path by means of the $\funTr$ function.
Next, in \cref{algline:ran-18}, the total transmission load on each edge, denoted by $\Load[cand]$, is obtained by adding the transmission demand of the current edge $\Load[temp]$ to the existing link utilization $\linkusage$. The resulting load $\Load[cand]$ is then compared against the corresponding link capacity $\linkC$. If the capacity constraint is satisfied, the candidate path is allocated in \cref{algline:ran-20}; otherwise, $\allocatedpath$ is assigned an empty value in \cref{algline:ran-22}. Furthermore, if no feasible candidate path $\pathh[cand]$ is identified, $\allocatedpath$ is also set to empty in \cref{algline:ran-24}.
In \cref{algline:ran-27}, each feasible $\allocatedpath$ is assigned a score according to the number of slices traversing the nodes along the path, such that paths shared by a larger number of slices receive higher scores. Then, in \cref{algline:ran-28}, the variable $\bestpath$ is updated with the candidate path that achieves the highest score. Finally, the selected path $\bestpath$ and the associated transmission load $\Load[cand]$ are returned in \cref{algline:ran-29}.

The \texttt{Energy Minimization} function, defined in \cref{alg:Energy}, aims to minimize the overall energy consumption of the RAN while satisfying all model constraints. The function receives as input the number of iterations $IT$, the initial solution structures $\overall$, $\ppusage$, and $\linkusage$, as well as the set of non-working configurations $\ignored$. Initially, the system energy consumption is computed in \cref{algline:en-2}. Subsequently, the algorithm iterates over the predefined number of iterations $IT$, sorting the PP utilization in descending order to obtain $\ppusage^{ord}$, thereby prioritizing the least utilized processing pools, as shown in \cref{algline:en-4}.

\begin{algorithm}[!t]
\footnotesize	
\caption{Placement and Routing Module}
\label{alg:Functions}
\SetAlgoLined
\SetKwInput{KwIni}{Init}
\KwIn{Network graph $\mathcal{G}$, RUs set $\jmset$, Slices set $\imset$, Virtual Configurations set $\kmset$, PP set $\mVP$, PP capacity $\compC$}
\KwOut{Placement, VC and routing result $\overall$,  resources utilization $\ppusage$, $\linkusage$, ignored configurations $\ignored$, Energy consumption }
\KwIni{$\ppusage$, $\linkusage$}
\SetKwFunction{Placement}{Placement}
\SetKwFunction{Routing}{Routing}
\SetKwFunction{Dijkstra}{Dijkstra}

\SetKwProg{Fn}{Function}{:}{}
\Fn{\Placement{$i,j,k, \mVP_{cand}, \unitt, \demand, \compC$}}{
    {$\ppaux \leftarrow$ Calculate unit processing load with $\funPrc$ \label{algline:ran-2}}\\
    
    \ForEach{$v^{cand}$ in $\mVP_{cand}$}{ \label{algline:ran-3}
        {$\PP[cand] \leftarrow \ppaux + \ppusage[v^{cand}]$} \label{algline:ran-4}\\
        \uIf{$\PP[cand] \leq \compC$ \tcp*{\label{algline:ran-5}}}{
            $\placednode = v^{cand} \label{algline:ran-6}$
        }
        \Else{
            $\placednode$ = None \label{algline:ran-8}
        }  
    }
    {return $\placednode, \PP[cand]$}\ \label{algline:ran-10}
}

\Fn{\Routing{$\gmset, i,j,k, \segment, \overall[(\pCU,\pDU,\pRU)])$}}{
    {$src, dest \leftarrow$ Retrieve source and destination from $\overall[(\pCU,\pDU,\pRU)]$} \label{algline:ran-11}\\
    {$\paths \leftarrow dijkstra(G,src,dest)$}\label{algline:ran-12}\\
    {$\bestpath \leftarrow$ None}\\
    
    \ForEach{$\pathh[cand]$ in $\paths$}{ 
        $\latS \leftarrow$ Calculate static latency \label{algline:ran-14} \\
        \uIf{$\latS \leq \latMax{max}$}{
            \ForEach{$e$ in $\pathh[cand]$}{
                $\Load[tmp] \leftarrow$ Calculate transmission load with $\funTr$ \label{algline:ran-17}\\
                $\Load[cand] \leftarrow \Load[temp] + \linkusage$ \label{algline:ran-18} \\
                \uIf{$\Load[temp] \leq \linkC$}{
                    $\allocatedpath = \pathh[cand]$ \label{algline:ran-20}
                }
                \Else{
                    $\allocatedpath$ = None \label{algline:ran-22}
                }
            }
        }
        \Else{
            $\allocatedpath$ = None \label{algline:ran-24}
        }  
        $\pathscore \leftarrow$  score($\allocatedpath$)  If the path edges are already used \label{algline:ran-27}\\
        $\bestpath \leftarrow \allocatedpath$ Path with best score \label{algline:ran-28}
    }
    {return $\bestpath, \Load[cand]$} \label{algline:ran-29}\
}

\end{algorithm}

\begin{algorithm}[!t]
\footnotesize	
\caption{Energy Minimization}
\label{alg:Energy}
\SetAlgoLined
\SetKwInput{KwIni}{Input}
\KwIni{Demand $\demand$, PP capacity $\compC$, Link capacity $\linkC$}
\KwOut{RAN configuration $\overall$, PP utilization $\ppusage$, Link utilization $\linkusage$, ignored configurations $\ignored$, Energy consumption $\energy$}

\SetKwFunction{EnergyMin}{EnergyMin}
\SetKwFunction{Configuration}{Configuration Handler}
\SetKwProg{Fn}{Function}{:}{}
\Fn{\EnergyMin{$IT, \overall, \ppusage, \linkusage, \ignored$}}{
    {Calculate overall RAN energy $\energy[]$} \label{algline:en-2} \\
    \ForEach{$it$ in $IT$}{ 
        $\ppusage^{ord} \leftarrow \ppusage$ Order descending less used PPs \label{algline:en-4}\\
        \ForEach{$v^{ord} \in \ppusage^{ord}$}{
            $\mathcal{S}_{i,j}^{ord} \leftarrow v^{ord}$ Retrieve slices $i,j$ in $v^{cord}$ \label{algline:en-6}\\
            \ForEach{$s_{i,j} \in \mathcal{S}_{i,j}^{ord}$}{
                \tcc{Save backup of working solutions}
                $\overall'' \leftarrow \overall, \quad \ppusage'' \leftarrow \ppusage,  \quad\linkusage'' \leftarrow \linkusage$\label{algline:en-8}\\
                $\ignored'' \leftarrow \ignored, \overall$ Copy and include slice configs. \label{algline:en-9}\\
                $\imset', \jmset' \leftarrow s_{i,j}$ Retrieve $i,j$ elements \label{algline:en-10}\\
                $\overall', \ppusage', \linkusage', \ignored'  \leftarrow$ \Configuration($\gmset, \jmset, \imset, \kmset$) \label{algline:en-11}\\
                \If{$\ignored'$ empty}{
                    $\overall \leftarrow \overall'$ Update complete solution \label{algline:en-13} \\
                    Calculate new RAN energy consumption $\energy'(\overall)$ \label{algline:en-14} \\
                    \If{$\energy' < \energy$}{
                        $\ppusage \leftarrow \ppusage', \linkusage \leftarrow \linkusage''$ Update PP and link utilization \label{algline:en-16}\\
                    }
                }
                \Else{ \label{algline:en-17}
                    \tcc{Roll back to backup}
                    $\overall \leftarrow \overall'', \quad \ppusage \leftarrow \ppusage'', \quad \linkusage \leftarrow \linkusage''$ \label{algline:en-18}\\
                }
            }
        } 
    }
    {return $\overall, \ppusage, \linkusage$ \label{algline:en-19}}
}
\end{algorithm}


For each PP $v^{ord}$ in the ordered PP list, \cref{algline:en-6} retrieves the set of gNB-slices $i,j$ allocated to that PP together with their current configurations, storing them in $\mathcal{S}_{i,j}^{ord}$. Then, for each retrieved slice $s_{i,j}$, \cref{algline:en-8} stores backup copies of the current solutions to enable rollback in case the reconfiguration process does not improve the objective value. Similarly, \cref{algline:en-9} creates a backup of $\ignored$ and inserts the current configuration into this structure to prevent repeated evaluations during subsequent configuration attempts. In \cref{algline:en-10}, the gNB-slice elements are retrieved as sets to be reconfigured. Next, \cref{algline:en-11} invokes the \texttt{Configuration Handler}, updating the structures $\overall$, $\ppusage$, and $\linkusage$. If no pending configurations remain during the process, the new solution $\overall'$ (considering only $\imset'$ and $\jmset'$) is merged into the previous solution in \cref{algline:en-13}. The updated energy consumption $\energy'$ is then computed in \cref{algline:en-14}. Whenever the new energy consumption is lower than the previous value, \cref{algline:en-16} updates the PP and link utilization structures accordingly.

If the reconfiguration process fails, as indicated in \cref{algline:en-17}, the algorithm restores the previous feasible state by rolling back to the backup configuration in \cref{algline:en-18}. The procedure continues until all possible slice placements have been evaluated. Finally, the algorithm returns the optimized configuration along with the updated PP and link-utilization structures, as presented in \cref{algline:en-19}.



\section{PERFORMANCE EVALUATION}
\label{sec:Results}

This section evaluates the proposed \ac{MILP} model and heuristic under two realistic network scenarios. It first describes the evaluation setup, including RAN resource capacities, RF parameters, traffic demand generation, and execution environment. It then compares the performance of the two approaches, optimal and heuristic, across two network topologies and three objective functions.

The numerical analysis proceeds as follows. First, the heuristic is assessed in terms of computational time and solution quality relative to the \ac{MILP} optimum. Next, RAN energy consumption and latency are analyzed across the three objective functions to examine their trade-offs. The evaluation then investigates VC selection and centralization ratio for each network slice, followed by an analysis of computational and transport resource utilization as a function of traffic demand.


\subsection{EXPERIMENTS DESCRIPTION}
\label{sec:Scenario}

\begin{figure}[!t]
    \centering
    \subfloat[Hierarchical topology]{%
        \includegraphics[width=0.2\textwidth]{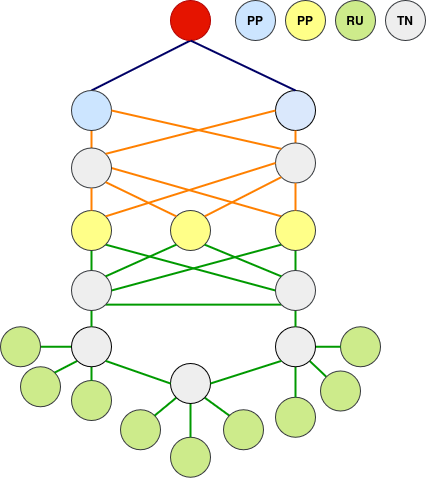}%
        \label{fig:hierarchical}
    }
    \subfloat[Mesh topology]{%
        \includegraphics[width=0.2\textwidth]{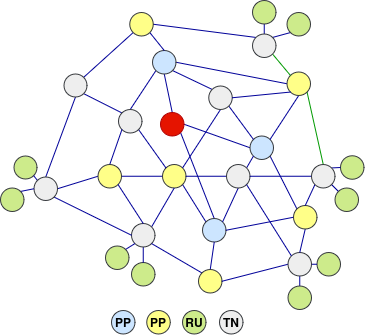}%
        \label{fig:mesh}
    }
    \caption{Evaluated topologies}
    \label{fig:topologies}
\end{figure}

Two types of network topologies are considered to evaluate the proposed model: hierarchical and mesh, as illustrated in Fig.~\ref{fig:topologies}. The hierarchical topology is derived from the PASSION project \cite{PASSION-D2}, whereas the mesh topology is based on \cite{klinkowsi2023} and reflects practical deployment scenarios. Both topologies are designed in accordance with the architectural guidelines for access and aggregation networks specified in \cite{ORAN-WG9_XHaul_transport}.

The network topologies are modeled as graphs, as illustrated in Fig.~\ref{fig:topologies}, where each node can operate either as PP or as a router. Green nodes correspond to cell sites that host RUs, which provide both RF and processing functions. Yellow and blue nodes represent PPs, while gray nodes denote routing devices. The red node corresponds to the 5G Core.
Each PP is equipped with a set of Intel Xeon Gold 6140 processors \cite{Moreira}, each providing a computational capacity of 864 GOPS \cite{intel2024appmetrics}. In the evaluated scenarios, blue nodes are provisioned with 5 processors, whereas yellow nodes and RUs are provisioned with 2 processors each. According to \cite{SPECpowerssj2008}, an Intel Xeon Gold 6140 processor exhibits an average power consumption of approximately 343 W under full load and 52.4 W in the idle state. In addition, the chassis power consumption is assumed to be 146 W, with line card consumption accounted for within the chassis \cite{daSilva_energy}, and an incremental power consumption of 3.5 W per active port \cite{ciscoASR9000guide}.

The FH and MH segments employ Ethernet-based links, with transmission capacities summarized in Table~\ref{tab:link_capacity}, following the assumptions in \cite{klinkowsi2023}. The optical link lengths are derived from the PASSION project specifications. The propagation delay is assumed to be $5\mu$s/km, while the store-and-forward delay is set to $5\mu$s. An Ethernet frame size of 1542 bytes is considered in the evaluation.

\begin{table}
    \centering
    \caption{Link capacities}
    \begin{tabular}{cc}
        \toprule
         \textbf{Link type} & \textbf{Capacity [Gbps]}\\
         \midrule
         Router - Router & 100 \\
         Router - PP & 200 \\
         PP - RU & 25 \\
         PP - PP & 200 \\
         \bottomrule
    \end{tabular}
    \label{tab:link_capacity}
\end{table}


The proposed model is evaluated in a realistic 5G RAN scenario using RF configurations derived from real-world deployments as in \cite{rochman2024-real_implementation} and standardized specifications such as 3GPP TS 36.213. Consistent with \cite{rochman2024-real_implementation}, standalone RAN implementations consider heterogeneous RF configurations across RUs. The key RF parameters include channel bandwidth (BW), modulation scheme (Mod), antenna configuration (NL), and numerology ($\mu$). The coding rate ($R$) and the number of PRBs are obtained from \cite{3GPP_PRB_TS_38_211}. The complete set of RF configurations adopted for the RUs in both topologies is summarized in Table~\ref{tab:5g_configurations}.

\begin{table}[!t]
\begin{center}
\caption{Configurations per RU}
\resizebox{\columnwidth}{!}{%
    \begin{tabular}{cccccccc} 
    \toprule
      \ \textbf{\% RUs} &  \textbf{BW} &  \textbf{Mod} &  \textbf{NL} &  \textbf{R} &  \textbf{PRB} &  \textbf{$\mu$} & \textbf{N$_{SC}$} \\ 
    \midrule
    50 & 100 & 264 & 8 & 1/3 & 273 & 1& 3276 \\ 
    40 & 40 & 64 & 4 & 3/4 & 100 & 1 & 1200 \\ 
    10 & 20 & 64 & 4 & 2/3 & 100 & 0 & 1200 \\
    \bottomrule
    \end{tabular}}
    \label{tab:5g_configurations}
\end{center}
\end{table}

Per-gNB traffic demand is generated through synthetic trace synthesis based on the Madrid dataset \cite{madrid_dataset}, which provides millisecond-resolution Radio Resource Control (RRC) traces collected from six 4G BSs in Madrid between 2020 and 2021. The selected BSs represent urban and rural deployments operating at carrier frequencies from 796 to 2650 MHz. To adapt the 4G traces to a 5G scenario, the procedure described in this section is applied. The original traces are aggregated into 15-minute intervals, and synthetic demand series are generated using an inverse empirical cumulative distribution function (eCDF) and a moving-block bootstrap, preserving both the marginal traffic distribution and temporal correlations.

\begin{figure*}[t]
    \centering
    \includegraphics[width=0.8\linewidth]{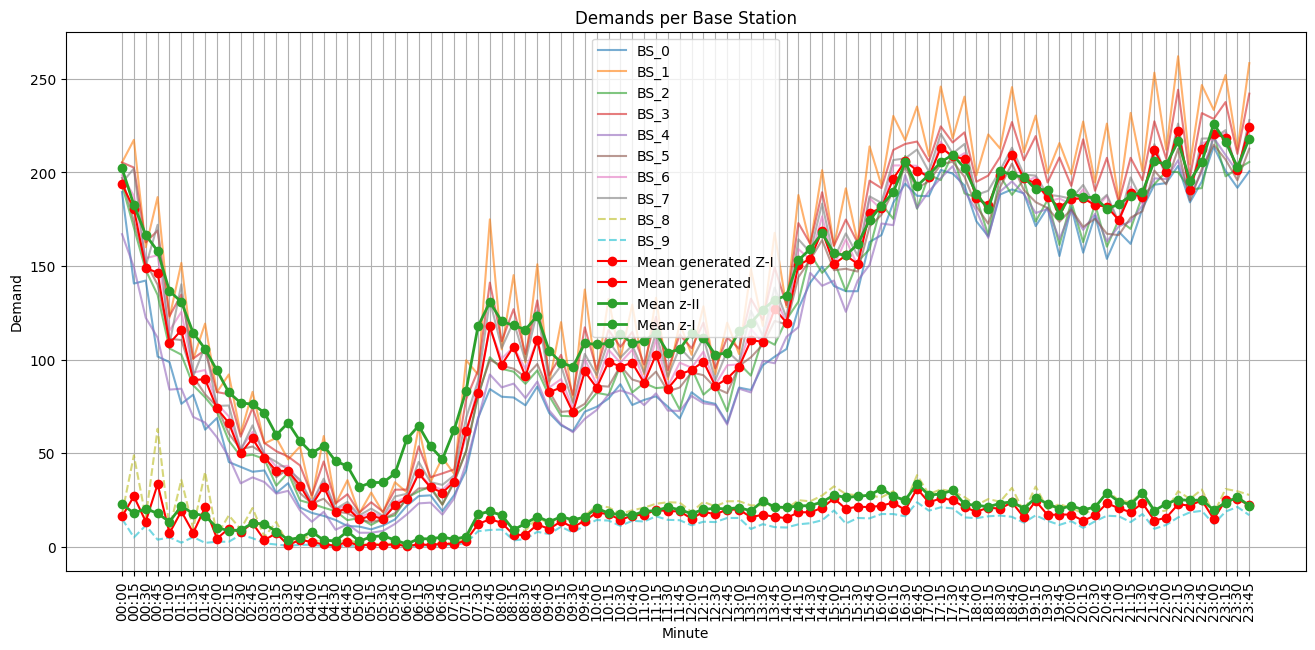}
    \caption{Synthetic data for nine BS.}
    \label{fig:data_synt}
\end{figure*}

The generation process first maps the aggregated traffic samples to percentile space using the eCDF computed for each time interval. Overlapping traffic blocks are then extracted and recombined using a moving-block bootstrap. Candidate blocks are selected by minimizing the mean squared error (MSE) over the overlap, ensuring temporal consistency between adjacent segments. The resulting normalized sequence is transformed back to traffic values using the inverse eCDF and scaled according to the peak data rates of different RF configurations \cite{3gpp_peak_38306_V18_4_0_2025}, representing both urban and suburban deployment scenarios. \Cref{fig:data_synt} illustrates examples of the synthesized traffic profiles. This procedure generates a full-day downlink demand profile with 15-minute resolution for each RU.

The MILP model and heuristic were implemented in Python. For topologies with fewer than 30 nodes, the MILP formulation was solved using Gurobi under an academic license on a Dell OptiPlex 3000 equipped with a 12th Gen Intel\textsuperscript{\textregistered} Core\textsuperscript{\texttrademark} i5-12500 processor (12 cores) and 16~GB of RAM. For larger topologies, the MILP experiments were executed on a server with an Intel\textsuperscript{\textregistered} Xeon\textsuperscript{\textregistered} E5-2660 v4 processor (56 cores) and 225~GB of RAM. The heuristic was evaluated on the former platform for all topologies. For each objective function and heuristic configuration, the algorithms were executed 96 times, corresponding to the 15-minute intervals of a 24-hour demand cycle. The implementation is publicly available at \url{https://github.com/lrc-unicamp/dynamic-disaggregated-ran}.


\subsection{HEURISTIC EVALUATION}

The heuristic is evaluated by its computational efficiency and solution quality, quantified by the solving time and the energy consumption deviation from the MILP-optimal solution.
Table~\ref{tab:heuristic_performance} compares the two approaches across six network instances. Each instance is identified by the topology type (hierarchical or mesh) and the corresponding number of nodes. The energy deviation is reported as the minimum, mean, and maximum percentage values

\begin{table}[t]
\caption{MILP vs Heuristic Performance}
\small
\begin{tabular}{lrcccc}
\hline
\multicolumn{1}{c}{\multirow{2}{*}{\textbf{Topology}}} & \multicolumn{2}{c}{\textbf{Solving time {[}s{]}}}      & \multicolumn{3}{c}{\textbf{Error {[}\%{]}}} \\
\cline{2-6}
\multicolumn{1}{c}{}                                   & \multicolumn{1}{c}{\textbf{MILP}} & \textbf{Heuristic} & \textbf{min}       & \textbf{mean}       & \textbf{max}       \\
\hline
Hierarchical22                                         & 343                             & \textless{}1       & 2.1                & 4.06               & 15.47              \\
Hierarchical28                                         & 1052                            & \textless{}1       & 1.65               & 5.02               & 13.92              \\
Hierarchical34                                         & 10307                           & 1.4                & 02.04              & 10.2                & 26.96              \\
Mesh29                                                 & 1471                            & 2.4                & 7.7                & 8.87                & 10.84              \\
Mesh35                                                 & 10794                           & 2.7                & 3.29               & 6.4                 & 11.72      \\
\hline
\end{tabular}
\label{tab:heuristic_performance}
\end{table}

For network instances with up to approximately 30 nodes, the MILP solver obtains the optimal solution within 30 min, which is acceptable for offline planning by \acp{MNO}. However, the computational time increases significantly for larger instances, exceeding 2 hours for Hierarchical34 and Mesh35, while no feasible solution is found for Hierarchical40 within the 10-hour time limit. In contrast, the heuristic solves all instances in less than 10 seconds on average, reducing the computational time by several orders of magnitude.

Despite the substantial reduction in computational time, the heuristic maintains high solution quality across all evaluated instances. For mesh topologies, the energy gap remains consistently below 9\% on average and does not increase with network size, demonstrating stable performance across Mesh29 and Mesh35. For hierarchical topologies, the mean gap increases from 4.06\% for the smallest instance to 10.2\% for Hierarchical34, indicating a moderate degradation in solution quality as the network scales. Although a maximum gap of 26.96\% is observed for Hierarchical34, this corresponds to a worst-case scenario rather than typical behavior, as reflected by the considerably lower mean deviation. Overall, the heuristic provides a practical and, for larger network instances, the only viable alternative, given the computational intractability of the MILP formulation beyond approximately 30 nodes.


\begin{figure*}[!ht]
    \centering
    \subfloat[Hierarchical Topology]{%
        \includegraphics[width=0.49\textwidth]{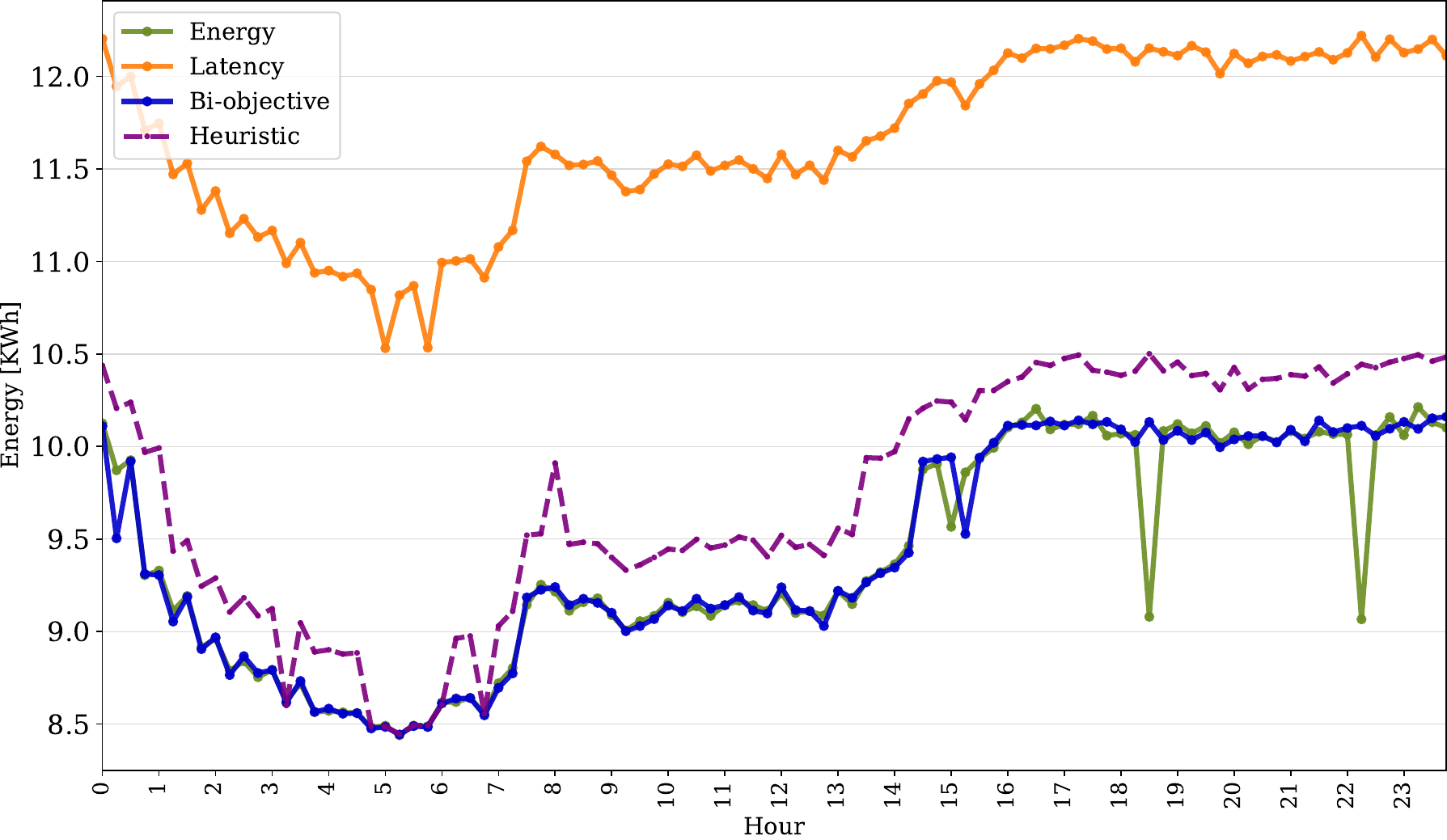}%
        \label{fig:energy_consumption_hierarchical}
    }\hfill
    \subfloat[Mesh Topology]{%
        \includegraphics[width=0.49\textwidth]{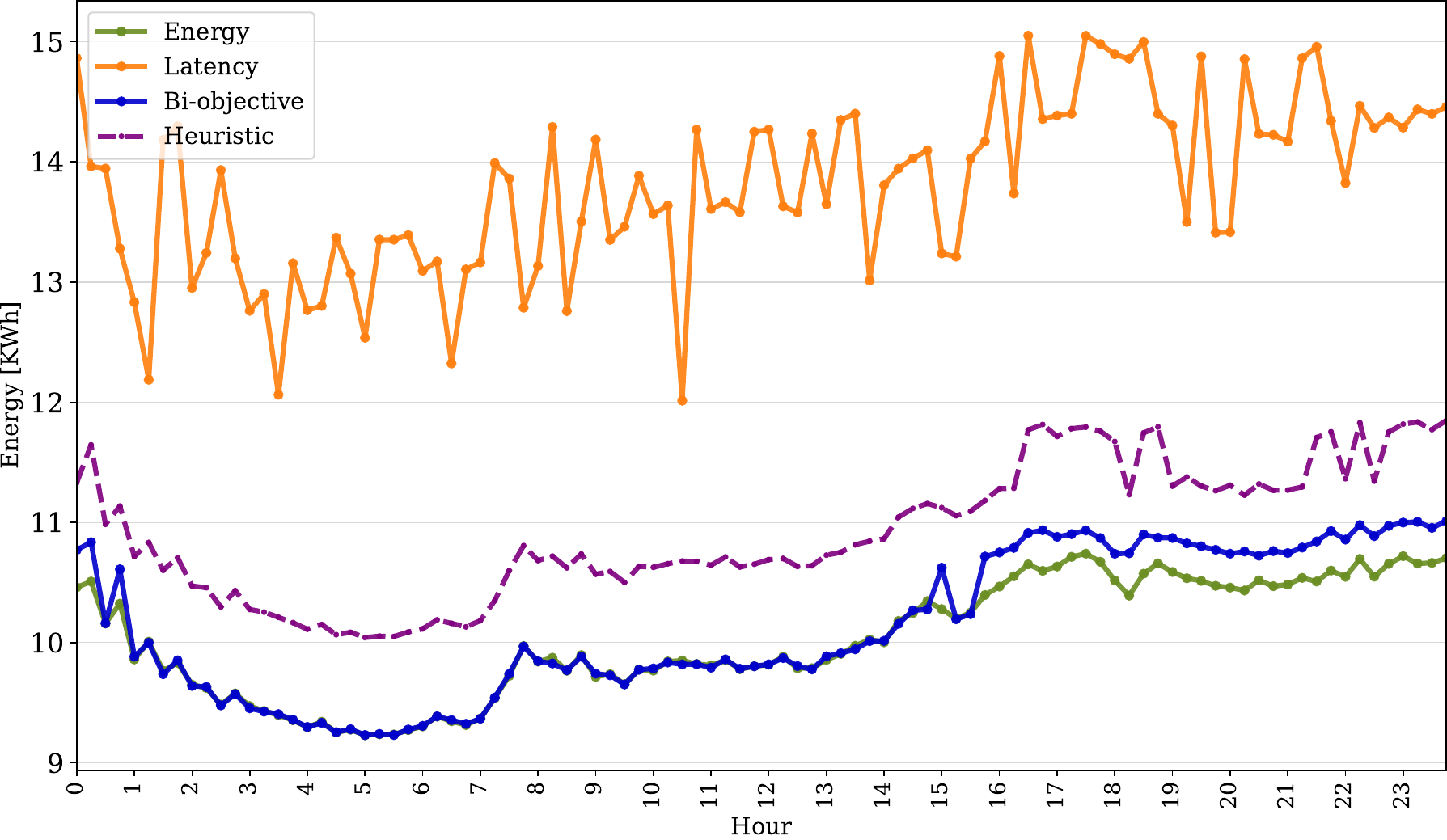}%
        \label{fig:energy_consumption_mesh}
    }
    \par\medskip

    \caption{Energy consumption}
    \label{fig:energy_consumption}
\end{figure*}


\begin{figure*}[!t]
    \centering
    \includegraphics[width=0.7\textwidth]{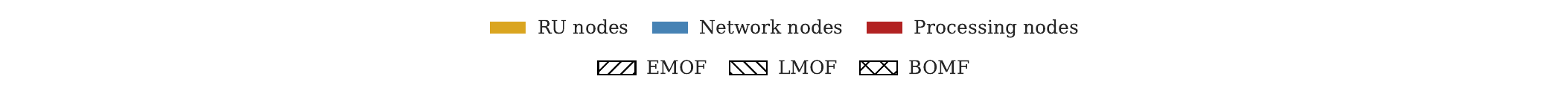}
    \subfloat[Hierarchical topology]{%
        \includegraphics[width=0.48\textwidth]{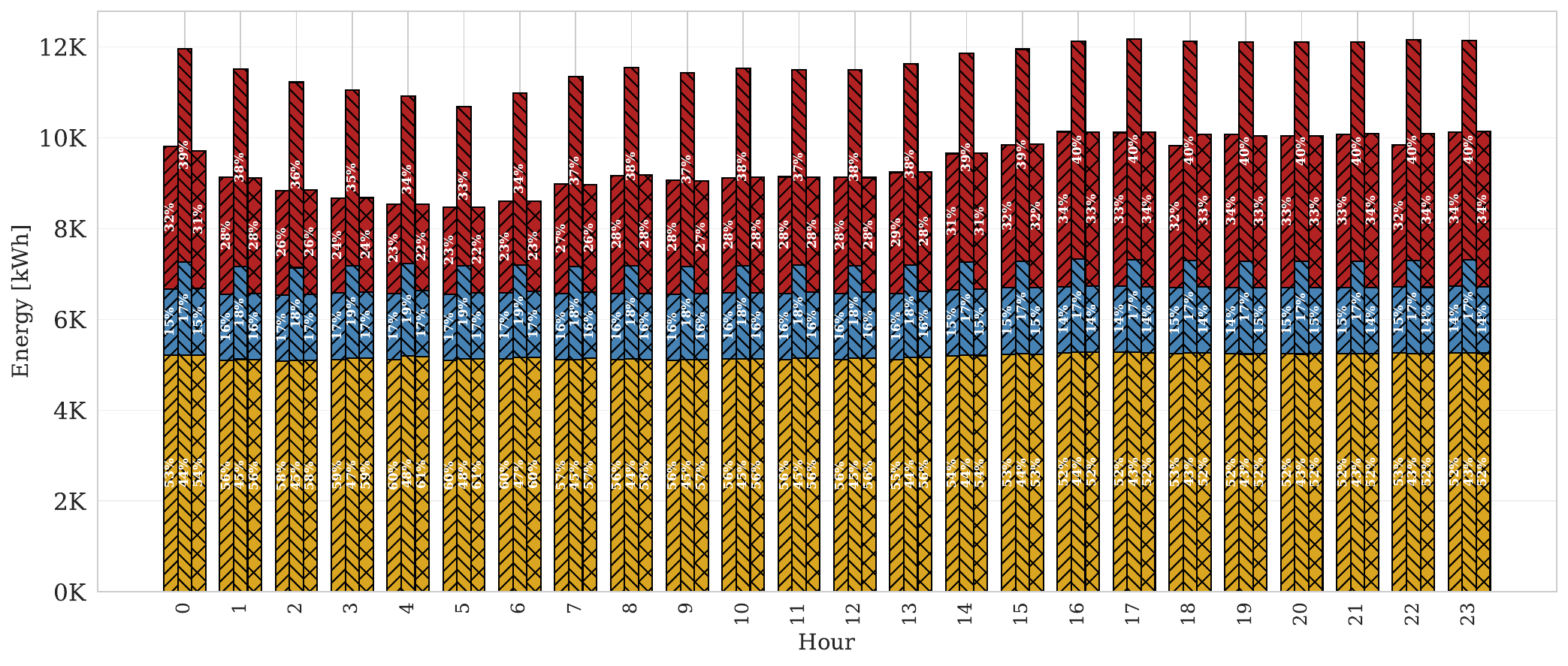}%
        \label{fig:energy_component_hierarchical}
    }
    \subfloat[Mesh Topology]{%
        \includegraphics[width=0.48\textwidth]{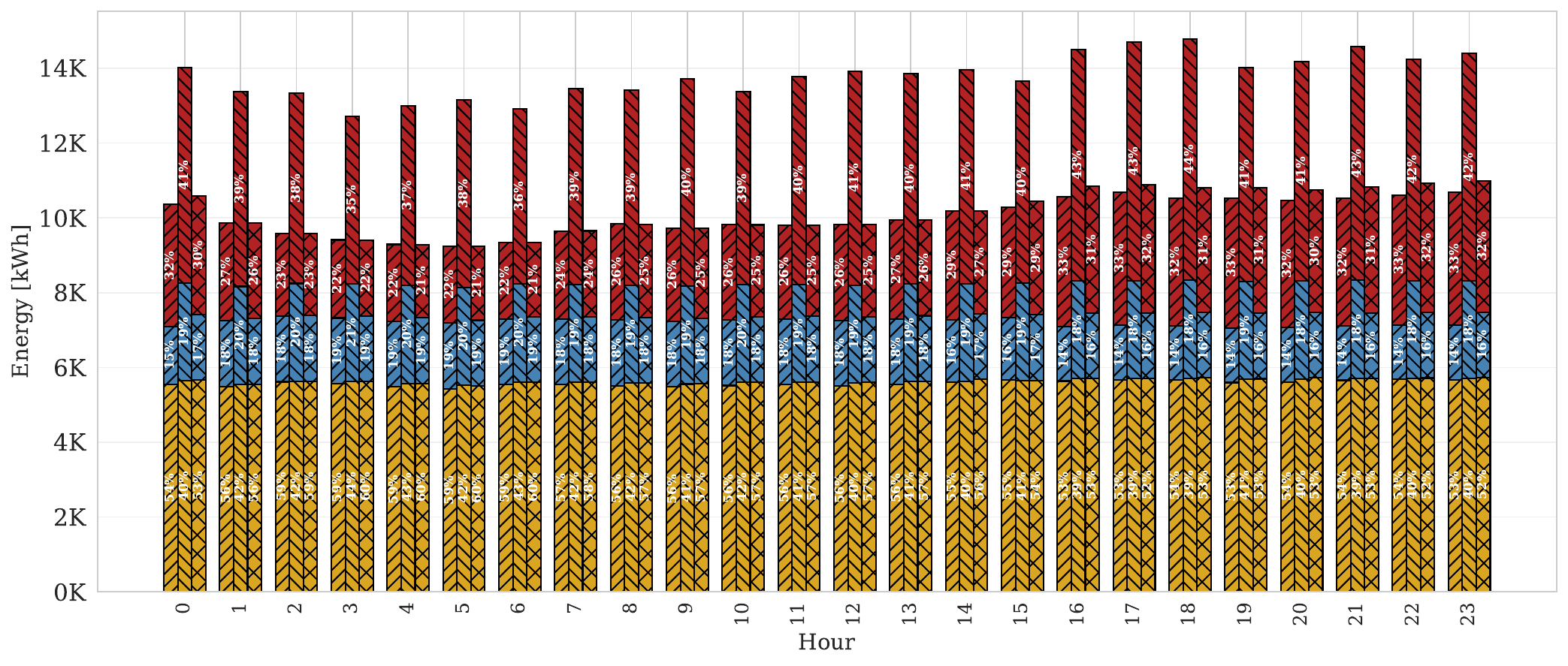}%
        \label{fig:energy_component_mesh}
    }  
    \caption{Energy consumption per component}
    \label{fig:energy_consumption_component}
\end{figure*}

\subsection{ENERGY CONSUMPTION}
\label{sec:energy}
The energy consumption analysis evaluates the temporal evolution of total RAN energy consumption for each optimization objective and the heuristic, including the contributions of individual components. This analysis enables the assessment of trade-offs among optimization strategies and quantifies the impact of RF, network, and processing components on the overall RAN energy consumption.

\Cref{fig:energy_consumption} compares the total RAN energy consumption achieved by Hierarchical topology (HT) and Mesh topology (MT) under \acs{EMOF}, \acs{LMOF}, \acs{BOMF}, and the heuristic approach. For both topologies, \acs{EMOF} consistently achieves the lowest energy consumption, whereas \acs{LMOF} yields the highest values. In contrast, \acs{BOMF} remains closely aligned with \acs{EMOF} throughout the evaluation period.

In the \acs{HT}, \acs{EMOF} reduces energy consumption by 19\% to 35\%, with an average reduction of 24\%, compared with \acs{LMOF} across all evaluated hours. Moreover, the performance of \acs{BOMF} nearly overlaps that of \acs{EMOF}. The hierarchical organization and limited routing alternatives help maintain a relatively stable set of active resources over time. The heuristic solution also closely follows the behavior of \acs{EMOF}, with an average energy gap of approximately 4\% and hourly deviations ranging from 2\% to 15\% above the \acs{EMOF} solution. This behavior is mainly attributed to its prioritization of VNF centralization.

For the \acs{MT}, the distinction among optimization objectives becomes more pronounced, particularly for \acs{LMOF}. The gap ranges from 22\% to 47\%, with a mean difference of 38\% relative to the \acs{EMOF}. This variability results from latency-oriented VC switching decisions, which modify the set of active PPs and consequently their associated power consumption profiles as traffic demand evolves. The \acs{BOMF} exhibits energy consumption values close to those obtained with \acs{EMOF}, with a maximum difference of only 3\%. The heuristic approach presents an average energy gap of 9\%, with hourly differences ranging from 7\% to 11\% above the \acs{EMOF} solution.

\Cref{fig:energy_consumption_component} presents the hourly energy-consumption breakdown by RF, network, and processing components for both topologies. For the two topologies, the RF subsystem represents the dominant contribution to total energy consumption and remains relatively stable over time. This behavior is mainly due to front-end hardware components, such as power amplifiers and converters, which establish a quasi-static energy baseline once the RU is activated. The PP constitutes the second-largest contributor and accounts for most of the observed variability, as it depends directly on traffic demand and VNF placement decisions that change the utilization of active processing resources. By comparison, the network-related energy component exhibits lower variability and is primarily influenced by the subset of active transport elements. Regarding the evaluated optimization objectives, \acs{EMOF} and \acs{BOMF} exhibit nearly overlapping energy profiles across all hours, whereas \acs{LMOF} consistently yields larger swings due to higher sensitivity to routing and placement choices.
As illustrated in \Cref{fig:energy_component_hierarchical}, the \acs{HT} exhibits a more stable behavior over time, with limited fluctuations between hours. Conversely, the \acs{MT}, shown in \Cref{fig:energy_component_mesh}, exhibits greater variation due to its greater sensitivity to routing and placement decisions.

\begin{figure*}[!t]
    \centering
    \subfloat[Energy Minimization]{%
        \includegraphics[width=0.33\textwidth]{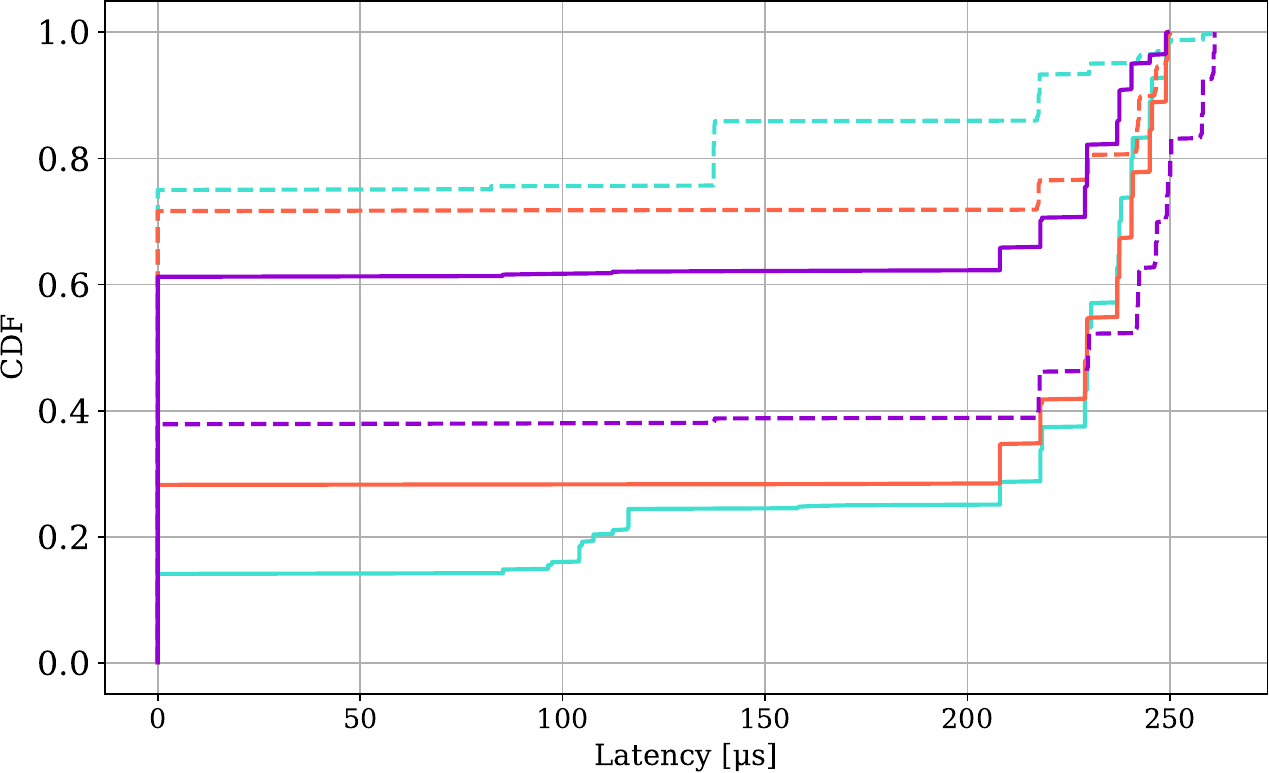}%
        \label{fig:CDF_hierarchical_energ_obj}
    }
    \subfloat[Latency Minimization]{%
        \includegraphics[width=0.33\textwidth]{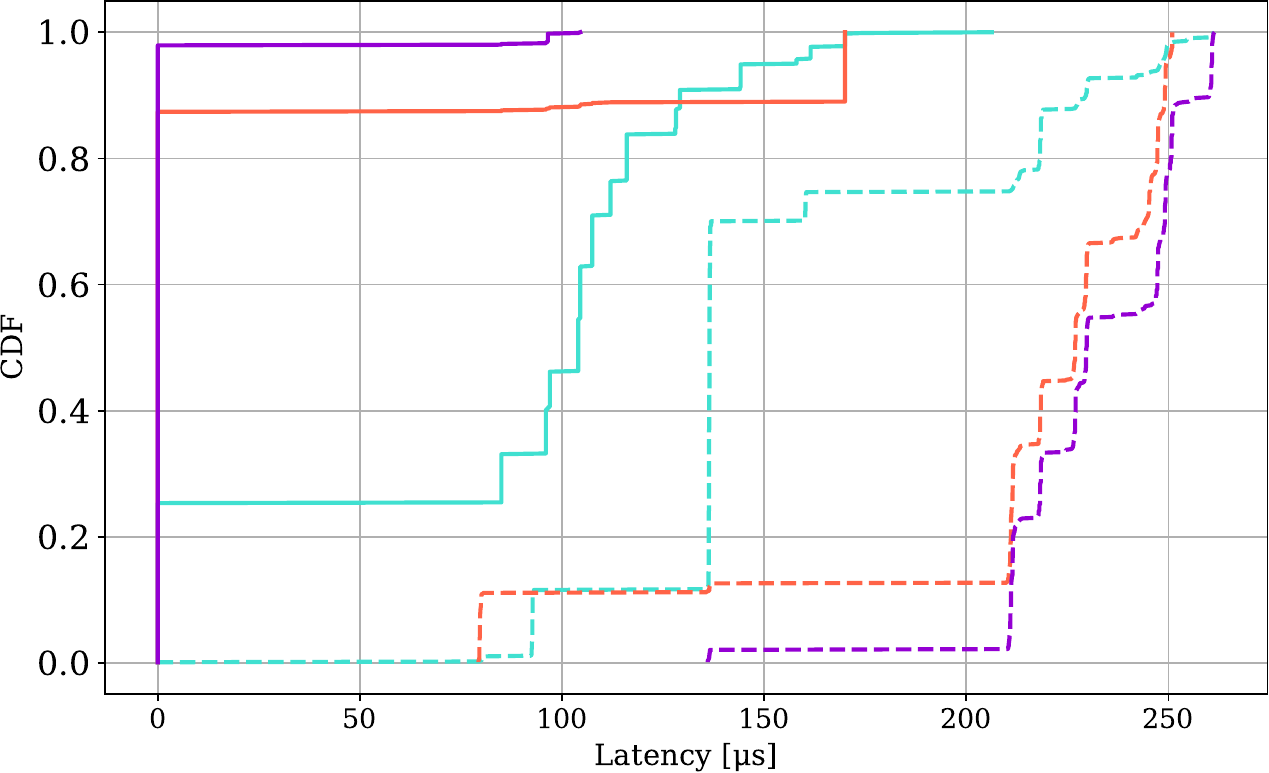}%
        \label{fig:CDF_hierarchical_latency_obj}
    }
    \subfloat[Bi-Objective Minimization]{%
        \includegraphics[width=0.33\textwidth]{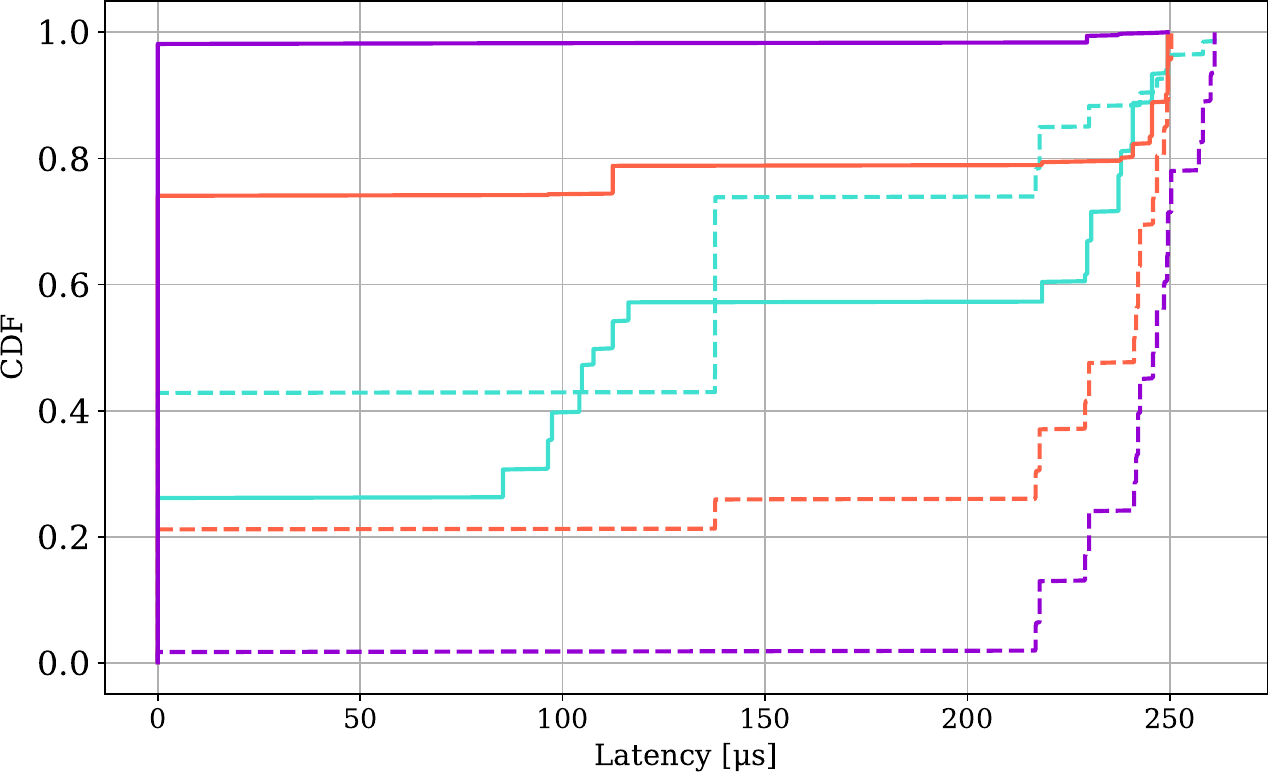}%
        \label{fig:CDF_hierarchical_bi_obj}
    }
    \par\medskip
    \includegraphics[width=0.65\textwidth]{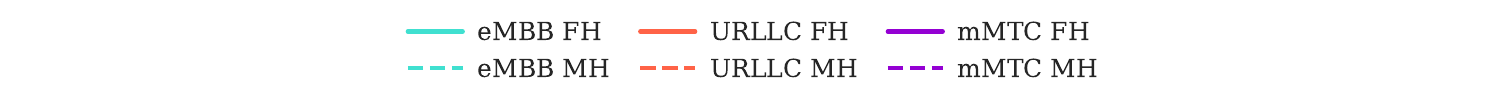}
    \caption{CDF of End-to-End Latency for the Hierarchical Topology}
    \label{fig:CDF_latency_hierarchical}
\end{figure*} 

\begin{figure*}[!t]
    \centering
    \subfloat[Energy Minimization]{%
        \includegraphics[width=0.33\textwidth]{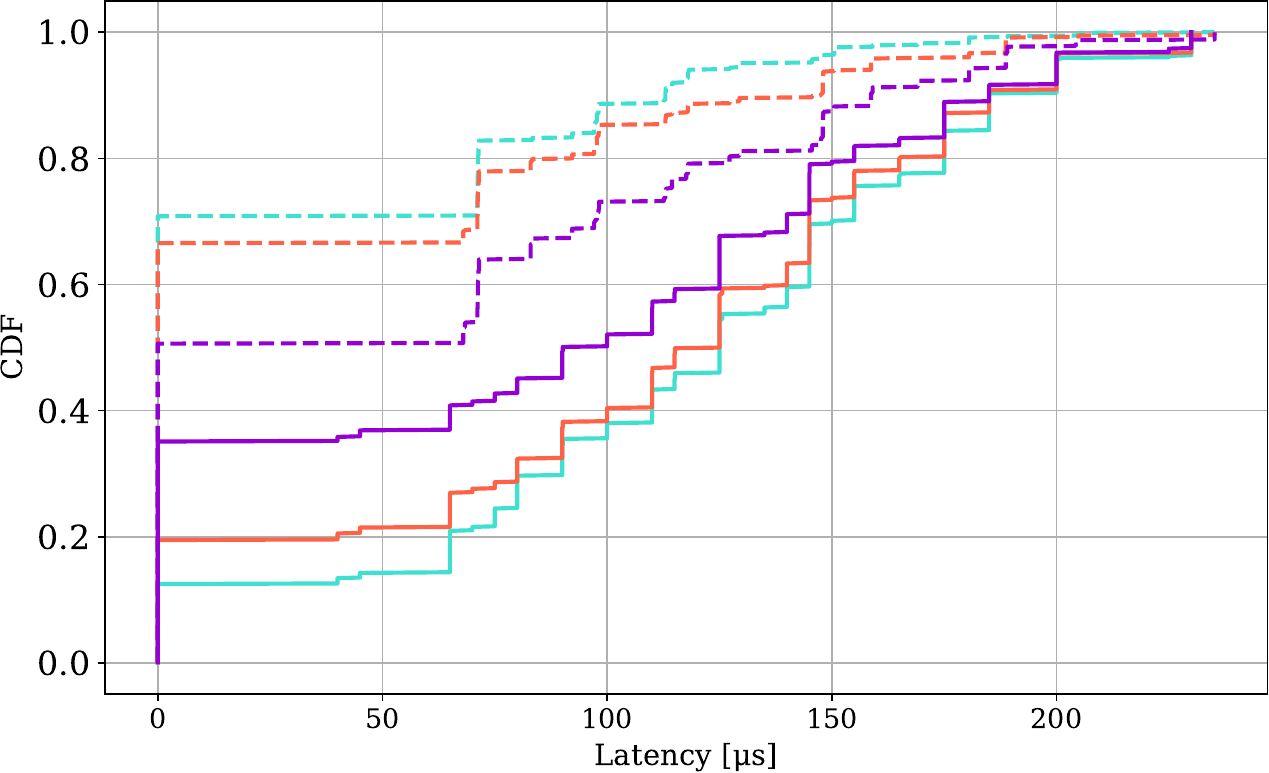}%
        \label{fig:CDF_mesh_energ_obj}
    }
    \subfloat[Latency Minimization]{%
        \includegraphics[width=0.33\textwidth]{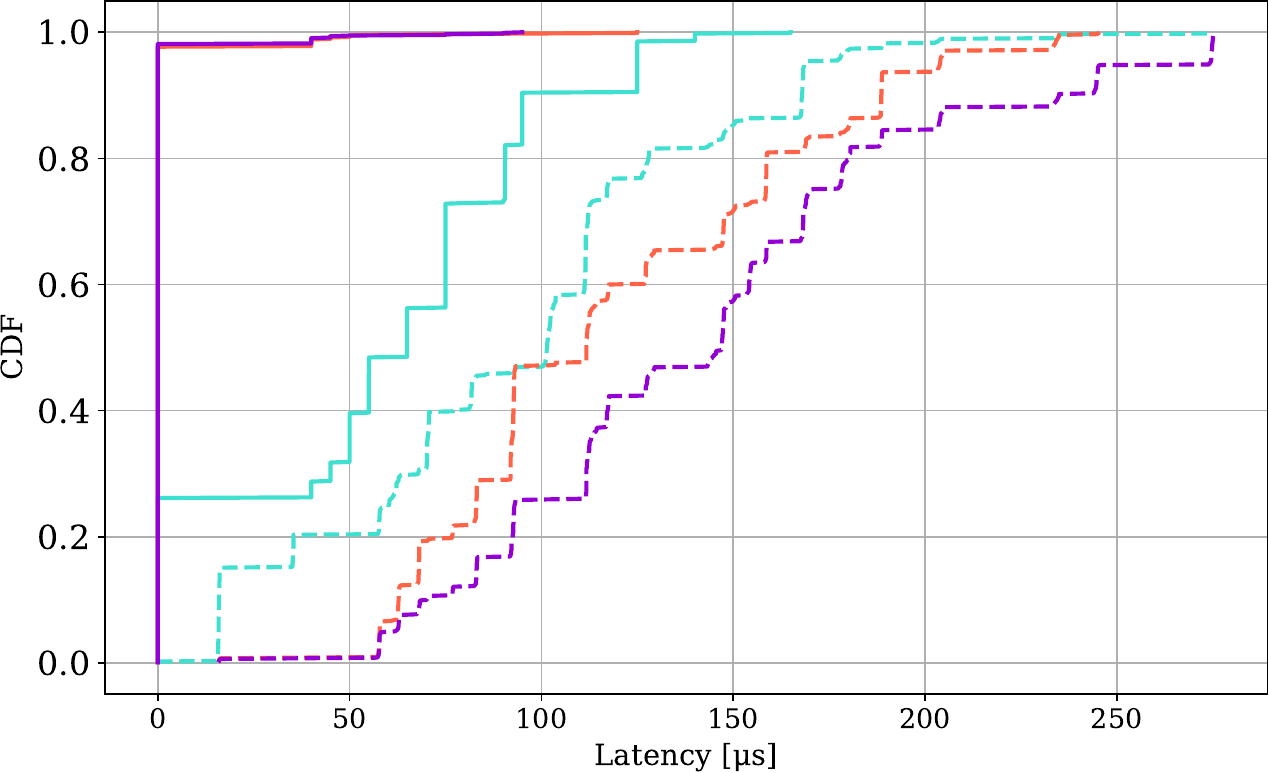}%
        \label{fig:CDF_mesh_latency_obj}
    }
    \subfloat[Bi-Objective Minimization]{%
        \includegraphics[width=0.33\textwidth]{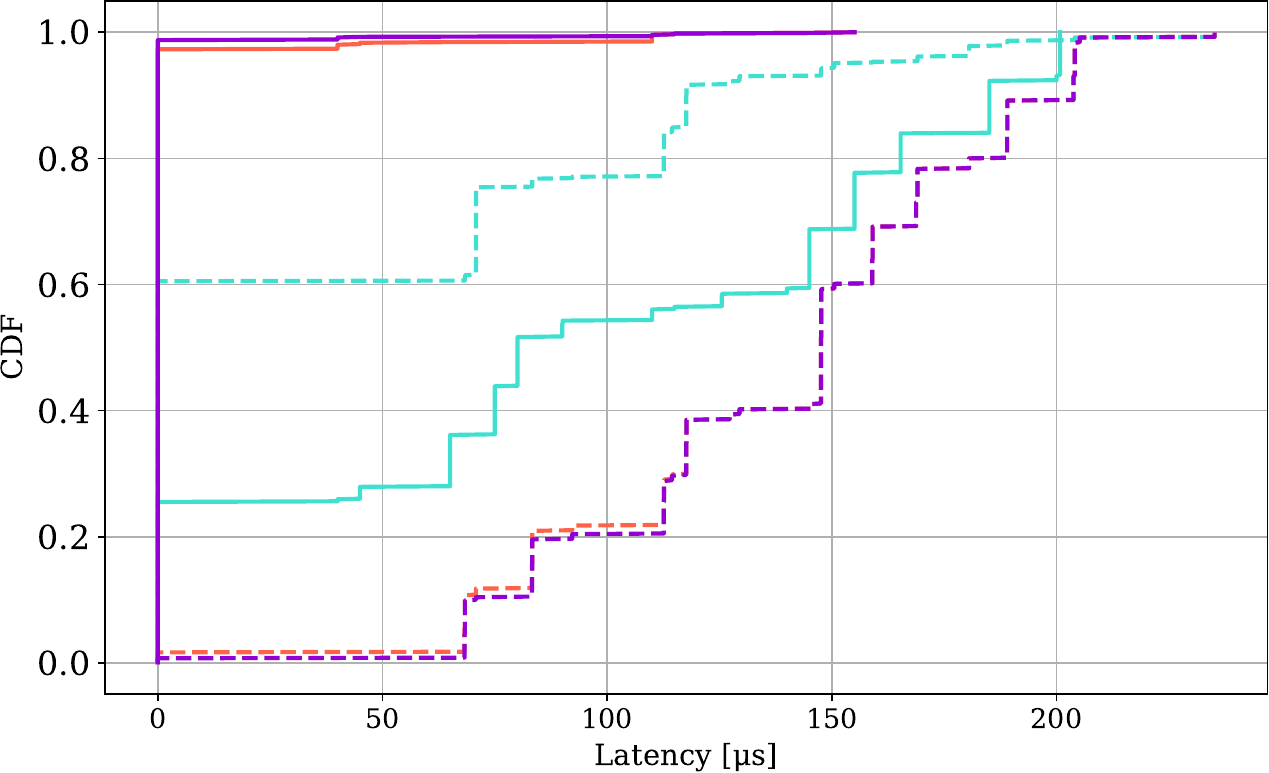}%
        \label{fig:CDF_mesh_bi_obj}
    }
    \par\medskip
    \includegraphics[width=0.65\textwidth]{results/legends/latency_cdf_slice_legend.pdf}
    \caption{CDF of end-to-end Latency for Mesh Topology}
    \label{fig:CDF_latency_mesh}
\end{figure*}

\begin{figure*}[!t]
    \centering
    \includegraphics[width=0.55\textwidth]{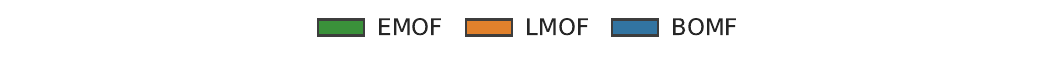}
    \subfloat[FH Boxplot for Hierarchical Topology]{%
        \includegraphics[width=0.48\textwidth]{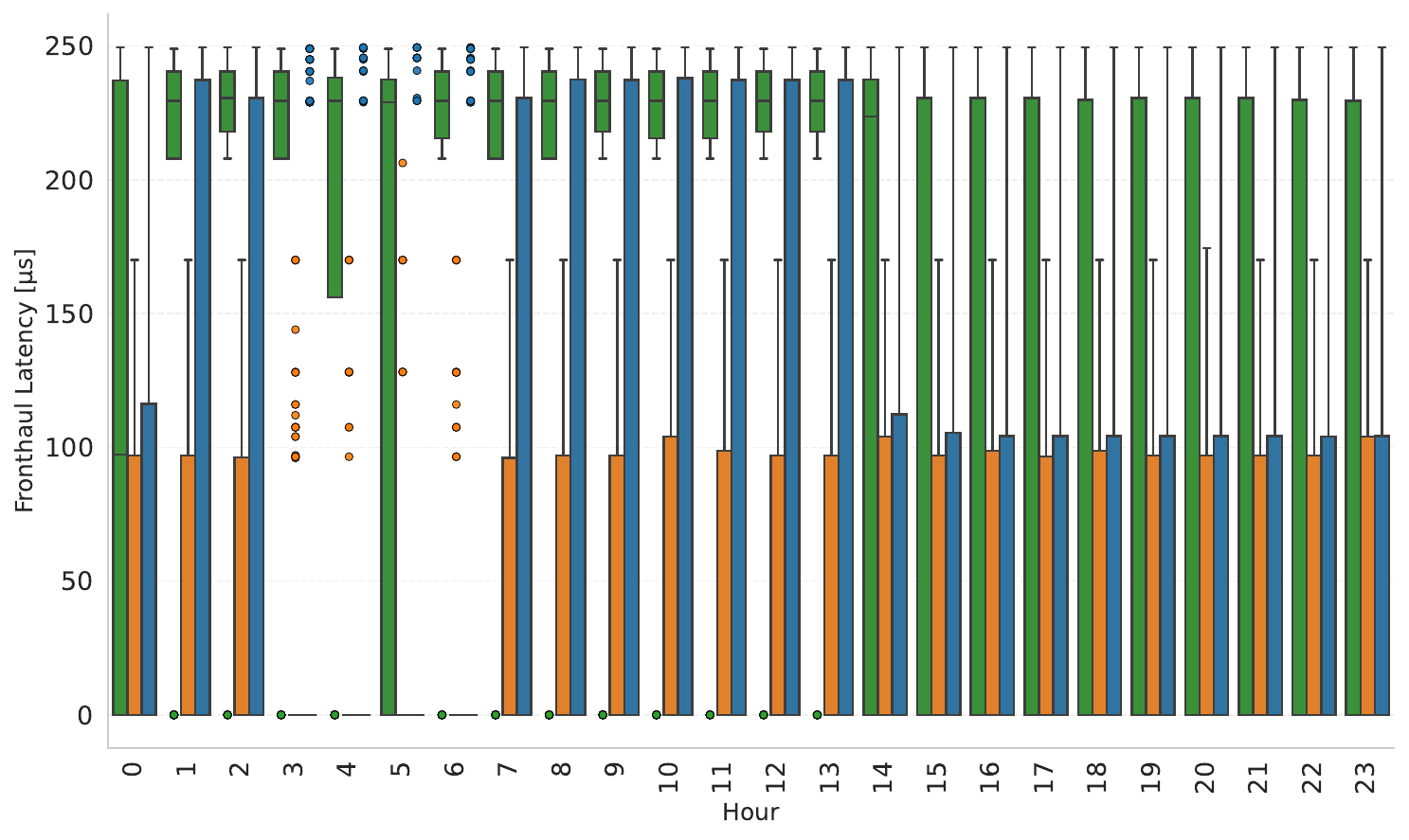}%
        \label{fig:latency_boxplot_FH_hierarchical}
    }
    \subfloat[MH Boxplot for Hierarchical Topology]{%
        \includegraphics[width=0.48\textwidth]{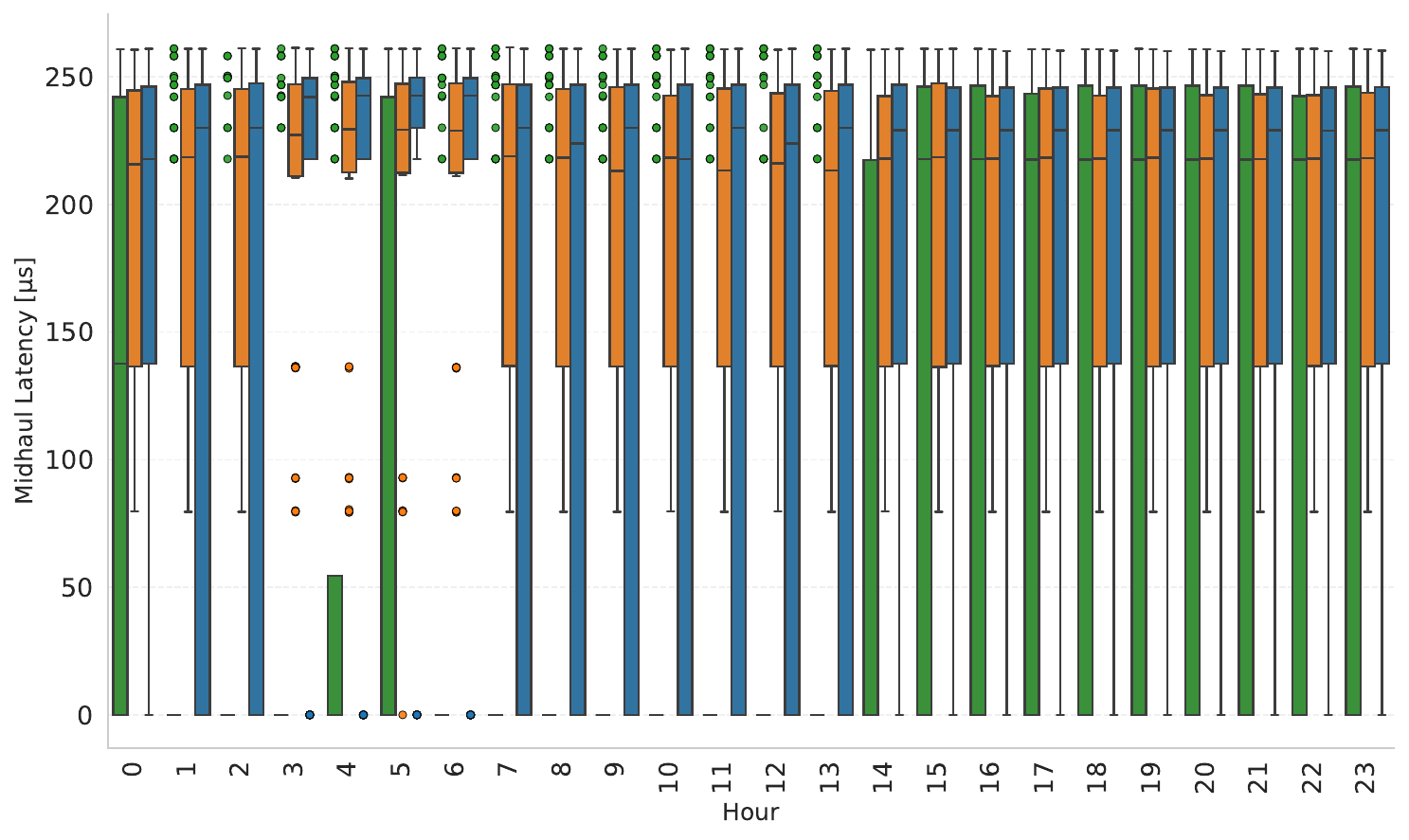}%
        \label{fig:latency_boxplot_MH_hierarchical}
    }
    \caption{Hourly boxplot for FH and MH latency.}
    \label{fig:latency_boxplot_hierarchical}
\end{figure*}

\begin{figure*}[!t]
    \centering
    \includegraphics[width=0.55\textwidth]{results/legends/latency_box_legend.pdf}
    \subfloat[FH Boxplot for Mesh Topology]{%
        \includegraphics[width=0.48\textwidth]{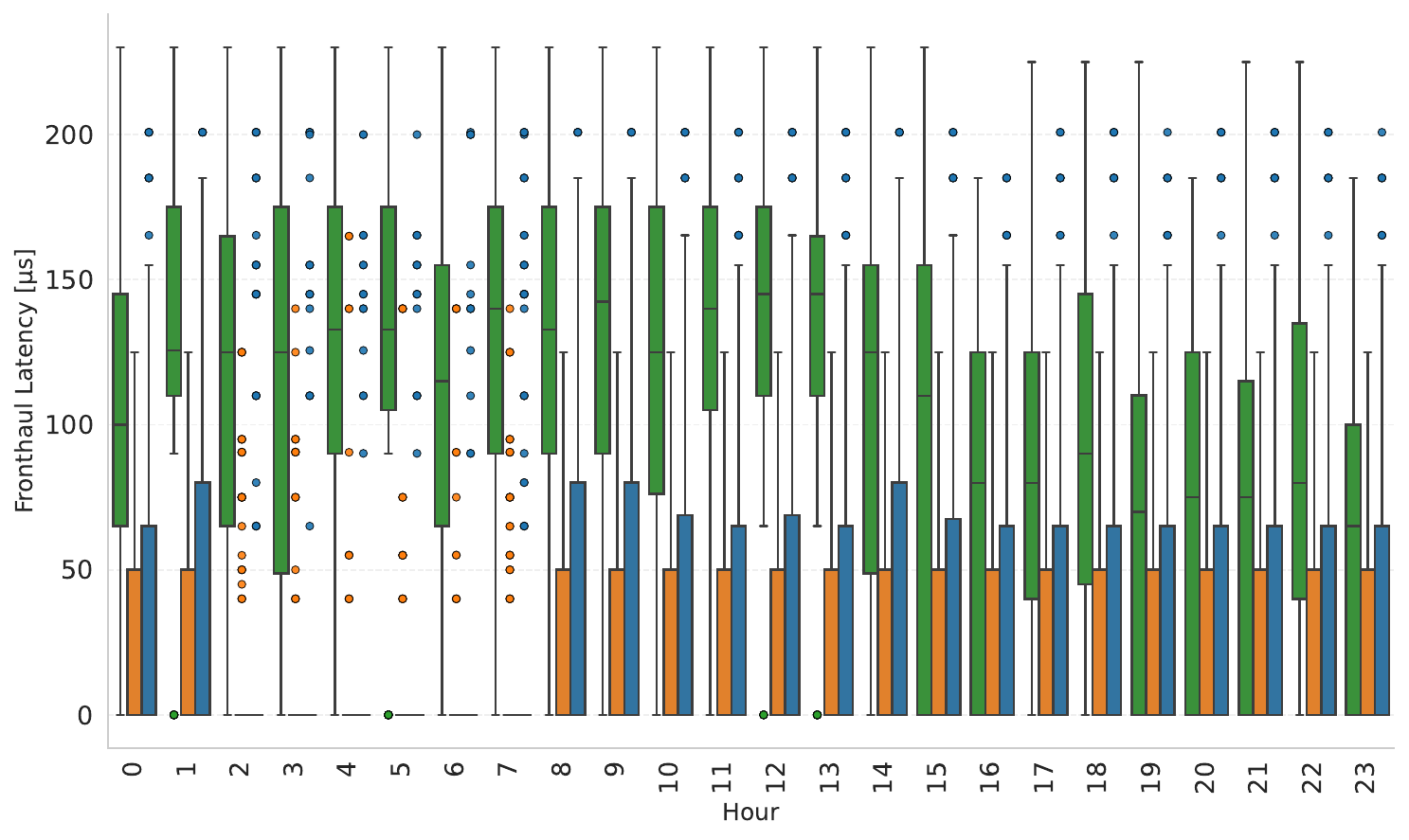}%
        \label{fig:latency_boxplot_FH_mesh}
    }
    \subfloat[MH Boxplot for Mesh Topology]{%
        \includegraphics[width=0.48\textwidth]{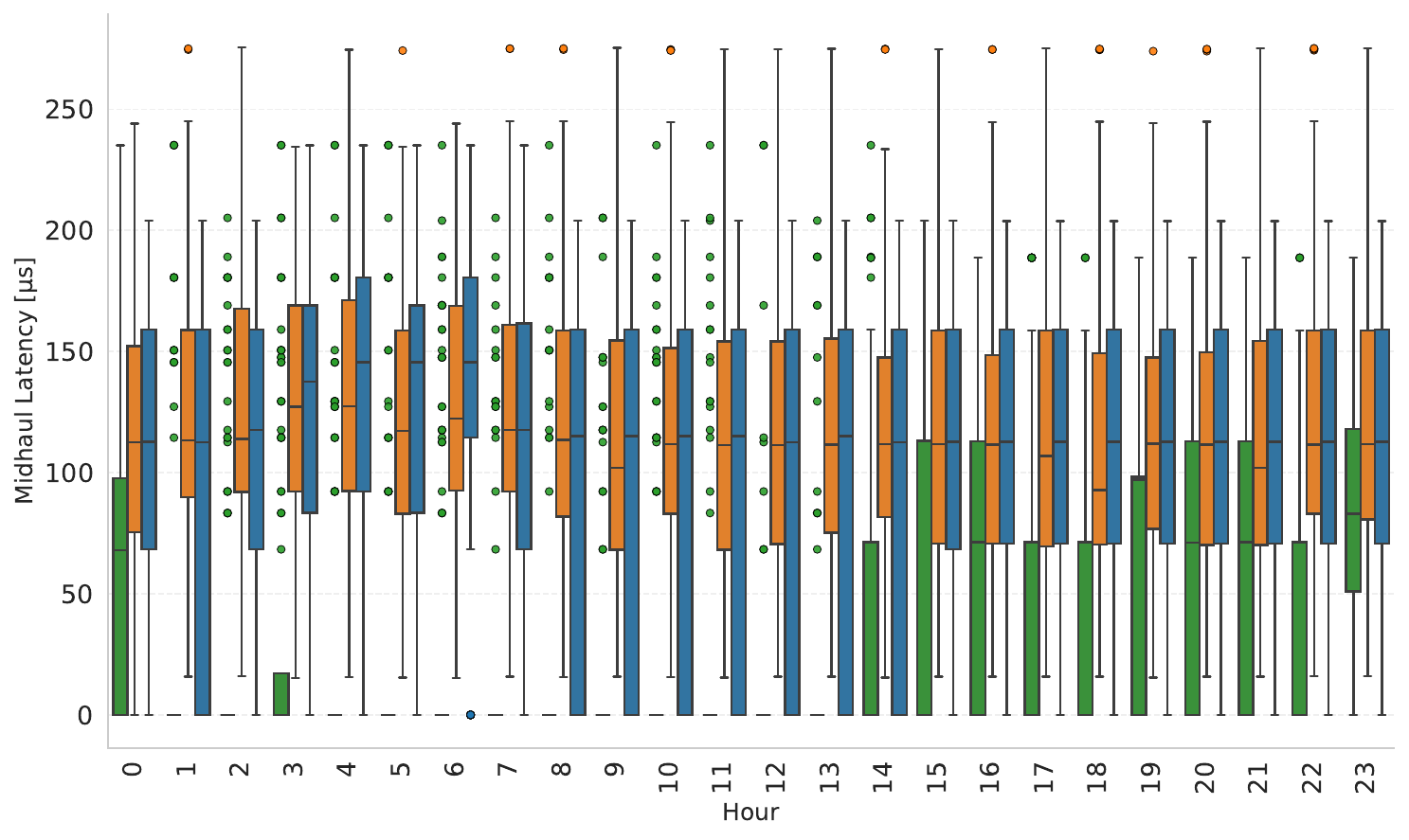}%
        \label{fig:latency_boxplot_MH_mesh}
    }
    \caption{Hourly boxplot for FH and MH latency.}
    \label{fig:latency_boxplot_mesh}
\end{figure*}

\subsection{LATENCY}

This subsection analyzes FH and MH latency from two complementary perspectives: slice requirements and optimization objectives over the course of the day. The slice-oriented analysis examines the impact of service-specific requirements on latency distribution, while the objective-oriented analysis evaluates energy-latency trade-offs under varying traffic conditions. 

\Cref{fig:CDF_latency_hierarchical,fig:CDF_latency_mesh} present the cumulative distribution functions (CDFs) of the FH (solid lines) and MH (dotted lines) latency for each slice under the \acs{HT} and \acs{MT}, aggregated across all RUs and hourly intervals. The three objective functions yield a consistent ordering in the FH segment: \acs{LMOF} achieves the lowest FH latency, \acs{EMOF} achieves the highest FH latency, and \acs{BOMF} achieves intermediate performance. This behavior is directly associated with the selected VCs. Specifically, \acs{EMOF} promotes stronger centralization to minimize the number of active nodes and interfaces, thereby increasing the distance between the RU and DU and driving FH latency close to the imposed constraint (up to approximately $250~\mu s$ in the \acs{HT} and $230~\mu s$ in the \acs{MT}).

Conversely, \acs{LMOF} places processing functions closer to the RU and frequently selects VC~3 (RU/DU co-location), making FH latency negligible for a substantial fraction of URLLC and mMTC samples. As a result, the 95th-percentile FH latency is reduced to approximately $170~\mu s$ in the \acs{HT} and $130~\mu s$ in the \acs{MT}. However, this reduction in FH latency comes at the expense of increased MH latency, since a larger portion of the processing remains at the cell site and the DU--CU segment becomes dominant, particularly for URLLC and mMTC slices. In turn, \acs{BOMF} combines centralized and distributed VC selections across RUs and time intervals, leading to intermediate FH and MH latency characteristics

The slice-level curves further highlight the heterogeneous computational requirements of the services under consideration. URLLC and mMTC frequently operate using VC~3 because their lower computational demands enable processing directly at the cell site, making zero FH latency a common outcome. By contrast, eMBB more frequently adopts centralized or partially distributed VCs (e.g., VC~1, VC~4, and VC~5), as its higher processing requirements limit RU--DU co-location and increase the need for adaptive VC selection.

Complementing the CDF analysis, \Cref{fig:latency_boxplot_hierarchical,fig:latency_boxplot_mesh} illustrate hourly FH and MH latency boxplots under \acs{EMOF}, \acs{LMOF}, and \acs{BOMF}, where each bar aggregates four-slot bins. In the \acs{HT}, \acs{EMOF} sustains high FH latency during low- and medium-demand periods (1h--13h), while MH latency remains close to zero due to DU/CU co-location. As traffic demand increases (14h--0h), FH latency decreases because the DU is progressively offloaded toward the edge. In contrast, \acs{LMOF} consistently maintains low FH latency throughout the day, shifting the dominant contribution to the MH segment. \acs{BOMF} generally follows the behavior of \acs{LMOF}, although it tolerates moderately higher FH latency during low-demand periods when additional centralization can improve energy efficiency. For the \acs{MT}, both FH and MH latency levels are lower and exhibit smaller dispersion under \acs{LMOF} and \acs{BOMF}, mainly because the richer path diversity of the mesh topology reduces transport delay even under more centralized deployments.

\begin{figure}[!t]

    \centering 
    \includegraphics[width=0.5\textwidth]{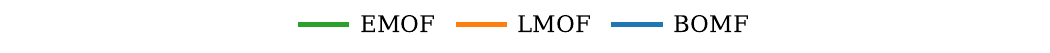}
    \subfloat[Hierarchical Topology]{%
        \includegraphics[width=0.24\textwidth]{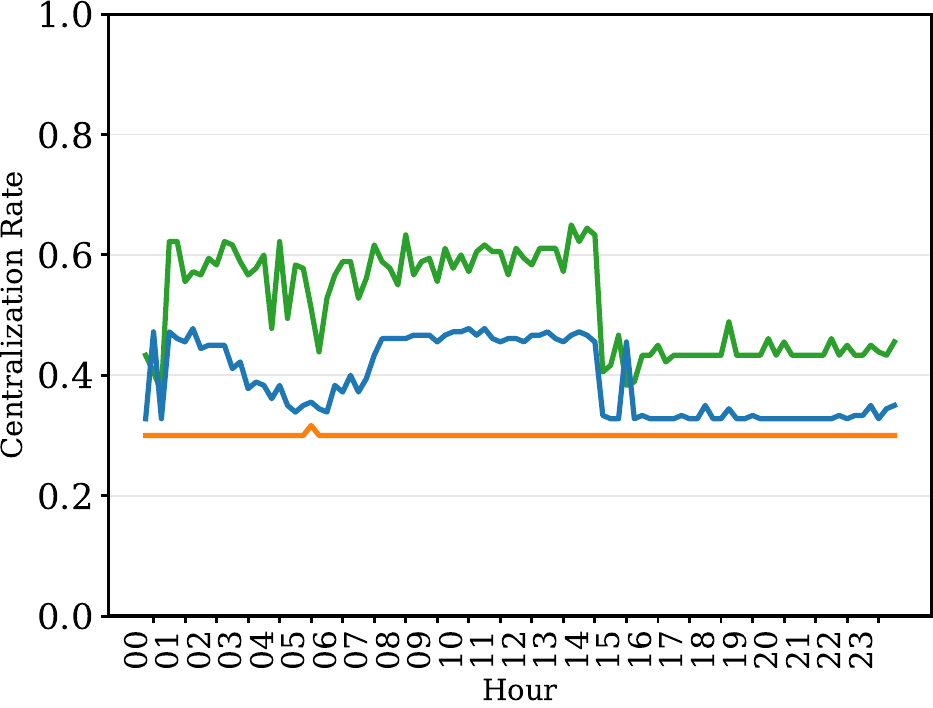}%
        \label{fig:centralization_hierarchical}
    }
    \subfloat[Mesh Topology]{%
        \includegraphics[width=0.235\textwidth]{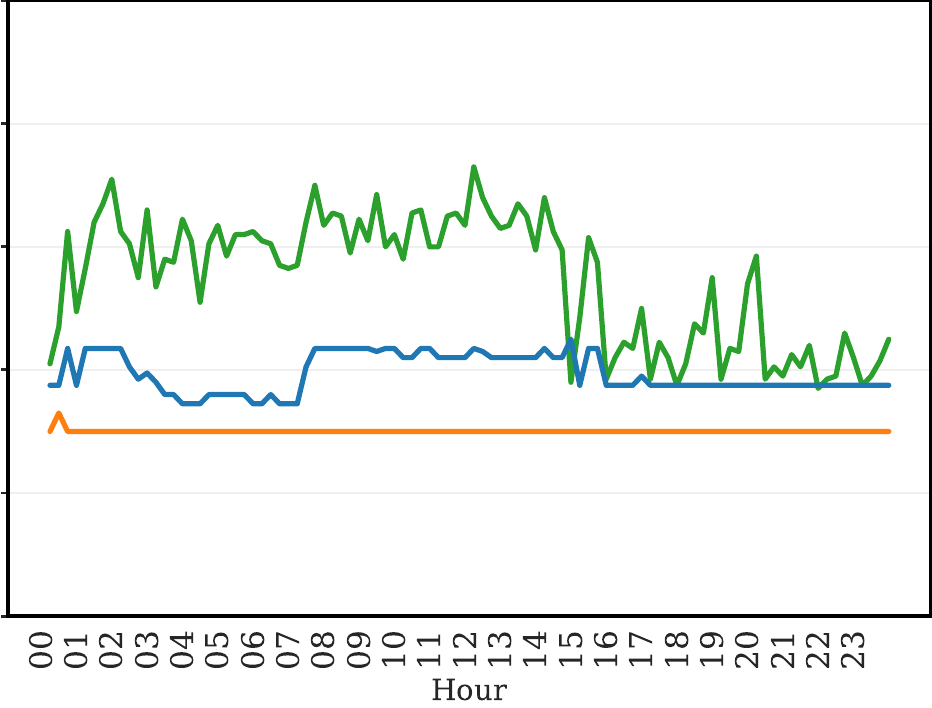}%
        \label{fig:centralization_mesh}
    }    
    \caption{Centralization ratio per objective function.}
    \label{fig:centralization_rate}
\end{figure}

\subsection{CENTRALIZATION RATIO AND VC SELECTION}

The centralization ratio quantifies the proportion of VNFs deployed on core-connected PPs (i.e., at the CU) relative to the total number of VNFs in the system. This metric complements the VC selection analysis by capturing the temporal evolution of RAN functional centralization. Figure~\ref{fig:centralization_rate} presents the centralization ratio obtained for the three objective functions in both hierarchical (Fig.~\ref{fig:centralization_hierarchical}) and mesh (Fig.~\ref{fig:centralization_mesh}) topologies.

Overall, both topologies exhibit a consistent trend across the three objective functions: the \acs{EMOF} achieves the highest centralization ratio, the \acs{LMOF} results in the lowest values, and the \acs{BOMF} yields intermediate behavior. Nevertheless, topology-specific differences emerge when each objective function is examined in greater detail.

With respect to the \acs{EMOF}, although the centralization ratio remains highest in both topologies, distinct behaviors emerge during peak-demand periods (15h–0h). The \ac{HT} exhibits a more stable centralization profile, whereas the \ac{MT} shows greater variability. This difference is attributed to the increased routing and placement flexibility in the mesh topology, which yields multiple feasible solutions with comparable objective values. Consequently, even minor fluctuations in traffic demand may lead to changes in the selected FS.
\begin{figure}[!t]
    \centering
    \includegraphics[width=0.55\textwidth]{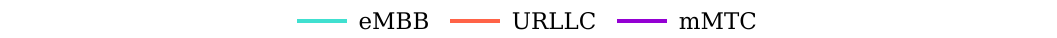}
    \subfloat[Hierarchical - Bi-Objective]{%
        \includegraphics[width=0.48\columnwidth]{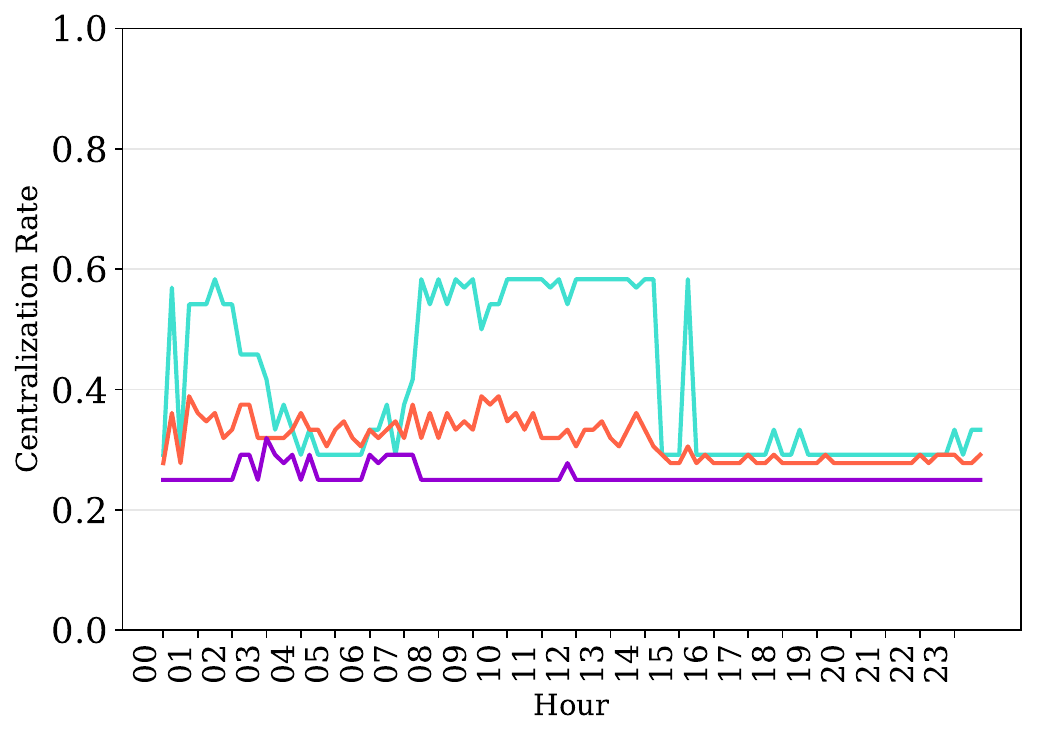}%
        \label{fig:centr_hier_bi}
    }
    \subfloat[Mesh - Bi-Objective]{%
        \includegraphics[width=0.47\columnwidth]{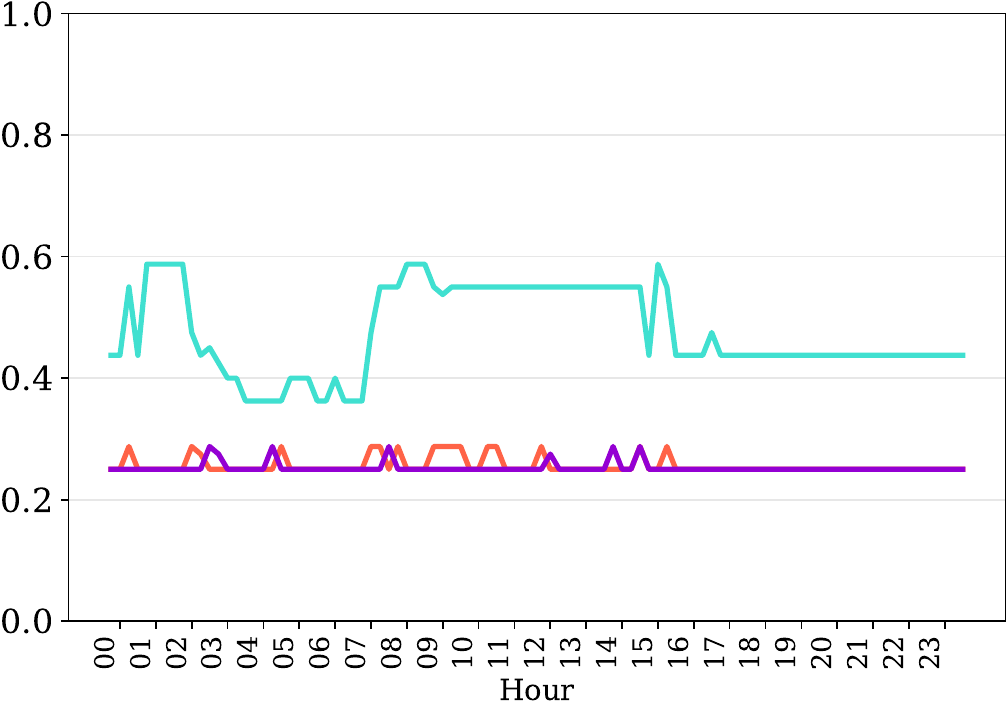}%
        \label{fig:centr_mesh_bi}
    }
    \caption{Centralization ratio per slice.}
    \label{fig:centr_rate_slices}
\end{figure}

\begin{figure*}[!t]
    \centering
    \subfloat[Energy Minimization]{%
        \includegraphics[width=0.96\textwidth]{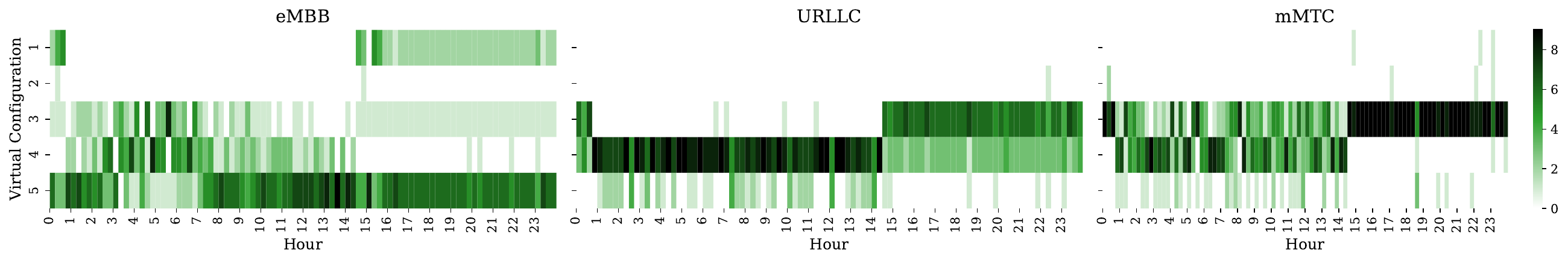}%
        \label{fig:hierarchical_vc_energy}
    }
    \par  
    \subfloat[Latency Minimization]{%
        \includegraphics[width=0.96\textwidth]{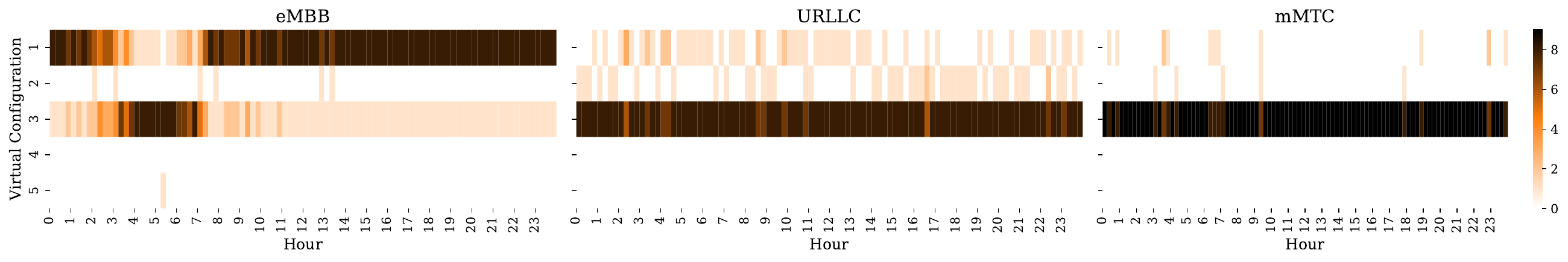}%
        \label{fig:hierarchical_vc_latency}
    }
    \par 
    \subfloat[Bi-Objective Minimization]{%
        \includegraphics[width=0.96\textwidth]{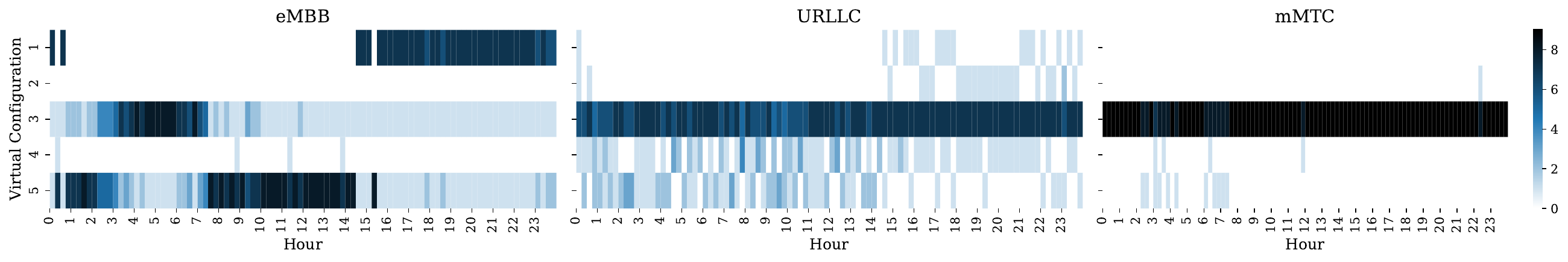}%
        \label{fig:hierarchical_vc_bi}
    }
    \caption{VC Selection for Hierarchical Topology}
    \label{fig:vc_selection_hierarchical}
\end{figure*}

\begin{figure*}[!t]
    \centering
    \subfloat[Energy Minimization]{%
        \includegraphics[width=0.96\textwidth]{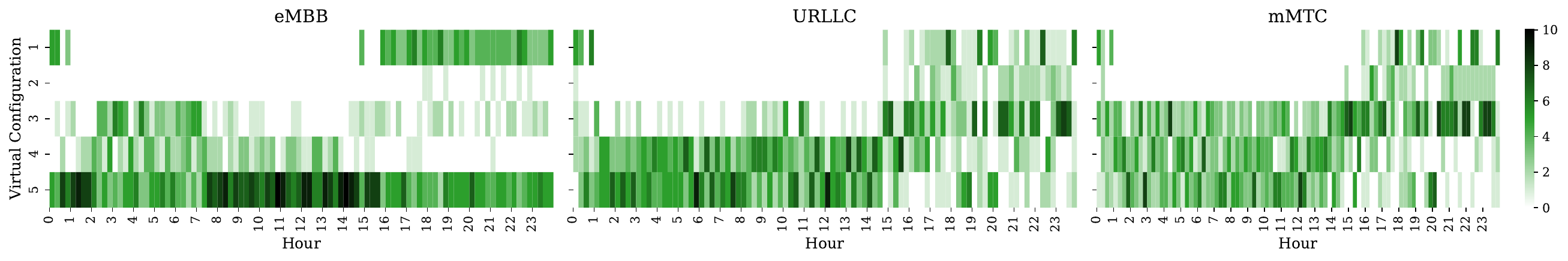}%
        \label{fig:mesh_vc_energy}
    }
    \par  
    \subfloat[Latency Minimization]{%
        \includegraphics[width=0.96\textwidth]{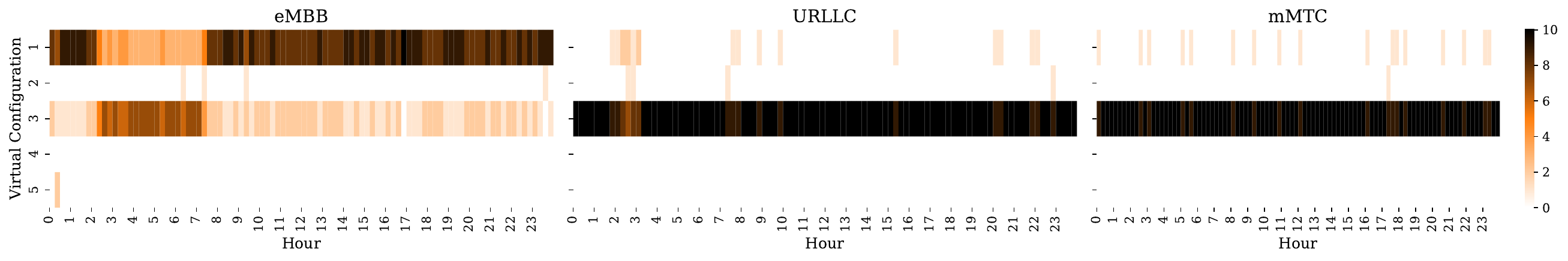}%
        \label{fig:mesh_vc_latency}
    }
    \par 
    \subfloat[Bi-Objective Minimization]{%
        \includegraphics[width=0.96\textwidth]{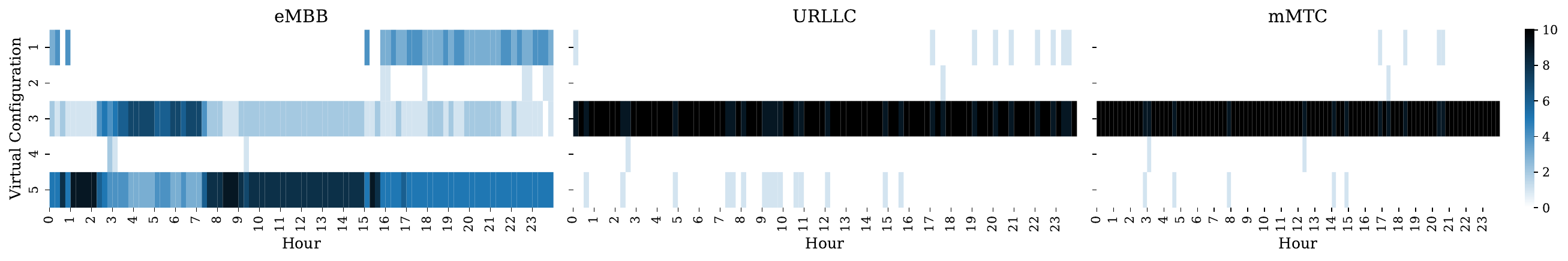}%
        \label{fig:mesh_vc_bi}
    }
    \caption{VC Selection for Mesh Topology}
    \label{fig:vc_selection_mesh}
\end{figure*}

The associated VC selection patterns are further illustrated in \cref{fig:hierarchical_vc_energy,fig:mesh_vc_energy}. In the \ac{HT}, the optimizer predominantly selects VC Options 5 and 4 during low- and medium-demand periods (1h--14h). As demand increases (15h--0h), the selection shifts toward Options 5 and 1 for \ac{eMBB} slices, while Options 4 and 3 are preferred for \ac{URLLC} and \ac{mMTC} traffic. In the \ac{MT}, VC Options 1 and 5 are more frequently selected for \ac{eMBB} during high-demand periods, whereas \ac{URLLC} and \ac{mMTC} slices exhibit selections spanning the complete set of VC Options, consistent with the higher centralization variability observed in the \ac{MT}.

Under the \acs{LMOF}, both topologies exhibit consistently low centralization ratio, as RAN functions are preferentially placed closer to — or directly at — the cell site, in accordance with VC Options 1 and 3. In the \ac{MT}, occasional peaks are observed when \ac{eMBB} slice functions are shifted to the CU due to resource saturation at the RU or DU.

Under the \acs{BOMF}, the centralization ratio exhibits a broadly similar overall behavior in both topologies, while revealing distinct patterns across three demand regimes.

During the low-demand period (1h–8h), the centralization ratio remains low, primarily due to the frequent selection of VC~3, especially by \ac{URLLC} and \ac{mMTC} slices. The relatively modest computational requirements of these slices enable most RAN functions to be deployed at the cell site, while the remaining functions are centralized on a single high-capacity PP.

During the medium-demand interval (8h–15h), the centralization ratio reaches its peak. The more centralized VC~5 configuration becomes increasingly prevalent for \ac{eMBB} slices, whereas \ac{URLLC} and \ac{mMTC} slices continue to predominantly adopt VC~3. The lower computational requirements of \ac{URLLC} and \ac{mMTC} allow these slices to remain deployed at the cell site, while the higher processing demands of \ac{eMBB} traffic drive centralization toward the CU.

During the high-demand period (15h–1h), the centralization ratio reaches its lowest level in the \ac{HT} and decreases noticeably in the \ac{MT}. This behavior is driven by the combined selection of VC~5 and VC~1 for \ac{eMBB} slices, alongside VC~3 for \ac{URLLC} and \ac{mMTC} slices. The adoption of VC~1 is primarily motivated by the increased computational requirements associated with higher traffic demand, necessitating the activation of additional processing resources.

Furthermore, \cref{fig:centr_hier_bi,fig:centr_mesh_bi} present the per-slice centralization ratio under the \acs{BOMF} for both topologies. These results confirm that \ac{eMBB} slices tend to require more centralized computational resources, whereas \ac{URLLC} and \ac{mMTC} slices are more amenable to distributed deployments and frequently remain at the RU due to their lower processing demands.

\begin{figure*}[t]
    \centering
    \includegraphics[width=1\linewidth]{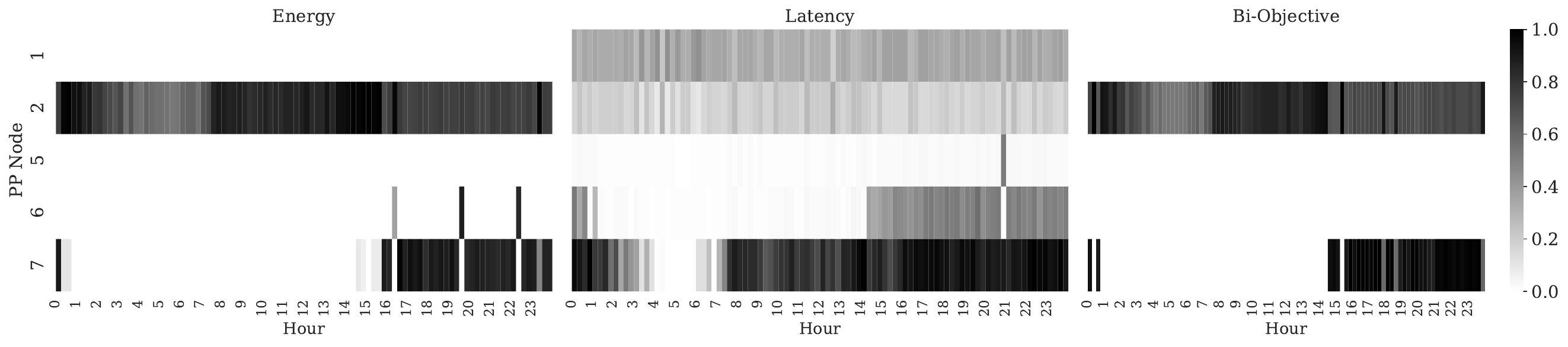}
    \caption{PP computing utilization for the Hierarchical Topology}
    \label{fig:PP_processing_hierarchical}
\end{figure*} 

\begin{figure*}[t]
    \centering
    \includegraphics[width=1\linewidth]{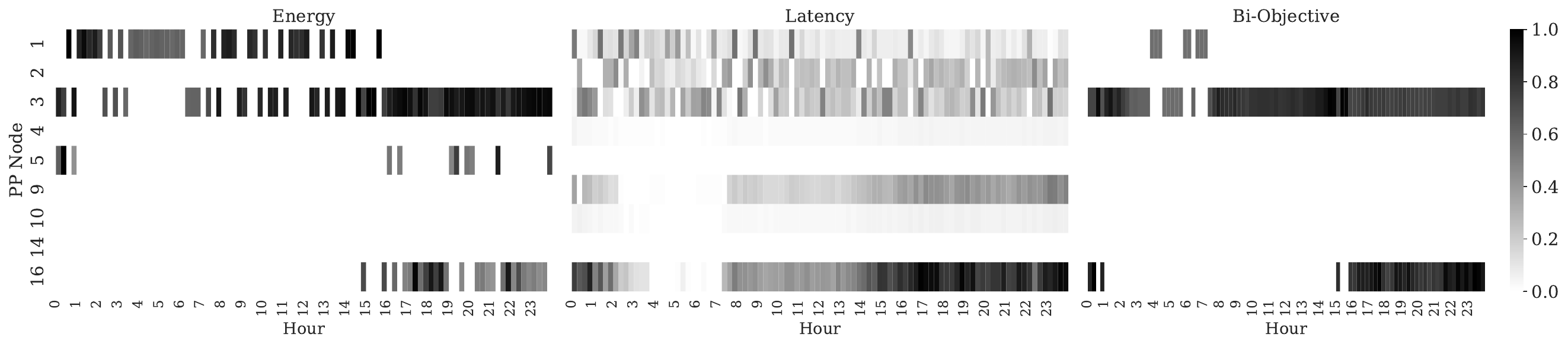}
    \caption{PP computing utilization for the Mesh Topology}
    \label{fig:PP_processing_mesh}
\end{figure*} 
\subsection{COMPUTING RESOURCE UTILIZATION}
The allocation of processing load across PP nodes reveals how each optimization objective exploits computational resources under different traffic demand conditions. Figure~\ref{fig:PP_processing_hierarchical} and Fig.~\ref{fig:PP_processing_mesh} illustrate the PP computational utilization over time for the \ac{HT} and \ac{MT}, respectively, under the three objective functions.

Under the \acs{EMOF}, the \ac{HT} relies on a single high-capacity PP to accommodate most of the processing load during low- and medium-demand periods. As demand approaches peak levels, a second PP is activated to offload the DU, thereby reducing the utilization of the primary node. This behavior aligns with the \acs{EMOF} objective, which prioritizes minimizing the number of active nodes while leveraging the higher computational capacity of central PPs before activating additional, lower-capacity resources.

In the \ac{MT}, a similar pattern is observed in terms of active node count: a single PP suffices under low-demand conditions, while two PPs are required during peak demand. However, due to the greater variability in centralization, the utilization of the high-capacity PP exhibits noticeable fluctuations, alternating with that of lower-capacity PPs. This behavior reflects a more dynamic and distributed allocation of processing resources across the mesh topology.

Under the \acs{LMOF}, nearly all PP nodes remain active in both topologies, as illustrated in Fig.~\ref{fig:PP_processing_mesh} and Fig.~\ref{fig:PP_processing_hierarchical}. This results in higher overall energy consumption compared to the \acs{EMOF}, primarily because VC~1 requires activating three PPs to support the full RAN protocol stack, as further evidenced by Fig.~\ref{fig:energy_consumption_hierarchical} and Fig.~\ref{fig:energy_component_mesh}.

Under the \acs{BOMF}, PP utilization reflects the three demand regimes identified in the centralization ratio analysis. During the low-demand period (1h–8h), only one PP node remains active in both topologies, consistent with the predominant selection of VC~3 and the consequent deployment of most RAN functions at the cell site. During the medium-demand interval (8h–15h), PP utilization increases as \ac{eMBB} functions are shifted toward the CU. During the high-demand period (15h–1h), a lower-capacity PP located closer to the cell site is activated — node~7 in the \ac{HT} and node~16 in the \ac{MT} — to accommodate the additional load while mitigating overall energy consumption.


\subsection{NETWORK UTILIZATION}
\begin{figure}[!t]
    \centering

    \includegraphics[width=0.55\textwidth]{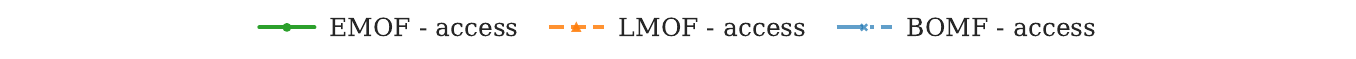}
    \subfloat[Hierarchical topology]{%
        \includegraphics[width=0.48\columnwidth]{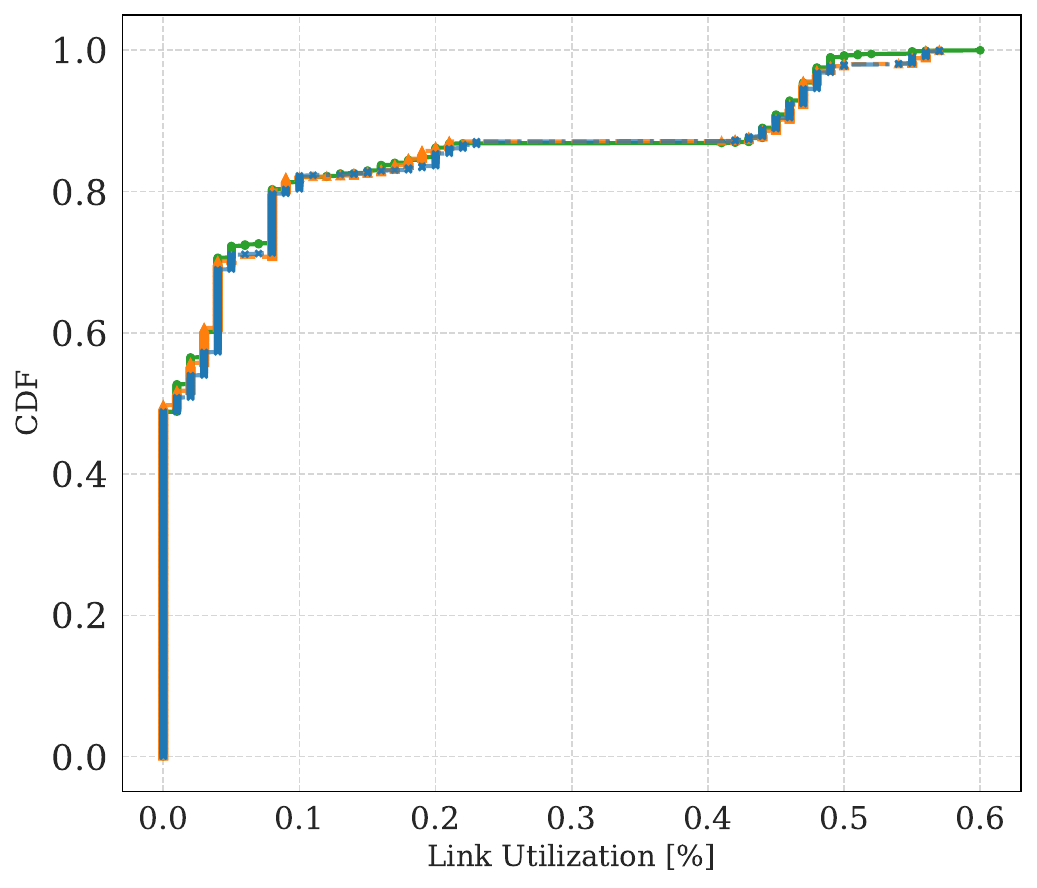}%
        \label{fig:ecdf_link_access_hierarchical}
    }
    \hfill
    \subfloat[Mesh topology]{%
        \includegraphics[width=0.48\columnwidth]{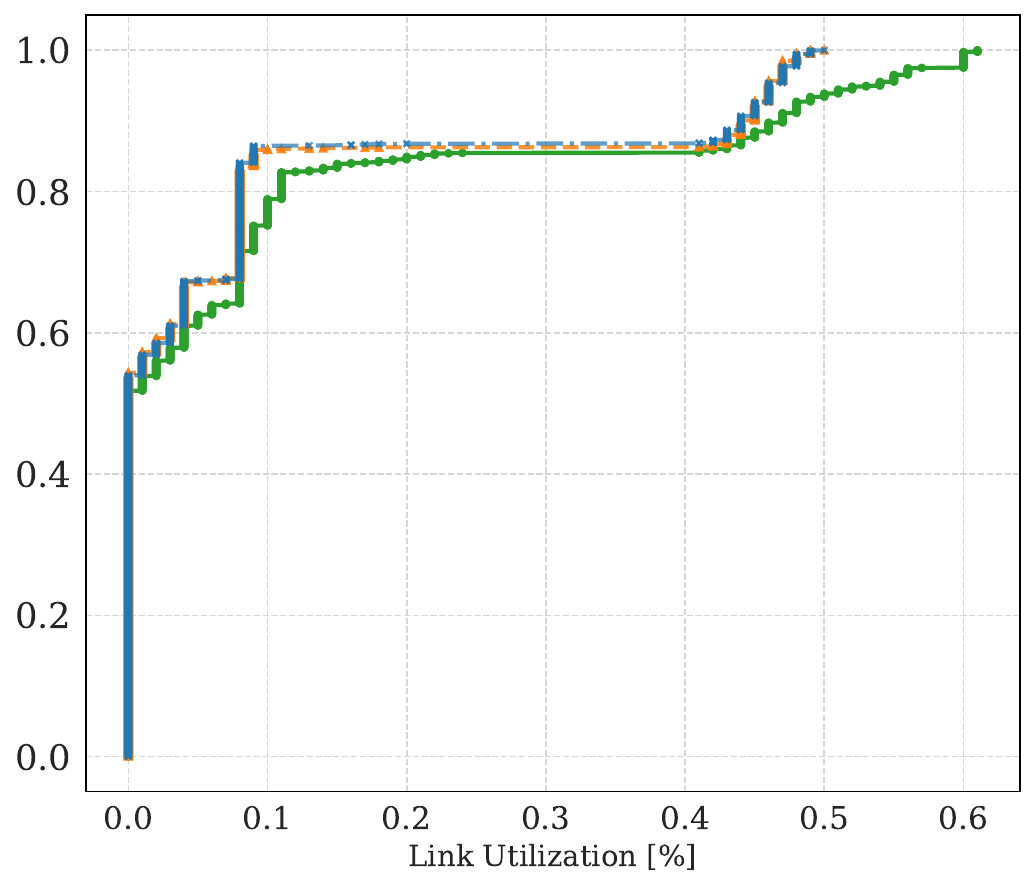}%
        \label{fig:ecdf_link_access_mesh}
    }

    \caption{Access link utilization for Hierarchical  and Mesh topologies under different optimization objectives.}
    \label{fig:ecdf_link_access}
\end{figure}

\begin{figure}[!t]
    \centering

    \includegraphics[width=0.55\textwidth]{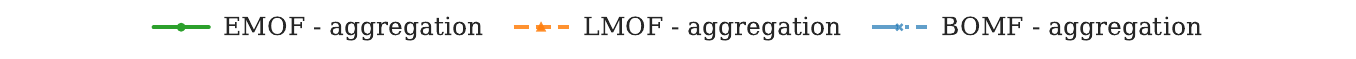}
    \subfloat[Hierarchical topology]{%
        \includegraphics[width=0.48\columnwidth]{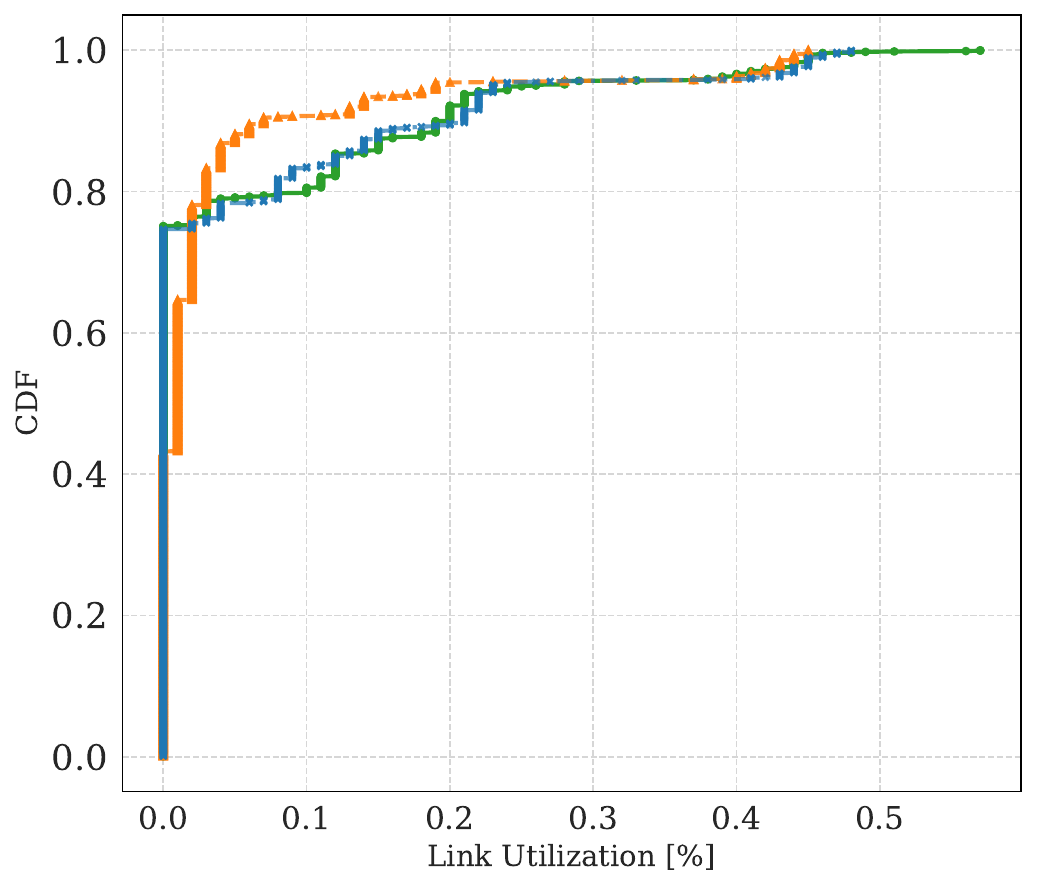}%
        \label{fig:ecdf_link_aggregation_hierarchical}
    }
    \hfill
    \subfloat[Mesh topology]{%
        \includegraphics[width=0.48\columnwidth]{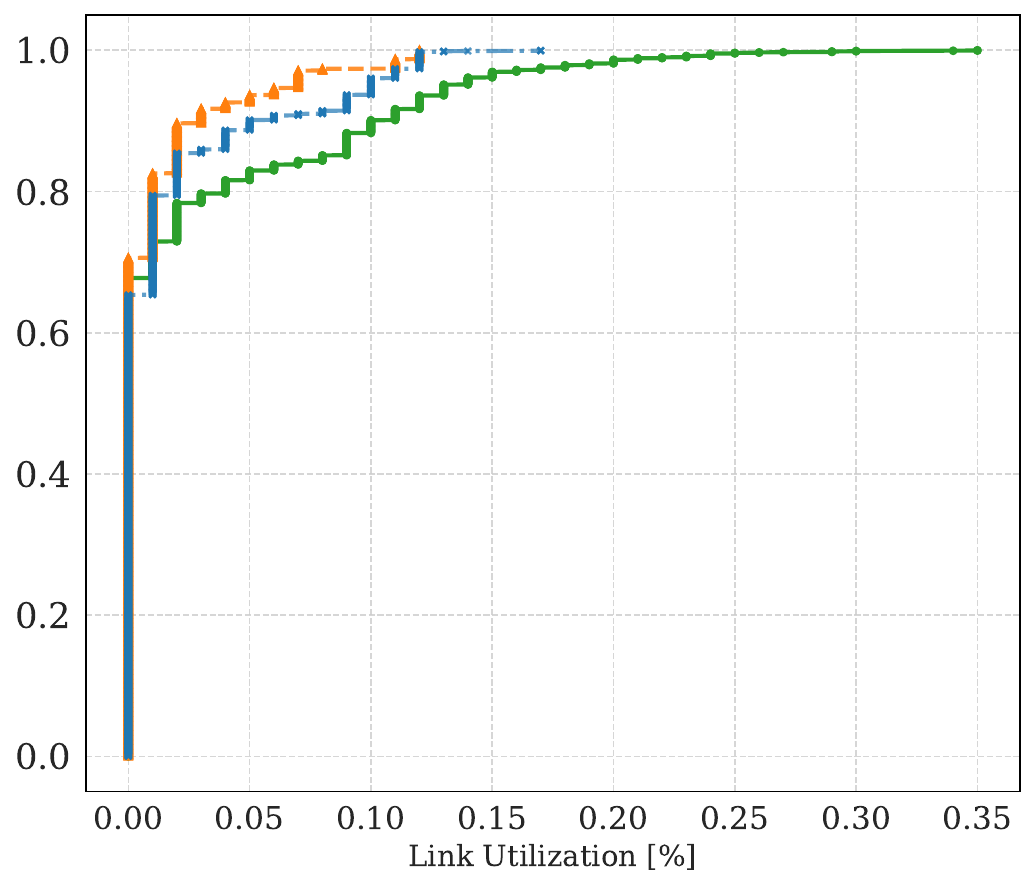}%
        \label{fig:ecdf_link_aggregation_mesh}
    }

    \caption{Aggregation link utilization for Hierarchical and Mesh topologies under different optimization objectives.}
    \label{fig:ecdf_link_aggregation}
\end{figure}

The network utilization analysis complements the energy and computing-resource evaluations by examining how VC selection, topology, and processing centralization shape traffic distribution across the transport network.
Link utilization is evaluated for each physical transport segment, distinguishing access links (cell–edge) in \cref{fig:ecdf_link_access} from aggregation links (edge–regional) in \cref{fig:ecdf_link_aggregation}, as illustrated in \cref{fig:system_model}. Utilization is defined as the ratio between carried traffic and link capacity, expressed as a percentage. The results are presented as CDFs, obtained by aggregating all links over all time slots for the three objective functions.

For the access segment, \cref{fig:ecdf_link_access_hierarchical,fig:ecdf_link_access_mesh} show that both topologies operate under light load conditions, with the CDF tails remaining below 0.6 utilization. A substantial fraction of access links remains unused, approximately 50\% in the \acs{HT} and slightly more in the \acs{MT}, mainly because ring interconnections between access routers are inactive for most of the time. \acs{EMOF} exhibits a marginally heavier right tail because it frequently selects VC~1 and VC~5, which rely on FSO~7.2x, consequently increasing FH bandwidth consumption due to encapsulation and coding overhead. In contrast, \acs{LMOF} and \acs{BOMF} often select FSO~2 (associated with VC~3), whose bandwidth demand closely follows the service traffic profile, resulting in lower access-link utilization.

For the aggregation segment, \cref{fig:ecdf_link_aggregation_hierarchical,fig:ecdf_link_aggregation_mesh} show that both topology and objective function significantly influence link-utilization behavior. The hierarchical topology presents higher aggregation-link utilization, whereas the mesh topology produces more compact utilization distributions and a clearer separation among the objective functions. Nevertheless, in both topologies, a large proportion of aggregation links remains inactive, corresponding to nearly 80\% in the \acs{HT} and approximately 60\% in the \acs{MT}. In hierarchical topologies, this behavior is primarily attributed to their tree-like structure, which offers limited routing diversity. Consequently, a small subset of aggregation links consistently carries most of the traffic, while many others remain idle.

Across both topologies, \acs{EMOF} consistently exhibits the heaviest right tail in the aggregation-link CDFs. This behavior is mainly associated with two factors: i) the preference for more centralized VC selections, as previously discussed; and ii) the tendency of \acs{EMOF} to consolidate traffic onto a reduced set of links to minimize the number of active nodes and interfaces.

By comparison, \acs{LMOF} achieves lower aggregation-link utilization, reducing peak utilization by approximately 10\% relative to \acs{EMOF} in the hierarchical topology and by nearly 25\% in the mesh topology. Furthermore, in the hierarchical topology, \acs{LMOF} activates more aggregation links because placing DUs closer to the RUs requires additional intermediate PPs and inter-site transport links; consequently, only about 40\% of aggregation links remain inactive.

As expected, \acs{BOMF} achieves intermediate behavior between the two extreme objectives by combining centralized and distributed VC configurations across RUs and time periods. Overall, these results highlight the inherent trade-off between traffic concentration and energy efficiency: objective functions favoring stronger centralization reduce the number of active resources but drive aggregation links closer to saturation.


\section{CONCLUSION}
\label{sec:Conclusion}
In this paper, we address the configuration of the RAN to minimize overall energy consumption while meeting stringent bandwidth and latency requirements. The proposed approach incorporates a detailed energy consumption model for both PPs and switching devices, ensuring compliance with service constraints. The problem is formulated as an \acs{MILP}  model, alongside a computationally efficient heuristic that jointly optimizes VC selection, VNF placement, and per-slice routing decisions.

The performance evaluation considers three representative network slice categories— \acs{eMBB}, \acs{URLLC}, and \acs{mMTC} — each characterized by distinct bandwidth and latency demands. Furthermore, FH and MH latency constraints are incorporated, accounting for both queuing and self-queuing delays in accordance with TSN standards.

The proposed heuristic achieves near-optimal energy efficiency while significantly reducing computational complexity compared to the MILP formulation. Extensive simulations are conducted across multiple network topologies and radio configurations to provide comprehensive insights into the impact of VC selection on RAN configuration. Two distinct topology classes are examined, and the MILP model is evaluated under three objective functions: \textit{(i)} energy minimization, \textit{(ii}) FH latency minimization, and \textit{(iii)} a combined bi-objective formulation.

The results indicate that VC selection significantly impacts computing resource and network utilization, and overall energy consumption. Furthermore, the interaction between VC selection and network topology exhibits distinct behavior depending on the optimization objective.

Under energy minimization conditions, RAN functions are predominantly centralized. In a hierarchical topology, this centralization remains prevalent; however, slices with higher processing demands frequently adopt dual-split VC, enabling the distribution of CU and DU workloads across two processing nodes. This allocation is typically governed by constraints such as inter-node distance and PP capacity.
In a mesh topology, RAN functions also exhibit a high degree of centralization, with deployments often placed on PPs near RUs. 
Furthermore, a subset of disaggregated gNBs, particularly those operating at 20 and 40 MHz bandwidths, tend to select VC 3, in which DU and RU functions are co-located. This configuration reflects their relatively modest processing requirements and supports efficient local execution.

In contrast, FH latency minimization yields markedly different behavior than energy-oriented optimization, resulting in a more disaggregated placement of RAN functions. Under this objective, most functions are deployed at the cell site using VCO 3. As traffic demand increases, the DU workload is progressively offloaded to an additional processing node, enabling improved latency performance through load distribution. While this approach reduces transport latency, it increases the computational burden on the PPs and results in comparatively low utilization of RU resources.

Finally, the bi-objective optimization achieves a balanced trade-off between energy consumption and FH latency, effectively reconciling computing resource utilization with latency constraints. The resulting solutions significantly reduce energy consumption relative to FH latency minimization while remaining closely aligned with the levels attained under pure energy minimization.
Across both topological configurations, VCO 3 emerges as the most frequently selected option, particularly for RUs operating at 20 and 40 MHz. In contrast, RUs with 100 MHz bandwidth requirements tend to centralize their VNFs under low demand, then transition to a more distributed deployment—offloading to an independent DU—as demand increases.

We analyze how the network dynamically reconfigures to satisfy slice demands while minimizing energy consumption. The proposed bi-objective formulation achieves a balanced trade-off between energy efficiency and FH latency reduction. However, this benefit comes at the cost of increased complexity at the RUs, which must be explicitly accounted for. Demand levels directly influence computational load and, consequently, the selection of VCs, as variations in channel bandwidth impact both traffic demand and the associated processing requirements. Importantly, the model satisfies latency constraints across all evaluated scenarios. These results highlight the need for adaptive RAN configurations that dynamically adjust to varying conditions to meet performance targets while optimizing energy usage.

\bibliographystyle{IEEEtran}
\bibliography{references}

\begin{IEEEbiography}[{\includegraphics[width=1in,height=1.25in,clip,keepaspectratio]{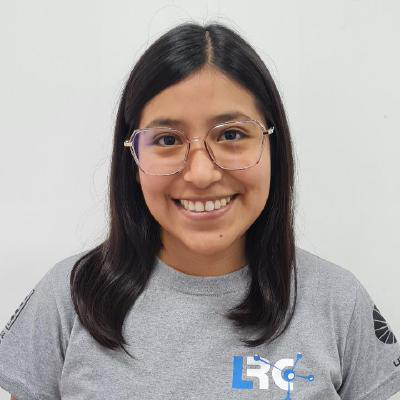}}]{Gabriela N. Caspa H. } (Student, IEEE) received the B.Sc. degree in Telecommunications Engineering from the Universidad Católica Boliviana “San Pablo,” in 2019. She is currently a Ph.D. student in Computer Science at the State University of Campinas (UNICAMP), Brazil. Her research interests include reconfigurable 5G RANs, network slicing, and linear programming. 
\end{IEEEbiography}

\begin{IEEEbiography}[{\includegraphics[width=1in,height=1.25in,clip,keepaspectratio]{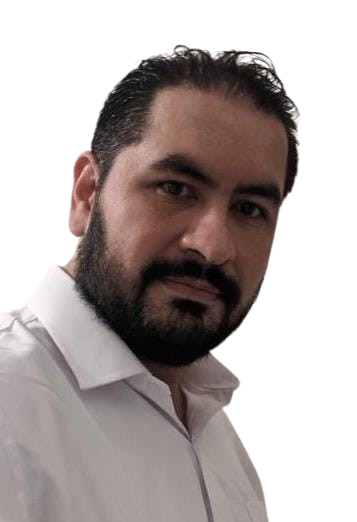}}]{Carlos A. Astudillo } obtained his Ph.D. degree in Computer Science from the University of Campinas (UNICAMP), Brazil, in 2022. He is an assistant professor at the Institute of Computing (IC), UNICAMP. He received the best paper award at the IEEE ISCC and the TAOS SAC-ICC, as well as top honors in thesis and dissertation contests in Brazil and Latin America. His research interests include network protocols and resource management in 5G/6G networks.



\end{IEEEbiography}

\begin{IEEEbiography}[{\includegraphics[width=1in,height=1.25in,clip,keepaspectratio]{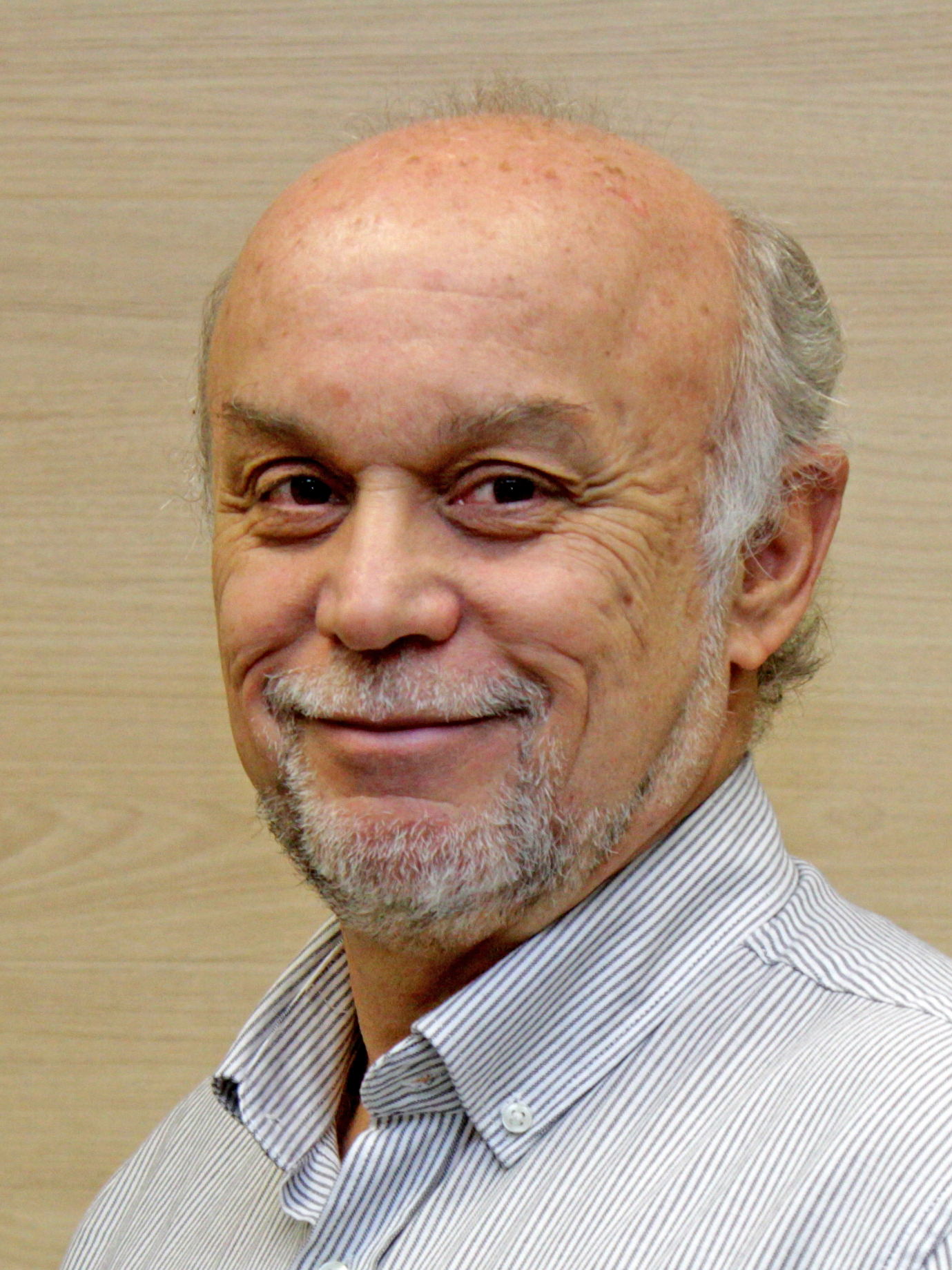}}]{Nelson L. S. da Fonseca} (M'88--SM'01) received the Ph.D. degree in computer engineering from the University of Southern California, Los Angeles, CA, USA, in 1994. He is currently a Full Professor with the Institute of Computing, State University of Campinas, Campinas, Brazil. He has authored or co-authored more than 450 papers and supervised over 80 graduate students. Nelson Fonseca is currently the IEEE ComSoc President. He served as VP Conferences, VP of Technical and Educational Activities, VP of Publications, VP Member Relations, the Director of Conference Development, Latin America Region, and Online Services. He is the former Editor-in-Chief of the IEEE COMMUNICATIONS SURVEYS AND TUTORIALS. He was a recipient of the 2023 ComSoc Education Award, 2020 IEEE ComSoc Harold Sobol Award for Exemplary Service to Meetings \& Conferences, 2012 IEEE ComSoc Joseph LoCicero Award for Exemplary Service to Publications, the Medal of the Chancellor of the University of Pisa in 2007, and the Elsevier Computer Networks Journal Editor of the Year 2001 Award.
\end{IEEEbiography}

\end{document}